\documentclass[structabstract]{aa} 
\usepackage[utf8]{inputenc}
\usepackage[T1]{fontenc}
\usepackage{graphicx}
\usepackage{txfonts}
\usepackage{amsfonts}
\usepackage{natbib}
\usepackage{float}
\usepackage{lscape}
\usepackage{nicefrac}

\begin{document}
   \title{Shockingly low water abundances in \textit{Herschel} / PACS \\ 
   observations of low-mass protostars in Perseus}

   \author{A.~Karska\inst{1,2,3}, L.~E.~Kristensen\inst{4}, E.~F.~van Dishoeck\inst{1,2}, 
   M.~N.~Drozdovskaya\inst{2}, J.~C.~Mottram\inst{2}, G.~J.~Herczeg\inst{5}, S.~Bruderer\inst{1}, 
   S.~Cabrit\inst{6}, N.~J.~Evans~II\inst{7}, D.~Fedele\inst{1}, A.~Gusdorf\inst{8}, J.~K.~J{\o}rgensen\inst{9,10}, M.~J.~Kaufman\inst{11}, \\
   G.~J.~Melnick\inst{3}, D.~A.~Neufeld\inst{12}, B.~Nisini\inst{13}, G.~Santangelo\inst{13},
   M.~Tafalla\inst{14}, S.~F.~Wampfler\inst{9,10}}

	\institute{$^{1}$ Max-Planck Institut f\"{u}r Extraterrestrische Physik (MPE),
   			  Giessenbachstr. 1, D-85748 Garching, Germany\\
   			  $^{2}$ Leiden Observatory, Leiden University, P.O. Box 9513,
          	  2300 RA Leiden, The Netherlands\\
          	  $^{3}$ Astronomical Observatory, Adam Mickiewicz University, Słoneczna 36, PL-60-268 Poznań, Poland \\
          	  $^{4}$ Harvard-Smithsonian Center for Astrophysics, 60 Garden Street, 
          	  Cambridge, MA 02138, USA\\
          	  $^{5}$ Kavli Institut for Astronomy and Astrophysics, Yi He Yuan Lu 5, HaiDian Qu, Peking University, Beijing,
          	  100871, PR China\\
          	  $^{6}$ LERMA, UMR 8112 du CNRS, Observatoire de Paris, \'{E}cole Normale Sup\'{e}rieure,
          	  Universit\'{e} Pierre et Marie Curie, Universit\'{e} de Cergy-Pontoise, 61 Av. de l'Observatoire,
          	  75014 Paris, France\\
          	  $^{7}$ Department of Astronomy, The University of Texas at Austin, 2515 Speedway, Stop C1400, Austin, TX 78712-1205, USA \\
         	  $^{8}$ LERMA, UMR 8112 du CNRS, Observatoire de Paris, \'{E}cole Normale Sup\'{e}rieure, 24 rue Lhomond, 75231 Paris Cedex 05, France \\
         	  $^{9}$ Niels Bohr Institute, University of Copenhagen, Juliane Maries Vej 30, DK-2100 Copenhagen {\O}., Denmark\\
          	  $^{10}$ Centre for Star and Planet Formation, Natural History Museum of Denmark, University of
				Copenhagen, {\O}ster Voldgade 5-7, DK-1350 Copenhagen K., Denmark\\
          	  $^{11}$ Department of Physics, San Jose State University, One Washington Square, San Jose, CA 95192-0106 \\
          	  $^{12}$ Department of Physics and Astronomy, Johns Hopkins University, 3400 North Charles Street, Baltimore, MD 21218, USA \\
          	  $^{13}$ INAF - Osservatorio Astronomico di Roma, 00040 Monte Porzio Catone, Italy \\
          	  $^{14}$ Observatorio Astron\'{o}mico Nacional (IGN), Calle Alfonso XII,3. 28014, Madrid, Spain\\
             \email{agata.karska@gmail.com}
             }

   \date{Received May 9, 2014; accepted September 15, 2014}
	\titlerunning{Water deficit in low-mass YSOs in Perseus}
	\authorrunning{A.~Karska et al. 2014}

  \abstract
  {Protostars interact with their surroundings through jets and winds impacting on the envelope and creating shocks, but the nature of these shocks is still poorly understood.}
  {Our aim is to survey far-infrared molecular line emission from a uniform and significant sample of 
  deeply-embedded low-mass young stellar objects (YSOs) in order to characterize shocks 
  and the possible role of ultraviolet radiation in the immediate protostellar environment.}
  {Herschel/PACS spectral maps of 22 objects in the Perseus molecular cloud were obtained 
  as part of the `William Herschel Line Legacy' (WILL) survey. Line emission from H$_\mathrm{2}$O, CO, and OH
  is tested against shock models from the literature. }
  {Observed line ratios are remarkably similar and do not show variations with source physical 
  parameters (luminosity, envelope mass). Most ratios are also comparable to those found at off-source outflow positions. Observations show good agreement with the shock models when line ratios of the same species are compared. 
   Ratios of various H$_\mathrm{2}$O lines provide a particularly good diagnostic of pre-shock gas densities, 
    $n_\mathrm{H}\sim10^{5}$ cm$^{-3}$, 
   in agreement with typical densities obtained from observations of the post-shock gas when a compression
   factor of order 10 is applied (for non-dissociative shocks). The corresponding shock velocities, 
   obtained from comparison with CO line ratios, are above 20 km\,s$^{-1}$. 
   However, the observations consistently show one-to-two orders of magnitude lower 
   H$_\mathrm{2}$O-to-CO and H$_\mathrm{2}$O-to-OH line ratios than predicted by the existing shock models.}
  {The overestimated model H$_\mathrm{2}$O fluxes are most likely caused by an overabundance of
   H$_\mathrm{2}$O in the models since the excitation is well-reproduced. 
   Illumination of the shocked material by ultraviolet photons produced either in the 
   star-disk system or, more locally, in the shock, would decrease the 
   H$_\mathrm{2}$O abundances and reconcile the models with observations. Detections of hot H$_\mathrm{2}$O 
   and strong OH lines support this scenario.}
   

   \maketitle
\section{Introduction}
Shocks are ubiquitous phenomena where outflow-envelope interactions take place in 
young stellar objects (YSO). Large-scale shocks are caused by the bipolar jets and 
protostellar winds impacting the envelope along the `cavity walls' carved by the passage 
of the jet \citep{Ar07,Fr14}. This important interaction 
needs to be characterized in order to understand and quantify the feedback from protostars 
onto their surroundings and, ultimately, to explain the origin of the initial mass function, 
  disk fragmentation and the binary fraction. 

Theoretically, shocks are divided into two main types based on a combination of magnetic field 
strength, shock velocity, density, and level of ionization \citep{Dr80,Dr83,Hol89,Hol97}. 
In `continuous' ($C-$type) shocks, in the presence of a magnetic field and low ionization, 
the weak coupling between the ions and neutrals results in a continuous change in the gas parameters. 
Peak temperatures of a few $10^3$ K allow the molecules to survive the passage of the shock, 
which is therefore referred to as non-dissociative. In `jump' ($J-$type) shocks, physical 
conditions change in a discontinuous way, leading to higher peak temperatures than in $C$ shocks
 of the same speed and for a given density. Depending on the shock velocity, $J$ shocks are 
 either non-dissociative (velocities below about $\sim30$ km\,s$^{-1}$, peak temperatures 
 of about a few $10^4$ K) or dissociative (peak temperatures even exceeding 10$^5$ K), but the 
 molecules efficiently reform in the post-shock gas. 
 
Shocks reveal their presence most prominently in the infrared (IR) domain, where 
the post-shock gas is efficiently cooled by numerous atomic and molecular emission lines.
Cooling from H$_2$ is dominant in outflow shocks \citep{Ni10b,Gi11}, but its mid-IR emission 
is strongly affected by extinction in the dense envelopes of young protostars \citep{Gi01,Ni02,Da08,Ma09}. 
In the far-IR, rotational transitions of water vapor 
(H$_2$O) and carbon monoxide (CO) are predicted to play an important role in the
 cooling process \citep{GL78,ND89,Hol97} and can serve as a diagnostic of the shock type, its velocity,
  and the pre-shock density of the medium \citep{Hol89,KN96,FP10,FP12}. 
  
The first observations of the critical wavelength regime to test these models ($\lambda\sim45-200$ $\mu$m) were 
taken using the Long-Wavelength Spectrometer \citep[LWS,][]{LWS} onboard the \textit{Infrared Space Observatory}
\citep[ISO,][]{ISO}. Far-IR atomic and molecular emission lines were detected toward several 
low-mass deeply-embedded protostars \citep{Ni00,Gi01,EvDISO},
but its origin was unclear due to the poor spatial resolution of the telescope \citep[$\sim80''$,][]{Ni02,Ce02}.

The sensitivity and spectral resolution of the Photodetector Array Camera and Spectrometer 
\citep[PACS,][]{Po10} onboard \textit{Herschel} allowed a significant increase in the number of young 
protostars with far-IR line detections compared with the early ISO results and has revealed
rich molecular and atomic line emission both at the protostellar
\citep[e.g.][]{vK10,Go12,He12,Vi12,Gr13,Ka13,Li14,Ma12,Wa13} and at pure outflow positions 
\citep{Sa12,Sa13,Cod12,Lef12,Va12,Ni13}. 

The unprecedented spatial resolution of PACS allowed detailed imaging 
of L1157 providing firm evidence that most of far-IR H$_2$O emission originates in the outflows \citep{Ni10}.
A mapping survey of about 20 protostars revealed similarities between the spatial extent of H$_2$O
and high$-J$ CO \citep{Ka13}. Additional strong flux correlations between
  those species and similarities in the velocity-resolved profiles \citep{Kr10,Kr12,IreneCO,Sa14} suggest 
  that the emission from the two molecules arises from the same regions. This is further confirmed by finely-spatially sampled PACS maps 
  in CO 16-15 and various H$_\mathrm{2}$O lines in shock positions of L1448 and L1157
  \citep{Sa13,Ta13}. On the other hand, the spatial extent of OH resembles 
 the extent of [\ion{O}{i}] and, additionally, a strong flux correlation between the two species is found \citep{Ka13,Wa13}. 
 Therefore, at least part of the OH emission most likely originates in a dissociative $J-$shock, together with [\ion{O}{i}] \citep{Wa10,Be12,Wa13}.

To date, comparisons of the far-IR observations with shock models have
been limited to a single source or its outflow positions
\citep[e.g.][]{Ni99,Be12,Va12,Sa12,Di13,Lee13,FP13}. Even in these
studies, separate analysis of each species (or different pairs of
species) often led to different sets of shock properties that needed
to be reconciled. For example, \citet{Di13} show that CO and H$_2$
line emission in Serpens SMM3 originates from a 20 km\,s$^{-1}$ $J$ shock
at low pre-shock densities ($\sim10^4$ cm$^{-3}$), but the H$_2$O and
OH emission is better explained by a 30-40 km\,s$^{-1}$ $C$ shock. In
contrast, \citet{Lee13} associate emission from both CO and H$_2$O in
L1448-MM with a 40 km\,s$^{-1}$ $C$ shock at high pre-shock densities
($\sim10^5$ cm$^{-3}$), consistent with the analysis of the L1448-R4
outflow position \citep{Sa13}. Analysis restricted to H$_2$O lines
alone often indicates an origin in non-dissociative $J$ shocks
\citep{Sa12,Va12,Bu14}, while separate analysis of CO, OH, and atomic
species favors dissociative $J$ shocks \citep{Be12,Le12}.  The question
remains how to break degeneracies between these models and how
typical the derived shock properties are for young protostars. Also,
surveys of high-$J$ CO lines with PACS have revealed two universal temperature
components in the CO ladder toward all deeply-embedded low-mass
protostars \citep{He12,Go12,Gr13,Ka13,Kr13,Ma12,Lee13}. The question
is how the different shock properties relate with the `warm'
($T\sim300$ K) and `hot' ($T\gtrsim700$ K) components seen in the CO
rotational diagrams.

In this paper, far-IR spectra of 22 low-mass YSOs observed as part of the 
`William Herschel Line Legacy' (WILL) survey (PI: E.F. van Dishoeck) are
compared to the shock models from \citet{KN96} and \citet{FP10}. All sources are 
confirmed deeply-embedded YSOs located in the well-studied Perseus molecular cloud 
spanning the Class 0 and I regime
\citep{KS00,En06,Jo06,Jo07b,Ha07b,Ha07a,Da08,En09,Ar10}.  
H$_2$O, CO, and OH lines are analyzed together for this uniform sample 
 to answer the following questions: Do far-IR line observations agree with the shock 
 models? How much variation in observational diagnostics of shock conditions is found between 
 different sources? Can one set of shock parameters explain all molecular species and transitions? 
 Are there systematic differences between shock characteristics inferred using the CO lines 
 from the `warm' and `hot' components? 
How do shock conditions vary with the distance from the powering protostar?
   
This paper is organized as follows. Section 2 describes our source sample, instrument with 
adopted observing mode, and reduction methods. \S 3 presents the results 
of the observations: line and continuum maps, and the extracted spectra. \S 4 shows comparison 
between the observations and shock models. \S 5 discusses results obtained in \S 4 and 
\S 6 presents the conclusions. 

\begin{table*}
\centering
\caption{Catalog information and source properties}
\label{catalog}
\renewcommand{\footnoterule}{}  
\begin{tabular}{lccrrrclll}
\hline \hline
Object & R.A. & Decl. & $T_\mathrm{bol}$ & $L_\mathrm{bol}$ & 
$M_\mathrm{env}$ & Region & Other names \\
~ & ($^\mathrm{h}$ $^\mathrm{m}$ $^\mathrm{s}$) & ($^\mathrm{o}$ $'$ $''$) & (K) & 
($\mathrm{L}_\mathrm{\odot}$) & ($\mathrm{M}_\mathrm{\odot}$) & ~ & ~ \\
\hline
Per01 & 03:25:22.32  & +30:45:13.9    & 44  &  4.5 &  1.14 &   L1448 & Per-emb 22, L1448 IRS2, IRAS03222+3034, YSO 1\\
Per02 & 03:25:36.49  & +30:45:22.2     & 50  & 10.6   &  3.17 &   L1448 & Per-emb 33, L1448 N(A), L1448 IRS3, YSO 2\\
Per03\tablefootmark{a} & 03:25:39.12  & +30:43:58.2     & 47  &  8.4 &  2.56 &   L1448 & Per-emb 42, L1448 MMS, L1448 C(N), YSO 3\\
Per04 & 03:26:37.47  & +30:15:28.1     &  61  &  1.2 &  0.29 &   L1451 & Per-emb 25, IRAS03235+3004, YSO 4\\
Per05 & 03:28:37.09  & +31:13:30.8     & 85  & 11.1   &  0.35 &   NGC1333 & Per-emb 35, NGC1333 IRAS1, IRAS03255+3103, YSO 11\\
Per06 & 03:28:57.36  & +31:14:15.9   & 85  & 6.9   &  0.30 &   NGC1333 & Per-emb 36, NGC1333 IRAS2B, YSO 16\\
Per07 & 03:29:00.55  & +31:12:00.8    &  37  &  0.7 &  0.32 &   NGC1333 & Per-emb 3, HRF 65, YSO 18\\
Per08 & 03:29:01.56  & +31:20:20.6    & 131  & 16.8  &  0.86 &   NGC1333 & Per-emb 54, HH 12, YSO 19\\
Per09 & 03:29:07.78  & +31:21:57.3    & 128  & 23.2   &  0.24 &  NGC1333   & Per-emb 50 \\
Per10 & 03:29:10.68  & +31:18:20.6    &  45  &  6.9 &  1.37 &   NGC1333 & Per-emb 21, HRF 46, YSO 23\\
Per11\tablefootmark{b} & 03:29:12.06  & +31:13:01.7    &  28  &  4.4 &  5.42 &  NGC1333 & Per-emb 13, NGC1333 IRAS4B', YSO 25\\
Per12 & 03:29:13.54   & +31:13:58.2     & 31  &  1.1 & 1.30 &   NGC1333 & Per-emb 14, NGC1333 IRAS4C, YSO 26\\
Per13 & 03:29:51.82   & +31:39:06.0    &  40  &  0.7 & 0.51 &   NGC1333 & Per-emb 9, IRAS03267+3128, YSO 31\\
Per14 & 03:30:15.14   & +30:23:49.4    &  88  &  1.8 &  0.14 &   B1-ridge & Per-emb 34, IRAS03271+3013\\
Per15 & 03:31:20.98   & +30:45:30.1     & 35  &  1.7 &  1.29 &   B1-ridge & Per-emb 5, IRAS03282+3035, YSO 32\\
Per16 & 03:32:17.96   & +30:49:47.5     & 30  &  1.1 &  2.75 &   B1-ridge & Per-emb 2, IRAS03292+3039, YSO 33\\
Per17\tablefootmark{c} & 03:33:14.38   & +31:07:10.9     & 43  &  0.7 & 1.20 &   B1 & Per-emb 6, B1 SMM3, YSO 35\\
Per18\tablefootmark{c} & 03:33:16.44   & +31:06:52.5    &  25  &  1.1 &  1.22 &   B1 & Per-emb 10, B1 d, YSO 36\\
Per19 & 03:33:27.29   & +31:07:10.2    &  93  &  1.1 &  0.23 &   B1 & Per-emb 30, B1 SMM11, YSO 40\\
Per20 & 03:43:56.52   & +32:00:52.8    &  27  &  2.3 &  2.05 &   IC 348 & Per-emb 1, HH 211 MMS, YSO 44\\
Per21 & 03:43:56.84   & +32:03:04.7    &  34  &  2.1 &  1.88 &   IC 348 & Per-emb 11, IC348 MMS, IC348 SW, YSO 43\\
Per22 & 03:44:43.96   & +32:01:36.2    &  43  &  2.6 &  0.64 &   IC 348 & Per-emb 8, IC348 a, IRAS03415+3152, YSO 48\\
\hline
\end{tabular}
\tablefoot{Bolometric temperatures and luminosities are determined including the PACS continuum 
values; the procedure will be discussed in Mottram et al. (in prep.).
Numbered Per-emb names come from \citet{En09}, whereas the numbered YSO names come from 
\citet{Jo06} and were subsequently used in \citet{Da08}. Other source identifiers were compiled 
using \citet{Jo07b}, \citet{Re07}, \citet{Da08}, and \citet{Ve14}. Observations obtained in 2011 
which overlap with Per03 and Per11 are presented in \citet{Lee13} and \citet{He12}, respectively. 
\tablefoottext{a}{Tabulated values come from \citet{Gr13}, where full PACS spectra are obtained. 
The WILL values for $T_\mathrm{bol}$ and $L_\mathrm{bol}$ are 48 K and 8.0 $\mathrm{L}_\mathrm{\odot}$, 
respectively, within 5\% of those of \citet{Gr13}.}
\tablefoottext{b}{Tabulated values come from \citet{Ka13} and agree within 2\% 
of the values 29 K and 4.3 $\mathrm{L}_\mathrm{\odot}$ obtained here. We prefer to use the previously published 
values because the pointing of that observation was better centered on the IRAS4B.}
\tablefoottext{b}{The off-positions of Per17 and Per18 were contaminated by other continuum 
sources, and therefore the $T_\mathrm{bol}$ and $L_\mathrm{bol}$
are calculated here without the Herschel / PACS points.}
}
\end{table*}
\section{Observations}
All observations presented here were obtained as part of the 
`William Herschel Line Legacy' (WILL) OT2 program on \textit{Herschel} (Mottram et al. in prep.). 
The WILL survey is a study of H$_\mathrm{2}$O lines and related species with PACS and the
Heterodyne Instrument for the Far-Infrared \citep[HIFI,][]{dG10} toward an unbiased
flux-limited sample of low-mass protostars newly discovered in the recent {\em Spitzer} \citep[c2d,][]{Gu09,Gu10,Ev09}
and \textit{Herschel} \citep{An10} Gould Belt imaging surveys. Its main aim is to study 
the physics and chemistry of star-forming regions in a statistically
significant way by extending the sample of low-mass protostars observed in the `Water in star-forming regions with Herschel' 
\citep[WISH,][]{WISH} and `Dust, Ice, and Gas in Time' \citep[DIGIT,][]{Gr13} programs.

This paper presents the \textit{Herschel}/PACS spectra of 22 low-mass deeply-embedded YSOs 
located exclusively in the Perseus molecular cloud (see Table \ref{catalog}) to ensure the 
homogeneity of the sample (similar ages, environment, and distance). The sources 
were selected from the combined SCUBA and \textit{Spitzer}/IRAC and MIPS catalog of \citet{Jo07b} and \citet{En09}, 
and all contain a confirmed embedded YSO \citep[Stage 0 or I,][]{Ro06,Ro07} in the center.

The WILL sources were observed using the line spectroscopy mode on
PACS which offers deep integrations and finely sampled spectral
resolution elements (minimum 3 samples per FWHM depending on the
grating order, PACS Observer's
Manual\footnote{http://herschel.esac.esa.int/Docs/PACS/html/pacs\_om.html})
over short wavelength ranges (0.5--2 $\mu$m). The line selection was
based on the prior experience with the PACS spectra obtained over the
full far-infrared spectral range in the WISH and DIGIT programs and is
summarized in Table \ref{lines}. Details of the observations of
the Perseus sources within the WILL survey are shown in Table \ref{log}.

PACS is an integral field unit with a 5$\times$5 array of spatial
pixels (hereafter \textit{spaxels}) covering a field of view of
$\sim47''\times47''$. Each spaxel measured $\sim9.4''\times9.4''$, or 2$\times$10$^{-9}$ sr,
and at the distance to Perseus \citep[$d=235$ pc,][]{Hir08} resolves emission down to $\sim$2,300 AU.
The total field of view is about 5.25$\times 10^{-8}$ sr and $\sim$11,000 AU.
The properly flux-calibrated wavelength ranges
include: $\sim$55--70 $\mu$m, $\sim$72--94 $\mu$m, and $\sim$105--187
$\mu$m, corresponding to the second ($<$ 100 $\mu$m) and first spectral orders ($>$ 100 $\mu$m).
Their respective spectral resolving power are $R\sim$2500--4500 (velocity resolution of $\Delta\varv\sim$70-120 km s$^{-1}$),
1500--2500 ($\Delta\varv\sim$120-200 km s$^{-1}$), and 1000--1500 ($\Delta\varv\sim$200-300 km s$^{-1}$). 
The standard chopping-nodding mode was used
with a medium (3$^\prime$) chopper throw. The telescope pointing
accuracy is typically better than 2$^\prime$$^\prime$ and can be evaluated to first
order using the continuum maps.

The basic data reduction presented here was performed using the
Herschel Interactive Processing Environment v.10
\citep[\textsc{hipe},][]{Ot10}. The flux was normalized to the telescopic
background and calibrated using observations of Neptune. Spectral
flatfielding within HIPE was used to increase the signal-to-noise
\citep[for details, see][]{He12,Gr13}.  The overall flux calibration
is accurate to $\sim 20\%$, based on the flux repeatability for
multiple observations of the same target in different programs,
cross-calibrations with HIFI and ISO, and continuum photometry.

Custom \textsc{idl} routines were used to extract fluxes using Gaussian fits
with fixed line width \citep[for details, see][]{He12}. The total,
5$\times$5 line fluxes were calculated by co-adding spaxels with
detected line emission, after excluding contamination from other
nearby sources except Per 2, 3, 10, and 18, where spatial separation
between different components is too small (see \S 3.1). For sources
showing extended emission, the set of spaxels providing the maximum
flux was chosen for each line separately. For point-like sources, the
flux is calculated at the central position and then corrected for the
PSF using wavelength-dependent correction factors (see PACS Observer's
Manual). Table \ref{det} shows the line detections toward  each source,
while the actual fluxes will be tabulated in the forthcoming paper for
all the WILL sources (Karska et al. in prep.). Fits cubes containing the 
spectra in each spaxel will be available for download in early 2015 at http://www.strw.leidenuniv.nl/WISH/.
\begin{figure*}[!tb]
  \begin{minipage}[t]{.33\textwidth}
  \begin{center}  
     \includegraphics[angle=90,height=6.5cm]{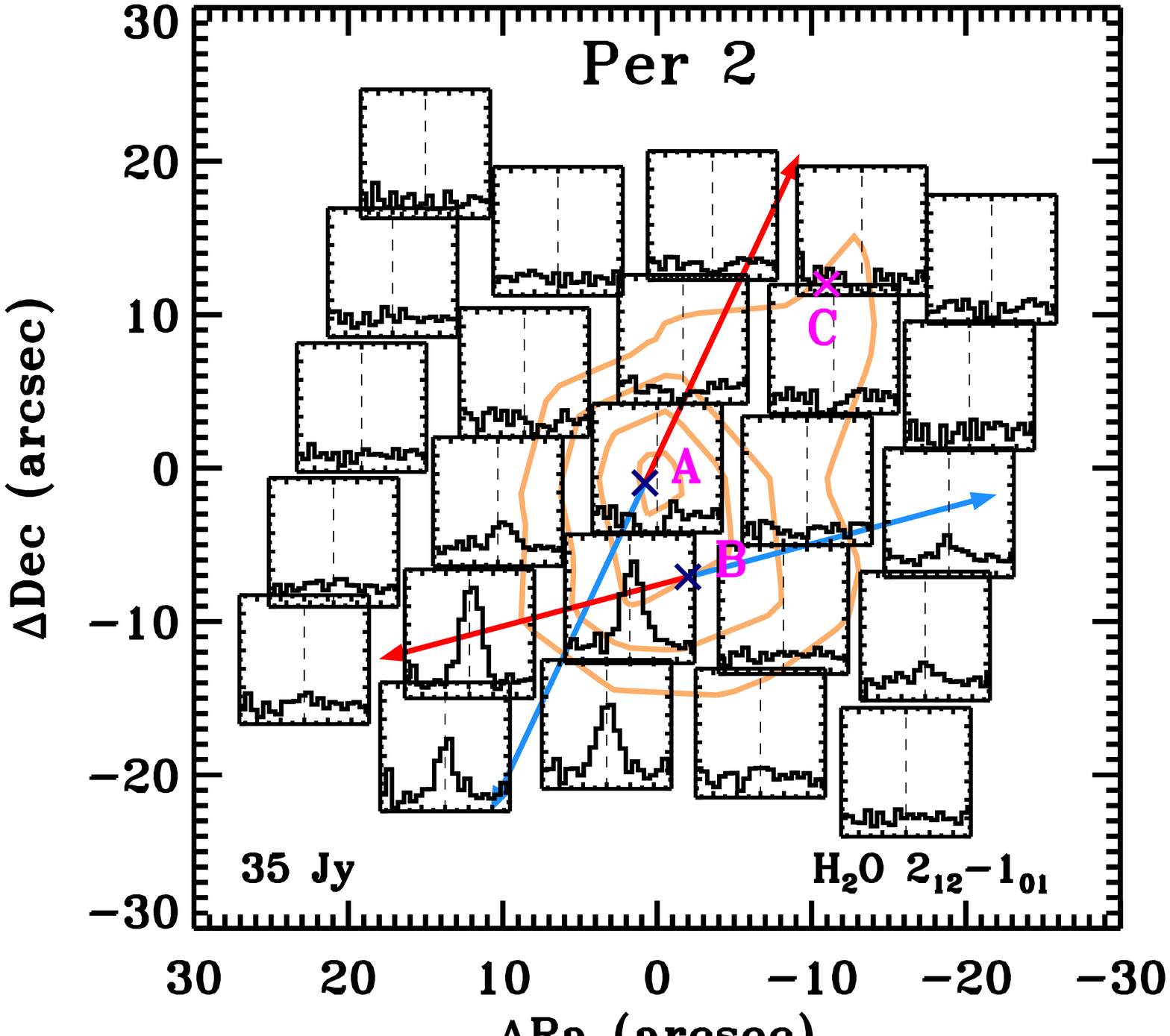}
                     \vspace{+0.5cm}
    \end{center}
  \end{minipage}
  \hfill
  \begin{minipage}[t]{.33\textwidth}
  \begin{center}         
     \includegraphics[angle=90,height=6.5cm]{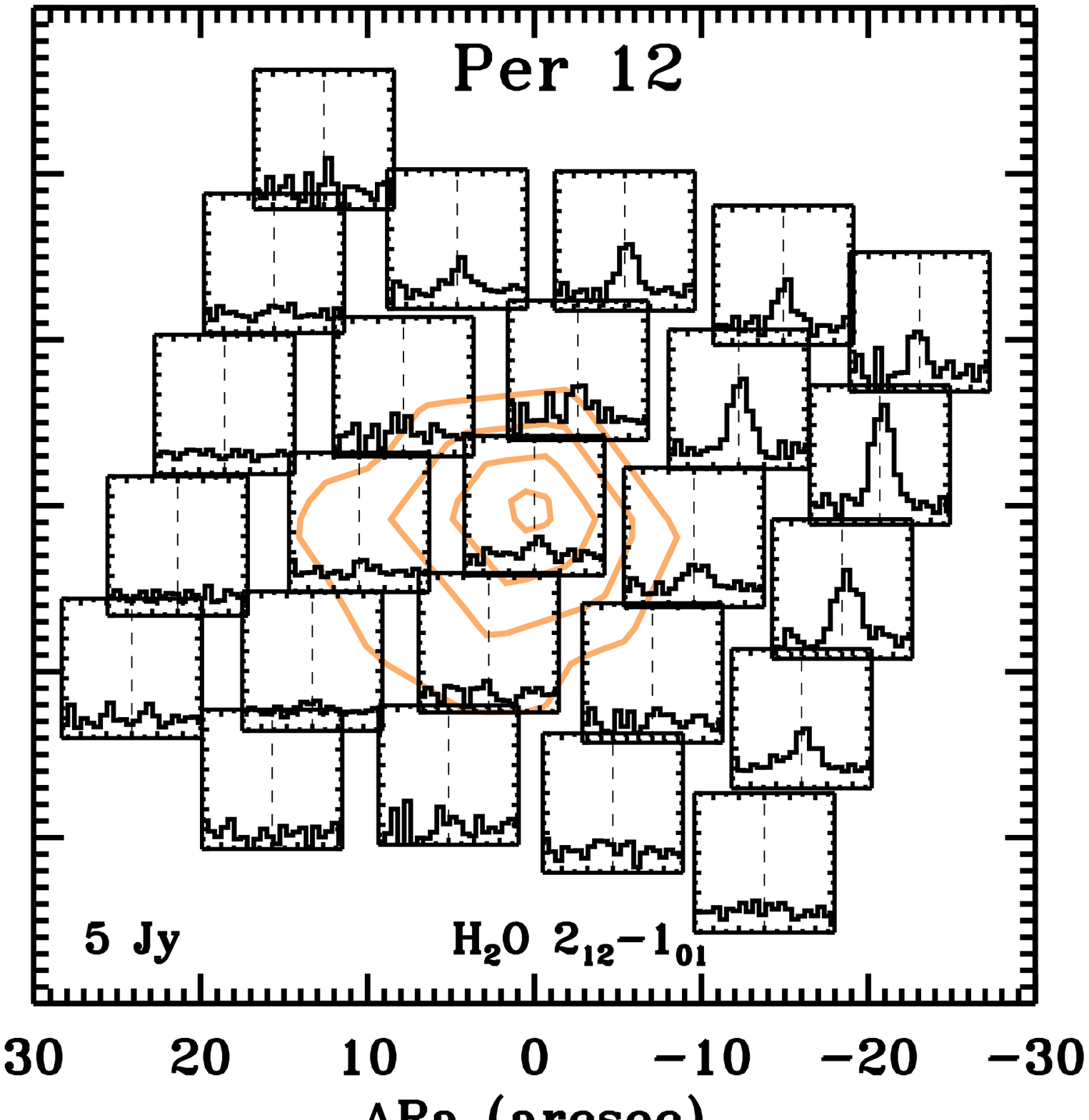}     
                     \vspace{+0.5cm}
    \end{center}
  \end{minipage}
    \hfill
  \begin{minipage}[t]{.33\textwidth}
  \begin{center}         
     \includegraphics[angle=90,height=6.5cm]{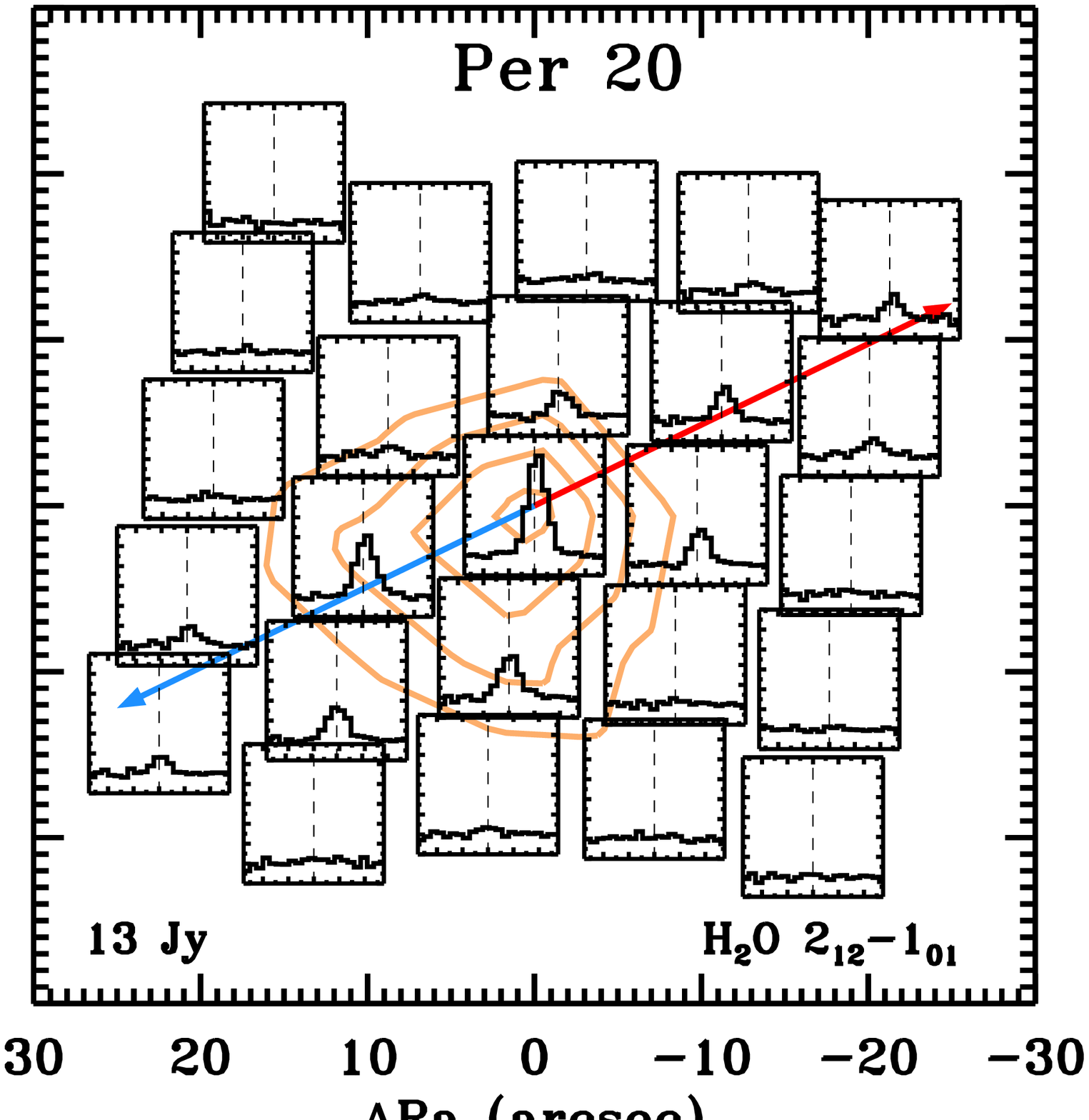}
                     \vspace{+0.5cm}
    \end{center}
  \end{minipage}
        \caption{\label{fig:maps}PACS spectral maps in the H$_2$O 2$_{12}$-1$_{01}$ 
        line at 179 $\mu$m illustrating sources with extended emission due to
        multiple sources in one field (Per 2), contamination by the outflow driven by another source 
        (Per 12), and associated with the targeted protostar (Per 20).
        Even though the emission on the maps seems to be extended in many sources in Perseus,
        the extended emission \textit{associated with the targeted protostar itself} 
        is detected only towards a few of them (see Table 2). The orange contours show continuum emission 
        at 30\%, 50\%, 70\%, and 90\% of the peak value written in the bottom left corner of each map.
        L1448 IRS 3A, 3B, and 3C sources and their CO 2--1 outflow directions are shown on the map of Per 2
        \citep{Kw06,Lo00}; the blue outflow lobe of L1448-MM also covers much of the observed field. 
        CO 2--1 outflow directions 
        of Per 20 / HH211 are taken from \citet{Gu99}. Wavelengths in  microns are translated to the velocity 
        scale on the X-axis using laboratory wavelengths (see Table \ref{lines}) of the
species and cover the range from -600 to 600 km\,s$^{-1}$. The Y-axis shows fluxes in Jy normalized to the 
    spaxel with the brightest line on the map in a range -0.2 to 1.2. }
\end{figure*}
\section{Results}
In the following sections, PACS lines and maps of the Perseus YSOs are
presented. Most sources in this sample show emission in just the
central spaxel. Only a few sources show extended emission and those
maps are compared to maps at other wavelengths to check for possible
contamination by other sources and their outflows. In this way, the
spaxels of the maps with emission originating from our objects are
established and line fluxes determined over those spaxels. The
emergent line spectra are then discussed.
\begin{figure*}[tb]
\begin{center}
\hspace{2cm}
\includegraphics[angle=0,height=26cm]{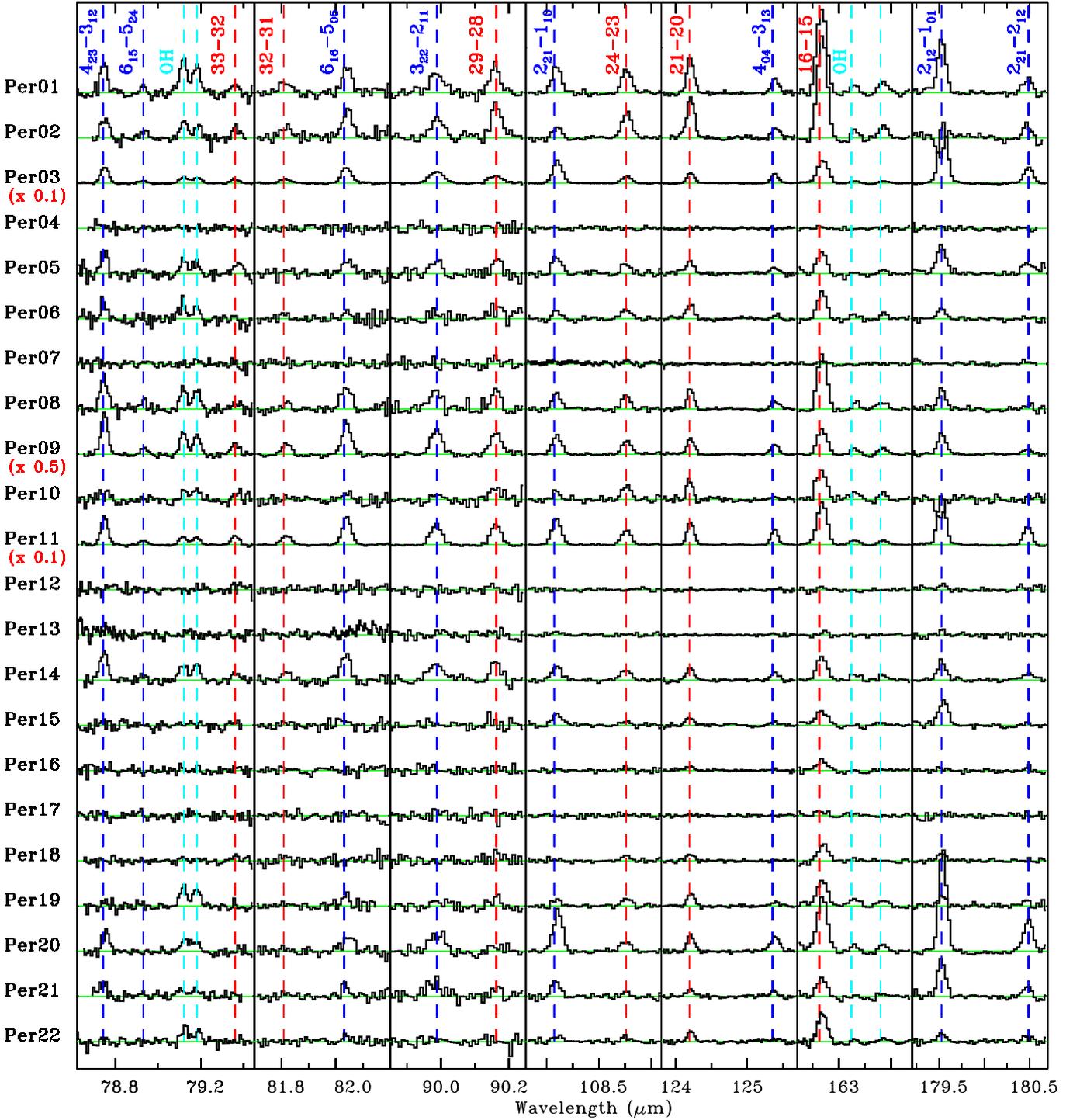}
\vspace{-6.5cm}

\caption{\label{spec} Line survey of deeply-embedded young stellar objects in Perseus 
at the central position on the maps. Spectra are continuum subtracted and not corrected 
for the PSF. Line identification of CO (red), H$_2$O (blue), and OH (light blue), are shown. 
Each spectrum is on a scale from 0 to 5 Jy in the y-axis, with the brightest sources
-- Per 3, Per 9, and Per 11 -- scaled down in flux density by a factor of 0.1, 0.5, and 0.1, respectively.}
\end{center}
\end{figure*}
\subsection{Spatial extent of line emission} 
\begin{table*}
\centering
\caption{Notes on mapped regions for individual sources}
\label{maps}
\renewcommand{\footnoterule}{}  
\begin{tabular}{l|cccc|cccc|ccc}
\hline \hline
Object  & \multicolumn{2}{c}{Continuum} & \multicolumn{2}{c}{Line} & \multicolumn{2}{c}{Sources} & \multicolumn{2}{c}{Outflows} & Remarks \\
   ~    & on & off &  on & off & single & multiple & single & multiple & \\
\hline
Per01 & X & - & X  & -    &  X & - & X & - & \\
Per02 & X & - & -  & X    &  - & X & - & Xc & three sources (3), contam. by Per03 (1,2) \\
Per03 & - & X & Xe  &      &  - & X & - & X & a binary (1,4)\\ 
Per04 & X & - & nd & nd   &  X & - & X & - & \\
Per05 & X & - & X  & -    &  X & - & X & - & \\
\hline
Per06 & - & X & - & X     &  - & X? & - & Xc & contam. by NGC1333 IRAS 2A in N-W (5)\\
Per07 & X & - & nd & nd     &  X & - & Xc & - & contam. in lines in N-W\\
Per08 & - & X & Xe & -     &  X & - & X & - & \\
Per09 & X & - & Xe & -     &  X? & - & X? & - & \\
Per10 & - & X & - & X    &  - & X & - & Xc & two sources (1,2), dominated by YSO 24 (1,2) \\
\hline
Per11 & - & X & Xe & -     &  - & X & - & Xc & NGC1333 IRAS4A in N-W (5)\\
Per12 & X & - & nd &  nd  &  X & - & Xc & - & contam. in lines by NGC1333 IRAS4A in N-W (5)\\
Per13 & X & - & nd &  nd  & X? & - & X? &  & \\
Per14 & X & - & X & -     & X? & - & X? &  & \\
Per15 & X & - & X & -     & X & - & X & - & \\
\hline
Per16 & X & - & nd &  nd     & X & - & X & - & \\
Per17 & - & X? & nd & nd & - & X & - & Xc? & Per18 in S-E, nodded on emission\\
Per18 & X & - & X & -     & X & - & - & Xc? & B1-b outflow? (2) \\
Per19 & X & - & X & -     & X & - & X & - & \\
Per20 & X & - & Xe & -     & X & - & X & - & \\
\hline
Per21 & X & - & Xe & -     & X? & - & X? & - & \\
Per22 & X & - & Xe & -     & X? & - & X? & - & \\
\hline
\end{tabular}
\tablefoot{
{Columns 2-5 indicate the location of the continuum at 179 $\mu$m and line emission peak of  
the H$_{2}$O 2$_{12}$-1$_{01}$ (179.527 $\mu$m) transition on the maps (whether on or off-center).
Columns 6-9 provide information about the number of sources and their 
outflows in the mapped region. Contamination by the outflows driven by sources 
outside the PACS field-of-view is mentioned in the last column. Extended emission associated with 
the targeted sources is denoted by 'e'. Non-detections are 
abbreviated by `nd', while `c' marks the contamination 
of the targeted source map by outflows from other sources. References: (1) \citet{Jo06}, 
(2) \citet{Da08}, (3) \citet{Lo00}, (4) \citet{Hi10,Lee13}, (5) Y{\i}ld{\i}z et al. (subm).}}
\end{table*}

 Table \ref{maps} provides a summary of the patterns of
 H$_\mathrm{2}$O 2$_{12}$-1$_{01}$ line and continuum emission at 179
 $\mu$m for all WILL sources in Perseus (maps are shown in
 Figs. \ref{specmap1} and \ref{specmap2}). The mid-infrared continuum
 and H$_\mathrm{2}$ line maps from \citet{Jo06} and \citet[][including
   CO 3--2 observations from \citealt{Ha07b}]{Da08} are used to obtain
 complementary information on the sources and their outflows. For a
 few well-known outflow sources, large-scale CO 6-5 maps from Y{\i}ld{\i}z
 et al. (subm.) are also considered.

As shown in Table \ref{maps}, the majority of the PACS maps toward Perseus YSOs 
do not show any extended line emission. The well-centered continuum
and line emission originates from a single object and an associated
bipolar outflow for 12 out of 22 sources. Among the sources with 
spatially-resolved extended emission on the maps, various reasons are 
identified for their origin as illustrated in Fig.~\ref{fig:maps}.
In the map of Per 2, contribution from three nearby
protostars and a strong outflow from the more distant L1448-MM source
cause the extended line and continuum pattern. Emission in the Per 12
map is detected away from the continuum peak, but the emission
originates from a large-scale outflow from NGC1333-IRAS4A, not the
targeted source. The H$_\mathrm{2}$O emission in the Per 20 map is
detected in the direction of the strong outflow and its extent is only
slightly affected by the small mispointing revealed by the asymmetric
continuum emission. 

Extended emission beyond the well-centered 
continuum, as in the case of Per 20, is seen clearly only in Per 9 and Per 21-22. 
Additionally, the continuum peaks for Per 3, 8, and 11 are off-center,
whereas the line emission peaks on-source, suggesting that 
some extended line emission is associated with the source itself and not 
only due to the mispointing.
\footnote{On-source observations that contained Per 3 and Per 11
  in the same field-of-view are discussed in separate papers by
  \citet{Lee13} and \citet{He12}, respectively.} Similar continuum
patterns are seen in Per 6 and Per 10, but here the line emission
peaks a few spaxels away from the map center. In both cases, contribution from
additional outflows / sources is the cause of the dominant off-source
line emission.

To summarize, when the contamination of other sources and their
outflows can be excluded, Perseus YSOs show that the H$_\mathrm{2}$O
2$_{12}$-1$_{01}$ line emission is either well-confined to the central
position on the map or shows at best weak extended line emission
(those are marked with \lq e' in Table 2). In total, 7 out of 22 sources show 
extended emission in the H$_\mathrm{2}$O 2$_{12}$-1$_{01}$ line associated with the targeted sources. 
Emission in CO, OH, and other H$_\mathrm{2}$O
lines follows the same pattern (see Figure \ref{specmapmix}). Similarly compact emission was seen in
a sample of 30 protostars surveyed in the DIGIT
program \citep{Gr13}. In contrast, the WISH PACS survey \citep{Ka13} revealed strong extended emission in
about half of the 20 low-mass protostars. There, the analysis of
patterns of molecular and atomic emission showed that H$_\mathrm{2}$O
and CO spatially co-exist within the PACS field-of-view, while OH and
[$\ion{O}{i}$] lines are typically less extended, but also follow each
other spatially and not H$_\mathrm{2}$O and CO. 
\subsection{Line detections}
\begin{figure}[tb]
\begin{center}
\includegraphics[angle=90,height=6.5cm]{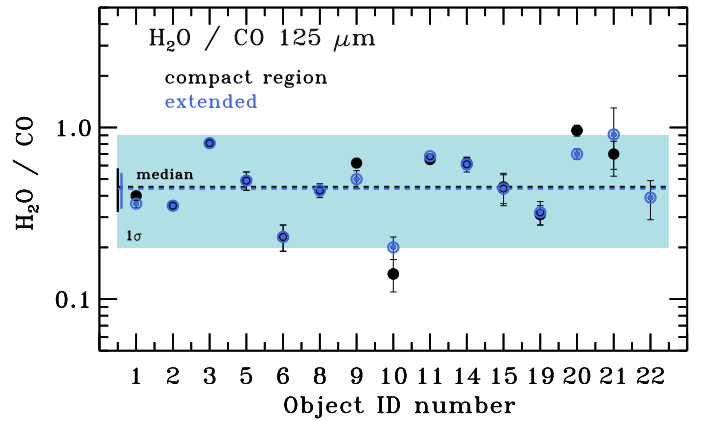}
\caption{\label{obsratio} Flux ratios of the H$_\mathrm{2}$O 4$_{04}$-3$_{13}$ and CO 21-20 lines at $\sim$125 $\mu$m 
calculated using \textit{compact} and \textit{extended} flux extraction regions (see text) corrected for 
contamination from other sources / outflows. Median values for the 
two configurations are shown by the dashed line. The light blue rectangle shows the parameter space between
the minimum and maximum values of line ratios in the more extended configuration. 
The error bars reflect the uncertainties in the measured fluxes of the two lines, excluding 
the calibration error, which is the same for those closely spaced lines.}
\end{center}
\end{figure}
\begin{table*}
\centering
\caption{Comparison of observed H$_2$O, CO, and OH line ratios with literature values}
\label{tab:comp}
\renewcommand{\footnoterule}{}  
\begin{tabular}{ccccccccc}
\hline
Object &  2$_{12}$-1$_{01}$ / 16-15 &  4$_{04}$-3$_{13}$ / 24-23 & 
16-15 / 24-23 & 2$_{12}$-1$_{01}$ / 4$_{04}$-3$_{13}$ & 2$_{21}$-1$_{10}$ / 4$_{04}$-3$_{13}$ & OH 84 / 79 & Ref.\\
\hline
\hline
\multicolumn{8}{c}{Perseus (this work)} \\
\hline
Perseus   & 0.2--2.4   & 0.2--1.1 &  1.2--4.6  & 1.3--6.3 & 1.4--5.5 & 1.1--2.4 & This work\\
\hline
\multicolumn{8}{c}{On--source (literature)} \\
\hline
SMM3 b   & 1.0$\pm$0.1   & 0.5$\pm$ 0.2 &  3.3$\pm$0.8  & 7.1$\pm$1.0 & 3.7$\pm$1.0 & 1.4$\pm$0.7 & (1) \\
     c & 0.9$\pm$0.2   & 0.9$\pm$ 0.3 & 4.6$\pm$1.0   & 4.9$\pm$1.0 & 2.3$\pm$0.8 & 2.4$\pm$1.2 & (1) \\
     r   & 1.0$\pm$0.1   & \textbf{2.0$\pm$0.7} &  \textbf{9.4$\pm$2.3}   & 4.7$\pm$1.0 & 3.1$\pm$0.9 & n.d. & (1) \\
SMM4 r   & 0.8$\pm$0.1   & 0.4$\pm$ 0.1 & 3.6$\pm$0.6  & 7.8$\pm$1.6 & 3.0$\pm$1.3 & 1.1$\pm$0.5 & (1) \\
L1448-MM & 2.3$\pm$1.3  & 1.2$\pm$0.7 & 2.1$\pm$1.2   & 4.1$\pm$2.3 & 2.6$\pm$1.5 & 1.4$\pm$0.9 & (2) \\
NGC1333 I4B & 1.0$\pm$0.1   & 1.0$\pm$0.1 & 1.9$\pm$0.1   & 1.9$\pm$0.1 & 2.0$\pm$0.1 & 1.2$\pm$0.2 & (3) \\
\hline
\multicolumn{8}{c}{Off--source (literature)} \\
\hline
L1157 B1  & \textbf{4.0$\pm$0.5}  & -- & --   & 11.0 & 2.9$\pm$1.2 & n.d. & (4,5)\\
      B1' & 2.1$\pm$0.2  & -- & --  & \textbf{9.2$\pm$2.2} & 2.8$\pm$1.0 & 0.9$\pm$0.5 & (4,5)\\
      B2  & \textbf{17.0$\pm$8.4}   & n.d. & $>$0.3   & \textbf{$>$7.7} & n.d. & -- & (6)\\
      R   & \textbf{5.3$\pm$2.4}  & $>$0.26  & $>$0.5  & \textbf{10.6$\pm$4.7} & 1.8$\pm$0.8 & -- & (6)\\
L1448 B2  & 2.3$\pm$0.4  & 0.4$\pm$0.3 & 2.8$\pm$0.9   & \textbf{15.0$\pm$6.5} & 3.7$\pm$1.6 & -- & (6)\\
      R4  & \textbf{8.2$\pm$3.0}  & $>$0.3 & $>$0.9 & \textbf{23.2$\pm$9.7} & 3.3$\pm$1.2 & -- & (6)\\
\hline
\hline
\end{tabular}
\tablefoot{Sources in the upper part of the table refer to protostellar positions within the PACS maps, 
sources in the lower part refer to shock positions away from the protostar. Ranges of line ratios calculated for Perseus sources 
are listed at the top. Line ratios of sources that exceed the Perseus values are shown in boldface.
The L1157 B1' position refers to the high-excitation CO emission peak close to the nominal position of the B1 shock spot
\citep{Be12}. Non-detections are abbreviated with n.d. OH 84 / 79 refers to the ratio of two OH doublets, at 84 and 79 $\mu$m 
respectively. The total flux of the 84 $\mu$m doublet is calculated by multiplying by two the 84.6 $\mu$m flux, due to the 
blending of the 84.4 $\mu$m line with the CO 31-30 line. References: (1) \citet{Di13}, (2) \citet{Lee13},
 (3) \citet{He12}, (4) \citet{Be12}, (5) \citet{Bu14}, (6) \citet{Sa13}.}
\end{table*}
In the majority of our sources, all targeted rotational transitions of CO, H$_2$O, and OH are detected, see Fig.~\ref{spec}
and Tables \ref{lines} and \ref{det}. 
The [\ion{O}{i}] line at 63 $\mu$m and the [\ion{C}{ii}] line at 158 $\mu$m 
will be discussed separately in a forthcoming paper including all WILL sources (Karska et al. in prep.) 
and are not included in the figure and further analysis.

The CO 16-15 line with an upper level energy
($E_\mathrm{u}/k_\mathrm{B}$) of about 750 K is seen in 17
($\sim$80\%), the CO 24-23 ($E_\mathrm{u}/k_\mathrm{B}\sim1700$ K) in
16 ($\sim$70\%), and the CO 32-31 ($E_\mathrm{u}/k_\mathrm{B}\sim3000$
K) in 8 ($\sim$30\%) sources. The most commonly detected ortho-H$_2$O
lines are: the 2$_{12}$-1$_{01}$ line at 179 $\mu$m
($E_\mathrm{u}/k_\mathrm{B}\sim110$ K) and the 4$_{04}$-3$_{13}$ line at
125 $\mu$m ($E_\mathrm{u}/k_\mathrm{B}\sim320$ K), seen in 15 sources
($\sim70$ \%), whereas the 6$_{16}$-5$_{05}$ line at 82 $\mu$m
($E_\mathrm{u}/k_\mathrm{B}\sim640$ K) is detected in 13 sources ($\sim60$ \%). 
The para-H$_2$O line 3$_{22}$-2$_{11}$ at 90 $\mu$m
($E_\mathrm{u}/k_\mathrm{B}\sim300$ K) is seen toward 9 sources
($\sim50$ \%).

Three OH doublets targeted as part of the WILL survey, the OH
${}^{2}\Pi_{\nicefrac{1}{2}}, J = \nicefrac{1}{2} - {}^{2}\Pi_{\nicefrac{3}{2}}, J = \nicefrac{3}{2}$
 at 79 $\mu$m, $^{2}\Pi_{\nicefrac{3}{2}}$ $J=\nicefrac{7}{2}-\nicefrac{5}{2}$ doublet at 84 $\mu$m
 ($E_\mathrm{u}/k_\mathrm{B}\sim290$ K), and $^{2}\Pi_{\nicefrac{1}{2}}$ $J=\nicefrac{3}{2}-\nicefrac{1}{2}$ doublet at 163 $\mu$m
 ($E_\mathrm{u}/k_\mathrm{B}\sim270$ K), are detected in 14, 15, and 13 objects, respectively
 ($\sim60-70$ \%). 

Sources without any detections of molecular lines associated with the targeted protostars 
are Per 4, 7, 12, and 17 (see Table \ref{det}). In Per 13 only two weak H$_2$O lines at 108 and
125 $\mu$m are seen, whereas in Per 16 only a few of the lowest-$J$
CO lines are detected. A common characteristic of this weak-line group
of objects, is a low bolometric temperature (all except Per 4) and a
low bolometric luminosity (see Table \ref{catalog}), always below 1.3
$L_{\odot}$. However, our sample also includes a few objects with
similarly low values of $L_\mathrm{bol}$, that show many more
molecular lines (in particular Per 3, but also Per 15 and 21), so low
luminosity by itself is not a criterion for weak lines. On the other
hand, low bolometric temperature and high luminosity is typically
connected with strong line emission \citep{Kr12,Ka13}.

\subsection{Observed line ratios}

Observed line ratios are calculated using the fluxes obtained from the
entire 5$\times$5 PACS maps in cases of extended emission where contamination by a nearby
source and / or outflows is excluded. For point sources,
a wavelength-dependent PSF correction factor is applied to fluxes
obtained from the central spaxel (for details, see \S 2). In
principle, ratios using lines that are close in wavelength could be
calculated using smaller flux extraction regions, and no PSF
correction would be required. However, the transition wavelengths of
H$_2$O are not proportional to the upper energy levels, as is the case
for CO, and comparisons of lines tracing similar gas have to rely on
lines that lie far apart in wavelength. We explore to what
extent the size of the extraction region affects the inferred line
ratios. Since our aim is to understand the influence of the 
extended emission associated with the source(s), the ratios of lines 
close in wavelength are studied to avoid the confusion due to the PSF variations.

Fluxes of the nearby H$_2$O 4$_{04}$-3$_{13}$ and CO 21-20 lines located
at 124--125 $\mu$m are calculated first using only the central spaxel
({\it compact region}) and then using all the spaxels with detected
line emission ({\it extended region}). Fig.~\ref{obsratio}
illustrates that the line ratios calculated in these two regions are
fully consistent, both for sources with compact and extended emission.
 Therefore, for the subsequent analysis the extended
region is used for comparisons with models. Table \ref{tab:ratios}
shows the minimum and maximum values of the observed line ratios,
their mean values, and the standard deviations for all sources
with detections.

Even taking into account the uncertainties in flux extraction, the
line ratios span remarkably narrow ranges of values, see Table
\ref{tab:comp} for a selection of H$_2$O, CO, and OH line ratios. The
largest range is seen in the H$_2$O 2$_{12}$-1$_{01}$/CO 16-15 line
ratio, which spans an order of magnitude. In all the other cases the
line ratios are similar up to a factor of a few. The most similar are
the OH line ratios which differ only by a factor of two, consistent
with previous studies based on a large sample of low-mass YSOs in
\citet{Wa13}. The observed similarities also imply that the line
ratios do not depend on protostellar luminosity, bolometric
temperature, or envelope mass (see Fig.~\ref{corr125}).
 
Our line ratios for Perseus sources are consistent with the previously
reported values for other deeply-embedded protostars observed in the
same way (`on source') as tabulated in Table~\ref{tab:comp}.  Some
differences are found for PACS observations of shock positions away
from the protostar (`off source'). Most notably, the ratios using the
low excitation H$_2$O 2$_{12}$-1$_{01}$ line at 179 $\mu$m are up to a
factor of two larger than those observed in the protostellar
vicinity. Such differences are not seen when more highly-excited
H$_2$O lines are compared with each other, for example
H$_2$O 2$_{21}$-1$_{10}$ and 4$_{04}$-3$_{13}$ lines, or with the
high$-J$ CO lines, for example CO 24--23. The ratios of two CO lines
observed away from the protostar, e.g.  the CO 16--15 and CO 24--23
ratios, are at the low end of the range observed toward the
protostellar position.
 
Spectrally-resolved profiles of the H$_2$O 2$_{12}$-1$_{01}$ line
observed with HIFI toward the protostar position reveal absorptions
at source velocity removing about 10\% of total line flux
\citep[e.g.][]{Kr10,Mo14}. Our unresolved PACS
observations therefore provide a lower limit to the H$_2$O emission in
the 2$_{12}$-1$_{01}$ line. This effect, however, is too small to
explain the differences in the line ratios at the `on source' and `off
source' positions.
 
\section{Analysis}

The fact that multiple molecular transitions over a wide range of
excitation energies are detected, points to the presence of hot, dense gas
and can be used to constrain the signatures of shocks created as a
result of outflow-envelope interaction. In particular, the line ratios
of H$_2$O, CO, and OH are useful probes of various shock types and
parameters that do not suffer from distance uncertainties.

Similarities between the spatial extent of different molecules (\S 3.1)
 coupled with similarities in the velocity-resolved line
profiles among these species \citep{Kr10,Yi13,IreneCO,Mo14} strongly
suggest that all highly-excited lines of CO and H$_2$O arise from the
same gas. Some differences may occur for OH, which is also
associated with dissociative shocks and can be affected by radiative
excitation (see \S 5). Modeling of absolute line fluxes requires
sophisticated two-dimensional (2D) physical source models for the proper
treatment of the beam filling factor \citep{Vi12}. Those models also
show that UV heating alone is not sufficient to account for the high
excitation lines. Hence, the focus in this analysis is on
shocks. Since the absolute flux depends sensitively on the assumed emitting area, 
in the subsequent analysis only the line ratios are compared.

In the following sections, properties of shock models and the
predicted line emission in various species are discussed (\S 4.1) and
observations are compared with the models, using line ratios of the
same species (\S 4.2), and different species (\S 4.3).  Special focus
will be given to $C-$type shocks where grids of model results are
available in the literature.  Observations suggest that most of the
mass of hot gas is in $C-$type shocks toward the central protostellar
positions, at least for H$_2$O and CO with $J$ $<$ 30
\citep[][Kristensen et al. in prep.]{Kr13}; higher-$J$ CO, OH and
      [\ion{O}{i}] transitions, on the other hand, will primarily
      trace $J-$type shocks \citep[e.g.][]{Wa13, Kr13}. The excitation
      of OH and [\ion{O}{i}] will be analyzed in a forthcoming paper;
      the $J>30$ CO emission is only detected toward $\sim$ 30\% of all
      sources and so is likely unimportant for the analysis and
      interpretation of the data presented here. Only limited discussion 
      of $J-$ type shocks is therefore presented below.

\begin{figure*}[tb]
\begin{center}
\vspace{+0.5cm}

\includegraphics[angle=0,height=19cm]{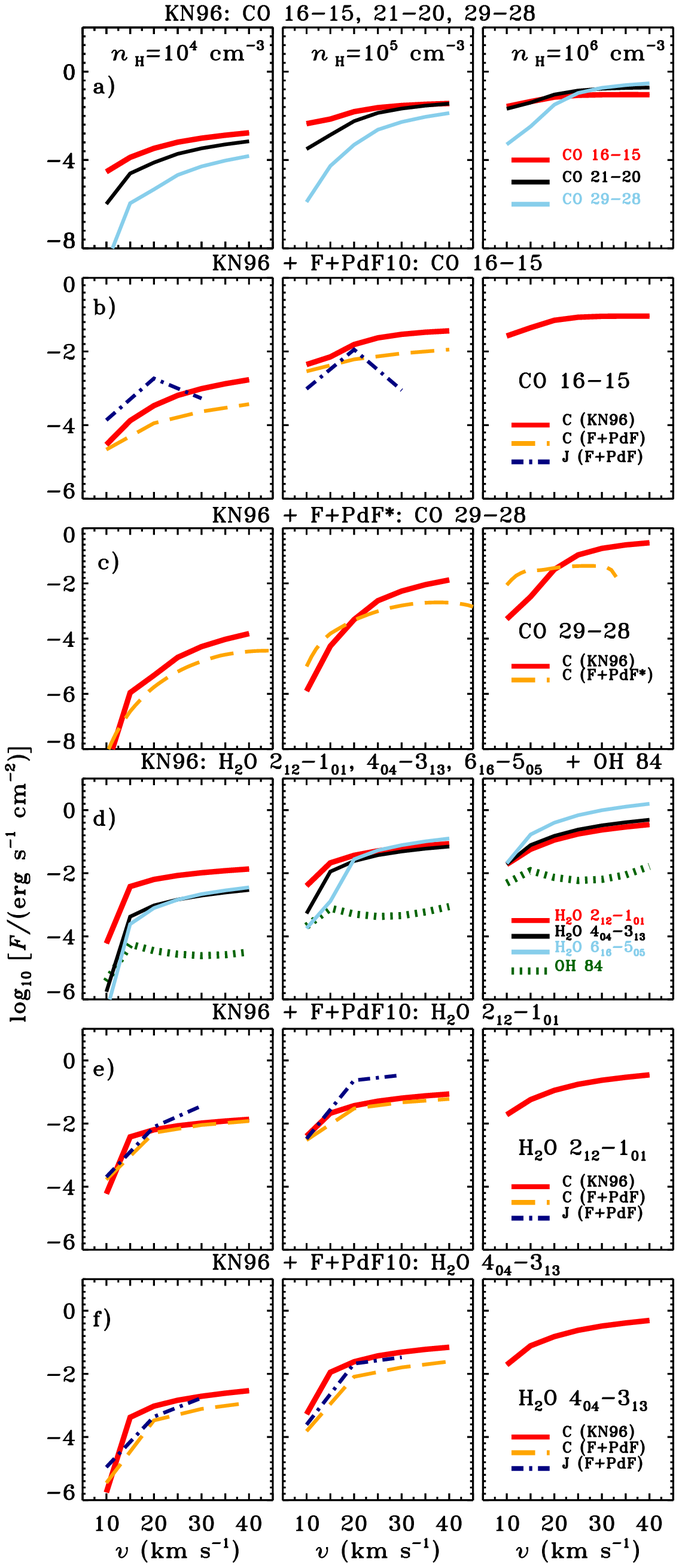}

  \vspace{1cm}
\caption{\label{vel} Absolute fluxes of selected CO, H$_\mathrm{2}$O, and OH lines 
predicted by Kaufman \& Neufeld and Flower \& Pineau des For\^ets models and shown as a function of shock velocity 
and for pre-shock densities of 10$^4$ cm$^{-3}$ (left),
10$^5$ cm$^{-3}$ (center), and 10$^6$ cm$^{-3}$ (right). The latter models 
are available only for the pre-shock densities of 10$^{4}$ cm$^{-3}$ and 10$^{5}$ cm$^{-3}$.
OH 84 refers to the OH $^{2}\Pi_{\nicefrac{3}{2}}$ $J=\nicefrac{7}{2}-\nicefrac{5}{2}$ doublet at 84 $\mu$m.}
\end{center}
\end{figure*}

\subsection{Model line emission}

Models of shocks occurring in a medium with physical conditions
typical for the envelopes of deeply-embedded young stellar objects
provide a valuable tool for investigating shock characteristics:
shock type, velocity, and the pre-shock density of (envelope)
material.

Model grids have been published using a simple 1D geometry either
for steady-state $C$ and $J$ type shocks \citep{Hol89,KN96,FP10} or
time-dependent $C-J$ type shocks \citep{Gus08,Gus11,FP12}. The latter are
non-stationary shocks, where a $J-$type front is embedded in a $C-$type
shock \citep{Ch98,Le04b,Le04a}. These shocks are intermediate between
pure $C-$ and $J-$type shocks and have temperatures and physical extents
in between the two extremes.

$C$-$J$ shocks may be required for the youngest outflows with ages less
than 10$^3$ yrs, \citep[][for the case of IRAS4B, Per 11]{FP12,FP13}. 
Here the dynamical age of the outflow is taken as an
upper limit of that of the shock itself, which may be caused by a more
recent impact of the wind on the envelope. The age of our sources is
of the order of 10$^5$ yrs \citep[][for Class 0 sources in Perseus]{Sad14}
 and they should have been driving winds and jets for the bulk
of this period, so this timescale is long enough for any shocks close
to the source position to have reached steady state. While we cannot
exclude that a few individual shocks have been truncated, our primary
goal is to examine trends across the sample. Invoking $C$-$J$ type shocks
with a single truncation age as an additional free parameter is
therefore not a proper approach for this study.
The focus is therefore placed on comparing $C-$type shock results from
\citet[][KN96 from now on]{KN96} and \citet[][F+PdF10 from now
  on]{FP10} with the observations.

All models assume the same initial atomic abundances and similarly low
degrees of ionization, $x_{\rm i}$ $\sim$ 10$^{-7}$ for $C$ shocks.
The pre-shock transverse magnetic field strength is parametrized as
$B_{0}=b\times \sqrt{n_\mathrm{H}(\mathrm{~cm}^{-3})}$ $\mu$Gauss,
where $n_\mathrm{H}$ is the pre-shock number density of atomic hydrogen and $b$ is the
magnetic scaling factor, which is typically 0.1-3 in the ISM
\citep{Dr80}. The value of $b$ in KN96 and F+PdF10 is fixed at a value
of 1.

The main difference between the two shock models is the inclusion of
grains in the F+PdF10 models \citep{FP03}. The latter models assume a standard 
MRN distribution of grain sizes \citep{Ma77} for grain radii between 
0.01 and 0.3 $\mu$m and a fractional abundance of the PAH in the gas phase of $10^{-6}$ 
\citep[the role of PAHs and the sizes of grains are discussed in][]{FP03,FP12}. 
As the electrons are accelerated in the magnetic precursor, they attach themselves to
grains thereby charging the grains and thus increasing the density of
the ionized fluid significantly. This increase in density has the
effect of enhancing the ion-neutral coupling \citep{Dr80}, thereby
effectively lowering the value of $b$ compared to the KN96 models.  As
a consequence, the maximum kinetic temperature is higher in the F+PdF10
models for a given shock velocity, $\varv$.  The stronger coupling
between the ions and neutrals results in narrower shocks 
\citep{FP10}, with shock widths scaling as $\propto b^2 (x_{\rm i}
n_\mathrm{H} \varv)^{-1}$ \citep{Dr80}. This proportionality does not 
capture the ion-neutral coupling exactly, as, for example, 
the grain size distribution influences the coupling \citep{Gu07,Gu11}.
The compression in $C$ shocks
also changes with the coupling since the post-shock density depends on
the magnetic field, $n_\mathrm{post}\sim 0.8 \varv n_\mathrm{H} b^{-1}$
\citep[e.g.][]{Ka13}. The column density of emitting molecules is a
function of both shock width and compression factor, and as a zeroth-order 
approximation the column density is $N\sim
n_\mathrm{post}\times L \sim b x_{\rm i}^{-1}$.  The ionization degree is not
significantly different between the KN96 and F+PdF10 models because it
is primarily set by the cosmic ray ionization rate ($\zeta=5\cdot10^{-17}$ s$^{-1}$, F+PdF10), and thus the
F+PdF10 models predict lower column densities than the KN96 models for
a given velocity and density.

Another important difference is that the F+PdF10 models take into
account that molecules frozen out onto grain mantles can be released
through sputtering when the shock velocity exceeds $\sim$ $15$ km\,s$^{-1}$ \citep{FP10,FP12,vL13}. 
Therefore, the gas-phase column densities of
molecules locked up in ices increase above this threshold shock
velocity with respect to the KN96 models, an effect which applies to
both CO and H$_2$O. Furthermore, H$_\mathrm{2}$O forms more abundantly
in the post-shock gas of F+PdF10 models, because H$_\mathrm{2}$
reformation is included, unlike in the KN96 models \citep{FP10}.

Molecular emission is tabulated by KN96 for a wide range of shock
velocities, from $\varv =5$ to 45 km\,s$^{-1}$ in steps of 5 km\,s$^{-1}$,
 and a wide range of pre-shock densities $n_\mathrm{H}$, from
10$^{4}$ to 10$^{6.5}$ cm$^{-3}$ in steps of 10$^{0.5}$ cm$^{-3}$. The
F+PdF10 grid is more limited in size, providing line intensities for
only two values of pre-shock densities, namely 10$^{4}$ and 10$^{5}$
cm$^{-3}$, and a comparable range of shock velocities, but calculated
in steps of 10 km\,s$^{-1}$. Calculations are provided for CO
transitions from $J=1-0$ to $J=60-59$ in KN96 and only up to $J=20-19$
in F+PdF10. The two sets of models use different collisional rate
coefficients to calculate the CO excitation. F+PdF10 show line
intensities for many more H$_\mathrm{2}$O transitions (in total
$\sim$120 lines in the PACS range, see \S 2) than in the older KN96
grid (18 lines in the same range), which was intended for comparisons
with the Submillimeter Wave Astronomy Satellite \citep[SWAS,][]{SWAS}
and ISO data. KN96 use collisional excitation rates
for H$_\mathrm{2}$O from \citet{Gr93} and F+PdF10 from \citet{Fa07}. 
Line intensities for OH are only computed by KN96,
assuming only collisional excitation and using the oxygen chemical 
network of \citet{Wa87}. The reaction rate coefficients in that network 
are within a factor of 2 of the newer values by \citet{Ba12} and tabulated in the 
UMIST database \citep[www.udfa.net,][]{UMIST}, see also a discussion 
in \citet{EvD13}.

We also use CO fluxes extracted from the grid of models presented
by \citet{Kr07}, since high$-J$ CO lines are missing in F+PdF10. 
This grid is based on the shock model presented in
\citet{FP03} and covers densities from 10$^4$--10$^7$ cm$^{-3}$ and
velocities from 10--50 km\,s$^{-1}$ (denoted as F+PdF* from now on). The main difference compared to
the results from F+PdF10 is that CO and H$_2$O level populations are
not calculated explicitly through the shock; instead analytical
cooling functions are used to estimate the relevant line cooling and
only afterwards line fluxes are extracted \citep{FG09}. Models with
$b$ = 1 are used. The CO line fluxes presented here are computed using
the 3D non-LTE radiative transfer code LIME \citep{Br10}, for levels
up to $J$ = 80--79. The CO collisional rate coefficients from
\citet{Ya10} extended by \citet{Ne12} are used.

In the following sections, the model fluxes of selected CO,
H$_\mathrm{2}$O, and OH lines are discussed for a range of shock
velocities and three values of pre-shock densities: 10$^{4}$, 10$^{5}$, and 10$^{6}$ cm$^{-3}$. 
Note that the post-shock densities traced by observations are related to the pre-shock densities via the compression factor
dependent on the shock velocity and magnetic field. In $C$ shocks, the compression factor is about 
10 \citep[e.g.][]{Ka13}. 

Figure~\ref{vel} compares model 
fluxes of various CO, OH, and H$_\mathrm{2}$O lines from the KN96 models
(panels a and d) and, for a few selected CO and H$_\mathrm{2}$O lines, 
compares the results with the F+PdF10 or F+PdF* models (panels b, c, e, and f).

\subsubsection{CO}
The KN96 model line fluxes for CO 16-15, CO 21-20, and CO 29-28
are shown in panel a of Fig. \ref{vel}. The upper energy levels of
these transitions lie at 750 K, 1280 K, and 2900 K, respectively,
while with increasing shock velocity, the peak $C-$shock temperature
increases from about 400 K to 3200 K (for 10 to 40 km\,s$^{-1}$) and is
only weakly dependent on the assumed density (see Fig. 3 of
KN96). Therefore, the CO 16-15 line is already excited at relatively
low shock velocities ($\varv \sim$10 km\,s$^{-1}$, for
$n_\mathrm{H}$=10$^{4}$ cm$^{-3}$), whereas the higher$-J$ levels
become populated at higher velocities. At a given shock velocity,
emission from the CO 16-15 line is the strongest due to its lower
critical density, $n_\mathrm{cr}\sim9\times10^5$ cm$^{-3}$ at $T$=1000
K \citep{Ne12}. This situation only changes for the highest pre-shock
densities, when the line becomes thermalized and the cooling in other
lines dominates. 

The CO 16-15 flux from the F+PdF10 $C-$type shock models is comparable
to the KN96 flux for the $\varv\sim$10 km\,s$^{-1}$ shock, but increases
less rapidly with shock velocity despite the sputtering from grain
mantles (panel b of Fig.~\ref{vel}). Because of the lower magnetic scaling 
factor $b$ in the models with grains (see \S 4.1), it is
expected that the column density and the corresponding line fluxes are
lower in the F+PdF10 models. For slow shocks the higher temperatures
in the latter models compensate for the smaller column density
resulting in a similar CO 16-15 flux. 

In $J-$type shocks, the peak temperatures of the post-shock gas are 1400,
5500, 12000, and 22000 K for the shock velocities of 10, 20, 30, and
40 km\,s$^{-1}$ respectively \citep{ND89,KN96}. For 10-20 km\,s$^{-1}$
shocks, these high temperatures more easily excite the CO 16-15 line with
respect to $C-$type shock emission. For shock velocities above 20 km\,s$^{-1}$,
 such high temperatures can lead to the collisional
dissociation of H$_\mathrm{2}$ and subsequent destruction of CO and
H$_\mathrm{2}$O molecules, resulting in the decrease of CO
fluxes. This effect requires high densities and therefore the CO flux
decrease is particularly strong for the pre-shock densities 10$^{5}$
cm$^{-3}$.
\begin{figure*}
\centering
\vspace{+0.5cm}

  \includegraphics[angle=0,height=19cm]{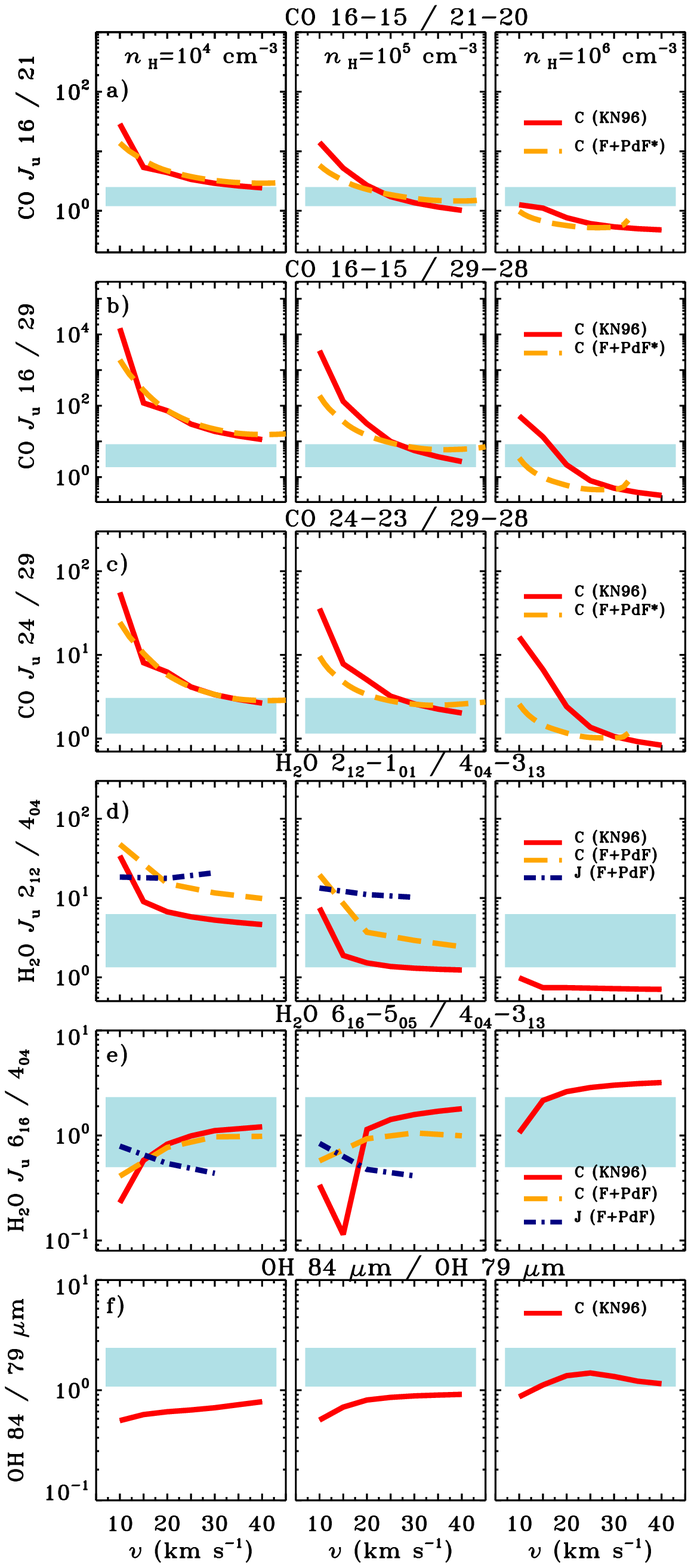}
  
  \vspace{1cm}
\caption{\label{allvel} Line ratios of the same species
using Kaufman \& Neufeld (1996) $C$ shock models (KN96, solid line) and 
Flower \& Pineau des For\^ets (2010) $C$ and $J$ shock models (F+PdF, dashed and dashed-dotted lines, respectively).
Ratios are shown as a function of shock velocity and for pre-shock densities of 10$^4$ cm$^{-3}$ (left),
10$^5$ cm$^{-3}$ (center), and 10$^6$ cm$^{-3}$ (right). Observed ratios are shown as blue rectangles.}
\end{figure*}
\begin{figure*}
\centering
  \vspace{0.5cm}
  \includegraphics[angle=0,height=19cm]{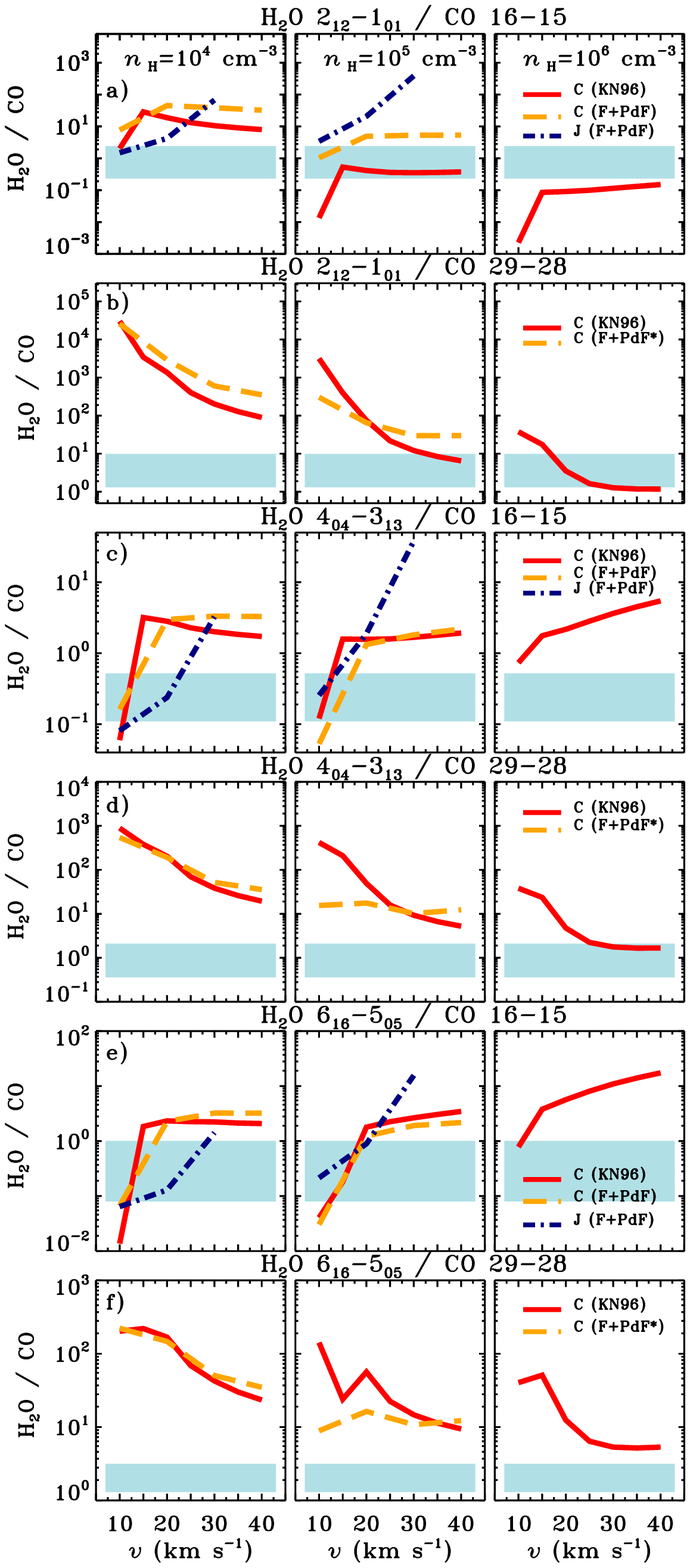}

  \vspace{1cm}
\caption{\label{allvelp2} H$_\mathrm{2}$O to CO line ratios 
using Kaufman \& Neufeld (1996) $C$ shock models (KN96, solid line) and 
Flower \& Pineau des For\^ets (2010) $C$ and $J$ shock models (F+PdF, dashed and dashed-dotted lines, respectively).
Ratios are shown as a function of shock velocity and for pre-shock densities of 10$^4$ cm$^{-3}$ (left),
10$^5$ cm$^{-3}$ (center), and 10$^6$ cm$^{-3}$ (right). The F+PdF models 
are available only for the pre-shock densities of 10$^{4}$ cm$^{-3}$ and 10$^{5}$ cm$^{-3}$.
Observed ratios are shown as blue rectangles.}
\end{figure*}
\subsubsection{H$_\mathrm{2}$O}

The H$_2$O fluxes show a strong increase with shock velocities above
$\varv\sim10-15$ km\,s$^{-1}$ in both models, especially at low
pre-shock densities (panels d-f of Fig.~\ref{vel}). At this velocity,
the gas temperature exceeds $\sim$400 K, so the high-temperature route of
H$_\mathrm{2}$O formation becomes efficient
\citep[][KN96]{EJ78,EJ78b,Be98}, which quickly transfers all gas-phase
oxygen into H$_\mathrm{2}$O via reactions with H$_\mathrm{2}$ (KN96).

In contrast to CO, the upper level energies of the observed
H$_\mathrm{2}$O lines are low and cover a narrow range of values,
$E_\mathrm{up}\sim200-600$ K. As a result, the effect of peak gas
temperature on the H$_\mathrm{2}$O excitation is less pronounced
(Fig.~3 of KN96) and after the initial increase with shock velocity,
the H$_\mathrm{2}$O fluxes in the KN96 models stay constant for all
lines. For high pre-shock densities, the higher lying levels are more
easily excited and, as a consequence, the fluxes of the
H$_\mathrm{2}$O 6$_{16}$-5$_{05}$ line become larger than those of the
H$_\mathrm{2}$O 2$_{12}$-1$_{01}$ line. The critical densities of
these transitions are about 2 orders of magnitude higher than for the
CO 16-15 line and the levels are still sub-thermally excited at densities of $10^6$-$10^7$ cm$^{-3}$
\citep[and effectively optically thin,][]{Mo14}.

The H$_\mathrm{2}$O 2$_{12}$-1$_{01}$ fluxes in the $C-$type F+PdF10
models are remarkably similar to those found by KN96 (panel e of
Fig.~\ref{vel}), while the H$_\mathrm{2}$O 4$_{04}$-3$_{13}$ fluxes
are lower by a factor of a few over the full range of shock velocities
and pre-shock densities in the F+PdF10 models (panel f of Fig.~\ref{vel}). Lower
H$_\mathrm{2}$O fluxes are expected due to smaller column of
H$_\mathrm{2}$O in the models with grains (\S 4.1). For the
lower$-J$ lines, the various factors (lower column density through a shock but
inclusion of ice sputtering and H$_2$ reformation) apparently conspire
to give similar fluxes as for KN96.

In $J-$type shocks, the fluxes of H$_\mathrm{2}$O 2$_{12}$-1$_{01}$ and
4$_{04}$-3$_{13}$ lines increase sharply for shock velocities 10--20
km\,s$^{-1}$ (panels e and f of Fig.~4). The increase is less steep at 30 km\,s$^{-1}$ shocks for
$n_\mathrm{H}$=10$^{5}$ cm$^{-3}$, when the collisional dissociation
of H$_\mathrm{2}$ and subsequent destruction of H$_\mathrm{2}$O
molecules occurs (fluxes for larger shock velocities are not computed in F+PdF10 and hence not shown). 
Below 30 km\,s$^{-1}$, line fluxes from
$J-$type shocks are comparable to those from $C-$type shock predictions
 except for the H$_\mathrm{2}$O 2$_{12}$-1$_{01}$ fluxes at
high pre-shock densities, which are an order of magnitude higher with
respect to the $C-$type shock predictions. The difference could be due
to smaller opacities for the low-excitation H$_\mathrm{2}$O line in the
$J$ shocks.

\subsubsection{OH}

The fluxes of the $^{2}\Pi_{\nicefrac{3}{2}}$
$J=\nicefrac{7}{2}-\nicefrac{5}{2}$ doublet at 84 $\mu$m
($E_\mathrm{u}/k_\mathrm{B}\sim290$ K) calculated with the KN96 models
are shown with the H$_2$O lines in panel d of Fig.~\ref{vel}. Not much
variation is seen as a function of shock velocity, in particular
beyond the initial increase from 10 to 15 km\,s$^{-1}$, needed to drive
oxygen to OH by the reaction with H$_2$. At about 15 km\,s$^{-1}$, the
temperature is high enough to start further reactions with H$_2$
leading to H$_2$O production. The trend with increasing pre-shock
density is more apparent, with OH fluxes increasing by two orders of
magnitude between the 10$^{4}$ cm$^{-3}$ to 10$^{6}$ cm$^{-3}$, as the
density becomes closer to the critical density of the transition.

\subsection{Models versus observations -- line ratios of the same species}
Comparison of observed and modeled line ratios of different pairs of
CO, H$_\mathrm{2}$O, and OH transitions is shown in Fig.~\ref{allvel}. 
The line ratios are a useful probe of molecular
excitation and therefore can be used to test whether the excitation in
the models is reproduced correctly, which in turn depends on density
and temperature, and thus shock velocity.

\subsubsection{CO line ratios}

Given the universal shape of the CO ladders observed toward 
deeply-embedded protostars (see \S 1 and the discussion on
 the origin of CO ladders in \S 5), three pairs of CO lines are
compared with models: (i) CO 16-15 and 21-20 line ratio, corresponding
to the `warm', 300 K component (panel a); (ii) CO 16-15 and 29-28 line
ratio, combining transitions located in the `warm' and `hot' ($>$ 700 K)
components (panel b); (iii) CO 24-23 and 29-28, both tracing the `hot'
component (panel c). All ratios are consistent with the $C-$type
shock models from both KN96 and F+PdF10 for pre-shock densities above
$n_\mathrm{H}$=10$^4$ cm$^{-3}$. For the CO 16-15/21-20 ratio, a
pre-shock density of $n_\mathrm{H}$=10$^5$ cm$^{-3}$ and shock
velocities of $20$-$30$ km\,s$^{-1}$ best fit the observations. Shock
velocities above $\sim25$ km\,s$^{-1}$ are needed to reproduce the
observations of the other two ratios at the same pre-shock
density. Alternatively, higher pre-shock densities with velocities
below 30 km\,s$^{-1}$ are also possible.

The KN96 $C-$shock CO line ratios for lower-to-higher$-J$ transitions
(panel b of Fig.~\ref{allvel}) decrease with velocity, due to the increase in peak temperature that
allows excitation of the higher-$J$ CO transitions. The effect is 
strongest at low pre-shock densities (see \S 4.1.1) and for
the sets of transitions with the largest span in $J$ numbers. The CO
16-15 / 29-28 line ratio ($\Delta J_\mathrm{up}=13$) decreases by
almost three orders of magnitude between shock velocities of 10 and 40
km\,s$^{-1}$ over the range of pre-shock densities. In contrast, the CO 16-15 / 21-20 and CO 24-23 / 29-28
line ratios show drops of about one order of magnitude with increasing velocity
(panel a and c of Fig.~\ref{allvel}). These model trends explain why the
observed CO line ratios are good diagnostics of shock velocity.

In absolute terms, the line ratios calculated for a given velocity are
inversely proportional to the pre-shock density. The largest ratios
obtained for $n_\mathrm{H}$=10$^4$ cm$^{-3}$ result from the fact that
the higher$-J$ levels are not yet populated at low shock-velocities,
while the lower$-J$ transitions reach 
LTE at high shock-velocities and do not show an increase of flux
with velocity. This effect is less prominent at higher pre-shock
densities, where the higher$-J$ lines are more easily excited at low
shock velocities.

The F+PdF* CO line ratios, extending the \citet{FP03} grid to higher$-J$ CO lines, are
almost identical to the KN96 predictions for pre-shock densities
$n_\mathrm{H}$=10$^4$ cm$^{-3}$. For higher densities, the
low-velocity $C-$shock models from F+PdF* are systematically lower than
the KN96 models, up to almost an order of magnitude for 10--15 km\,s$^{-1}$
 shocks at $n_\mathrm{H}$=10$^6$ cm$^{-3}$. Therefore, the
pre-shock density is less well constrained solely by CO lines.

For densities of $10^{5}$ cm$^{-3}$, shock velocities of 20-30 km\,s$^{-1}$ best reproduce the 
ratios using only transitions from the \lq warm' component, while shock velocities above 
25 km s$^{-1}$ match the ratios using the transitions from the \lq hot' component.
Velocities of that order are observed in CO $J=16-15$ HIFI line profiles \citep[][in prep.]{Kr13},
but higher-$J$ CO lines dominated by the hot component have not been obtained with sufficient velocity resolution.

\subsubsection{H$_\mathrm{2}$O line ratios}

Two ratios of observed H$_\mathrm{2}$O lines are compared with the $C-$
and $J-$type shock models: (i) the ratio of the low excitation H$_\mathrm{2}$O
2$_{12}$-1$_{01}$ and moderate excitation 4$_{04}$-3$_{13}$ lines
(panel d of Fig.~5) and (ii) the ratio of the highly-excited H$_\mathrm{2}$O 6$_{16}$-5$_{05}$
and 4$_{04}$-3$_{13}$ lines (panel e).  Similar to the CO ratios,
$C-$type shocks with pre-shock densities of 10$^5$ cm$^{-3}$ reproduce
the observations well. Based on the observations of ratio (i), C
shocks with a somewhat larger (F+PdF10) or smaller (KN96) pre-shock
density are also possible for a broad range of shock velocities.  On
the other hand, no agreement with the $J-$type shocks is found for this
low-excitation line ratio. Observations of ratio (ii) indicate
a similar density range as ratio (i) for the KN96 models, but extend to 10$^4$ cm$^{-3}$ for
the F+PdF10 models, with agreement found for both $C-$ and $J-$type.

The model trends can be understood as follows. For 10-20 km\,s$^{-1}$
shocks, increasing temperature in the $C$ shock models from KN96 allows
excitation of high-lying H$_\mathrm{2}$O lines and causes the
H$_\mathrm{2}$O 2$_{12}$-1$_{01}$/4$_{04}$-3$_{13}$ line ratio to
decrease and the H$_\mathrm{2}$O 6$_{16}$-5$_{05}$/4$_{04}$-3$_{13}$
line ratio to increase. At
higher shock velocities, the former ratio shows almost no dependence
on shock velocity, while a gradual increase is seen in the ratio using
two highly-excited lines in the KN96 models. At high pre-shock
densities ($n_\mathrm{H}$=10$^6$ cm$^{-3}$), the upper level
transitions are more easily excited and so the changes are even smaller.

The H$_\mathrm{2}$O 2$_{12}$-1$_{01}$/4$_{04}$-3$_{13}$ line ratios
calculated using the $C-$type shock models from F+PdF10 are a factor of
a few larger than the corresponding ratios from the KN96 models (see
the discussion of absolute line fluxes in Sec. 4.1.2). As a result,
when compared to observations, the F+PdF10 models require pre-shock
densities of at least $n_\mathrm{H}$=10$^5$ cm$^{-3}$, while the KN96
models suggest a factor of few lower densities. Overall, the best fit
to both the CO and H$_2$O line ratios is for pre-shock densities
around $10^5$ cm$^{-3}$.

\subsubsection{OH line ratios}

Comparison of the observed OH 84 and 79 $\mu$m line ratio with the
KN96 $C-$type models (panel f of Fig.~5) indicates an order of magnitude higher
pre-shock densities, $n_\mathrm{H}$=10$^6$ cm$^{-3}$, with respect to
those found using the CO and H$_\mathrm{2}$O ratios. However, the KN96
models do not include any far-infrared radiation, which affects the
excitation of the OH lines, in particular the 79 $\mu$m
\citep{Wa10,Wa13}. Additionally, part of OH most likely originates in
a $J-$type shock, influencing our comparison
\citep{Wa10,Be12,Ka13,Kr13}.

Similar to the absolute fluxes of the 84 $\mu$m doublet discussed
in \S 4.1.3, not much variation in the ratio is seen with shock
velocity. The ratio increases by a factor of about two between the
lowest and highest pre-shock densities.

\subsection{Models and observations - line ratios of different species}

Figure~\ref{allvelp2} compares observed line ratios of various
H$_\mathrm{2}$O and CO transitions with the C and $J-$type shock
models. The line ratios of different species are sensitive both to the
molecular excitation and their relative abundances.
 
\begin{figure}
\begin{center}
\vspace{0.5cm}
\hspace{-1.5cm}
\includegraphics[angle=90,height=7cm]{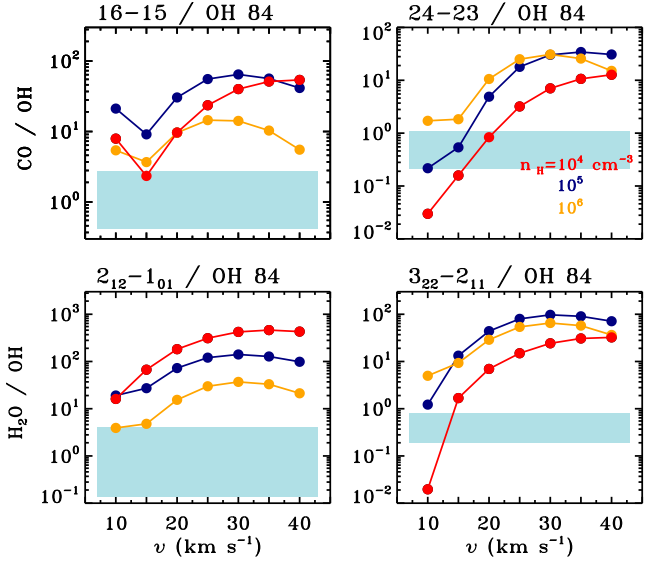}
\caption{\label{oh} CO to OH and H$_\mathrm{2}$O to OH line ratios
as a function of shock velocities using KN96. The ratios are shown for pre-shock densities
 of 10$^4$ cm$^{-3}$ (red), 10$^5$ cm$^{-3}$ (blue), and 10$^6$ cm$^{-3}$ (yellow).
The range of line ratios from observations is shown as filled 
rectangles.}
\end{center}
\end{figure}

\subsubsection{Ratios of H$_\mathrm{2}$O and CO}

Comparison of observations to the KN96 and F+PdF10 models shows that
the $C-$type shocks at pre-shock density $n_{\mathrm{H}}=10^5$
cm$^{-3}$, which best reproduces the line ratios of same species, fail
to reproduce the observed line ratios of different species
(Fig.~\ref{allvelp2}). There are only a few cases where the
observations seem to agree with the models at all.  For $n_{\rm
  H}=10^5$ cm$^{-3}$, a few H$_2$O/CO line ratios fit at low velocities ($<$ 20
km\,s$^{-1}$) (panels a, c and e) but this does not hold for all ratios. 
Moreover, such low shock velocities have been excluded in the previous section.
Higher densities, $n_{\mathrm{H}}=10^6$ cm$^{-3}$, are needed to
reconcile the observations of the H$_\mathrm{2}$O 2$_{12}$-1$_{01}$/CO
29-28 line ratio. Observations of all the other ratios, using more
highly-excited H$_\mathrm{2}$O lines, are well below the model
predictions.

The patterns seen in the panels in Fig.~\ref{allvelp2} can be
understood as follows. The KN96 C type shock models show an initial
rise in the H$_\mathrm{2}$O 2$_{12}$-1$_{01}$ and CO 16-15 line ratios
from 10 to 15 km\,s$^{-1}$ shocks (panel a), as the temperature reaches
the 400 K and enables efficient H$_\mathrm{2}$O formation. Beyond this velocity,
the line ratios show no variations with velocity. The decrease in this line ratio for
higher densities, from 10 at 10$^4$ cm$^{-3}$ to 0.1 at
$n_\mathrm{H}$=10$^6$ cm$^{-3}$ is due to the larger increase of the
column of the population in the $J_{\rm u}$=16 level with density
compared to the increase in the H$_\mathrm{2}$O $2_{12}$ level (see Fig.~\ref{vel} above).

Line ratios of H$_\mathrm{2}$O 2$_{12}$-1$_{01}$ and higher $-J$ CO
lines (e.g. 29-28, panel b) show more variation with velocity.  A
strong decrease by about an order of magnitude and up to two orders of
magnitude are seen for the ratios with CO 24-23 and CO 29-28,
respectively (the ratio with CO 24-23 is not shown here). 
These lines, as discussed in \S 4.1.1 and 4.2.1, are more sensitive
than H$_\mathrm{2}$O to the increase in the maximum temperature
attained in the shock that scales with shock velocities and therefore
their flux is quickly rising for higher velocities (Fig.~\ref{vel}).
The decrease is steeper for models with low pre-shock densities, since
$n_\mathrm{H}\sim10^6$ cm$^{-3}$ allows excitation of high$-J$ CO
lines at lower temperatures.  At this density, the H$_\mathrm{2}$O/CO
line ratios are the lowest and equal about unity.

Due to the lower CO 16-15 fluxes in the $C-$type shock models from
F+PdF10 and similar H$_\mathrm{2}$O 2$_{12}$-1$_{01}$ fluxes (Fig.~\ref{vel}),
 the H$_\mathrm{2}$O-to-CO ratios are generally larger than
in the KN96 models. The exceptions are the ratios with higher$-J$ CO
which are more easily excited, especially at low shock velocities, in the
hotter $C-$type shocks from F+PdF*.

For the same reason, the increasing ratios seen in the $J$ shock models
are caused by the sharp decrease in CO 16-15 flux for shock velocity
$\varv$ = 30 km\,s$^{-1}$, rather than the change in the H$_\mathrm{2}$O
lines. At such high-velocities for $J$ shocks, a significant amount of
CO can be destroyed by reactions with hydrogen atoms
\citep{FP10,Su14}. Since the activation barrier for the reaction of
H$_\mathrm{2}$O with H is about 10$^4$ K, the destruction of
H$_\mathrm{2}$O does not occur until higher velocities.

Similar trends to the line ratios with H$_\mathrm{2}$O
2$_{12}$-1$_{01}$ are seen when more highly-excited H$_\mathrm{2}$O
lines are used (panels c-f of Fig.~\ref{allvelp2}), supporting the
interpretations that variations are due to differences in CO rather than
H$_\mathrm{2}$O lines.

\subsubsection{Ratios of CO and H$_\mathrm{2}$O with OH}

Fig.~\ref{oh} shows line ratios of CO or H$_2$O with
the most commonly detected OH doublet at 84 $\mu$m. The ratios are
calculated for three values of pre-shock densities (10$^4$, 10$^5$,
and 10$^6$ cm$^{-3}$) using exclusively the KN96 models, because the
F+PdF10 grid does not present OH fluxes.

In general, the observed CO/OH, and H$_\mathrm{2}$O/OH ratios are similar for 
all sources but much
lower than those predicted by the models assuming that a significant fraction
of the OH comes from the same shock as CO and H$_\mathrm{2}$O (see Sec. 5.1.). The only exception is the
CO 24-23/OH 84 $\mu$m ratio where models and observations agree for
densities 10$^{4}-10^{5}$ cm$^{-3}$ and shock velocities below 20 km\,s$^{-1}$. 
For any other set of lines discussed here, the observations
do not agree with these or any other models.

As discussed in previous sections, the trends with shock velocity are
determined mostly by the changes in the CO lines, rather than the OH
itself, as seen in Fig.~\ref{vel} (panel d). 
At shock velocities below 20 km\,s$^{-1}$ the OH model flux exceeds
that of CO due to the abundance effect: not all OH has been transferred to
H$_\mathrm{2}$O yet at low temperatures.
Due to the lower critical densities of the CO lines
($n_\mathrm{cr}\sim10^{6}-10^{7}$ cm$^{-3}$) compared with the OH line
($n_\mathrm{cr}\sim10^{9}$ cm$^{-3}$), the lines for various pre-shock
densities often cross and change the order in the upper panels of
Fig.~\ref{oh}. The corresponding trends in the H$_\mathrm{2}$O/OH
line ratios are similar to those of CO/OH, except that the variations
with shock velocity are smaller and the critical densities are more
similar.

\section{Discussion}
\begin{figure*}[tb]
\begin{center}
\includegraphics[angle=90,height=10cm]{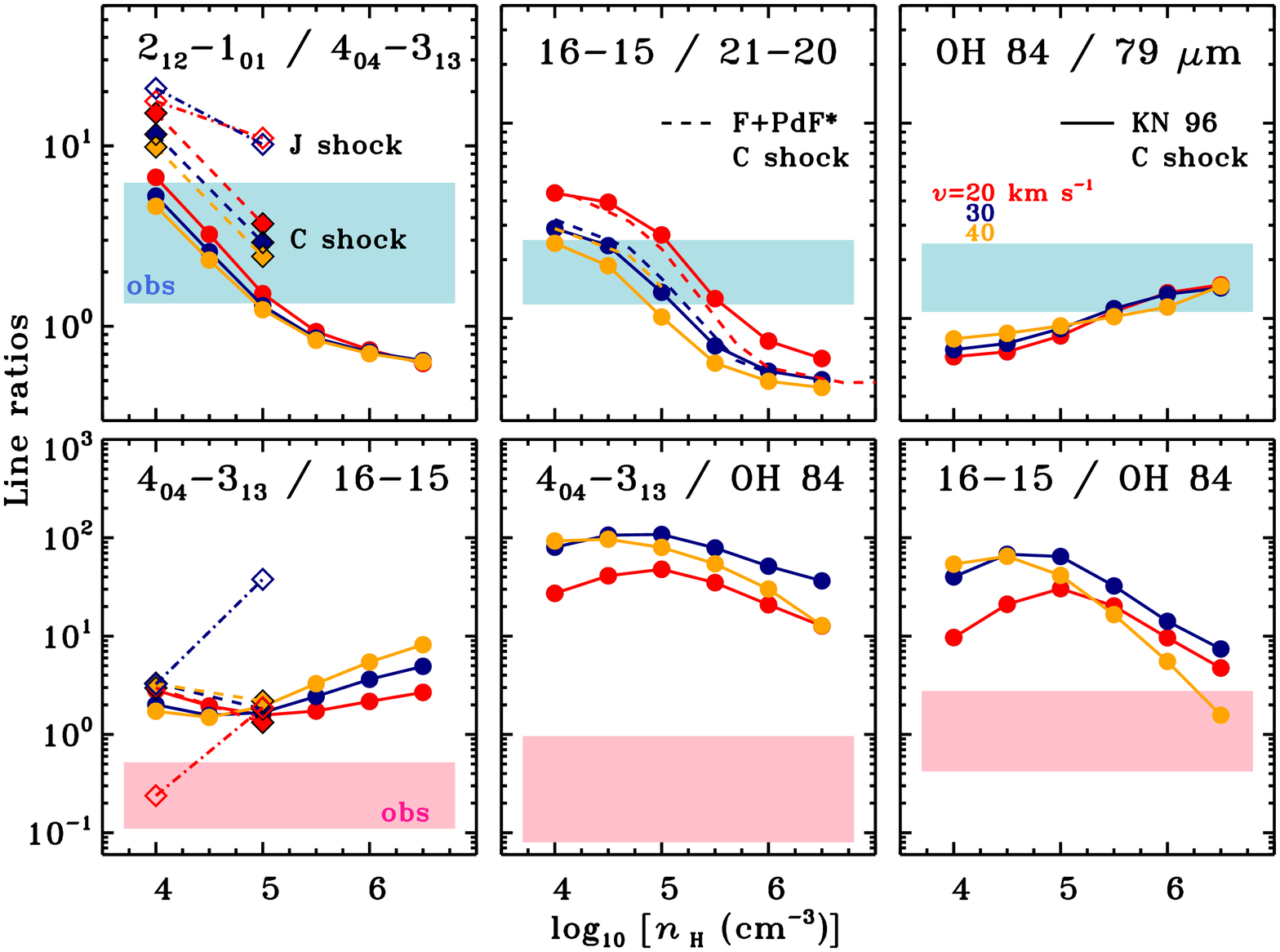}
\caption{\label{summary} Ratios of line fluxes in units of erg cm$^{-2}$ s$^{-1}$ 
as a function of logarithm of density of the pre-shock gas, $n_{H}$. Ratios of different 
transitions of the same molecules are shown at the top row and line ratios comparing different 
species are shown at the bottom. Filled symbols and full lines show models of $C$ shocks (circles -- from 
Kaufman \& Neufeld, diamonds -- from Flower \& Pineau des For\^ets), whereas the empty symbols and 
dash-dotted lines show models of $J$ shocks (Flower \& Pineau des For\^ets 2010). Colors distinguish shock velocities 
-- 20 km\,s$^{-1}$ shocks are shown in red, 30 km\,s$^{-1}$ in blue, and 40 km\,s$^{-1}$ in orange.}
\end{center}
\end{figure*}

\subsection{Shock parameters and physical conditions}

Spectrally resolved HIFI observations of the CO 10--9 and 16--15 line
profiles (Kristensen et al. 2013 and in prep., Yildiz et al. 2013) as
well as various H$_2$O transitions (Kristensen et al. 2012, Mottram et
al. 2014) reveal at least two different kinematic shock
components: non-dissociative C-type shocks in a thin layer along the
cavity walls (so-called `cavity shocks') and J-shocks at the base of
the outflow (also called `spot shocks'), both caused by interaction of
the wind with the envelope. Both shocks are different from the much
cooler entrained outflow gas that is observed in the low-$J$ CO line
profiles \citep{Yi13}.

One possible physical explanation for our observed lack of variation
is that although the outflow structure depends on the mass entrainment
efficiency and the amount of mass available to entrain (the envelope
mass), the wind causing the shocks does not depend on these
parameters. Instead the cavity shock caused by the wind impinging on
the inner envelope depends on the shock velocity and the density of
the inner envelope (Kristensen et al. 2013, Mottram et al. 2014). Thus, the lack of 
significant variation in the line ratios
suggests that the shock velocities by the oblique impact of the wind
are always around 20--30 km s$^{-1}$. 

In Section 4 the observed emission was compared primarily to models of $C-$type
 shock emission. Although $J-$type shocks play a role on small spatial scales in 
 low-mass protostars \citep{Kr12,Kr13,Mo14}
  their contribution to CO emission originating in levels with $J_{\rm up} \lesssim 30$
   is typically less than $\sim$ 50\%. Since higher$-J$ CO emission is only detected
    toward 30\% of the sources, the $J-$type shock component is ignored for CO. For 
    the case of H$_2$O, spectrally resolved line profiles observed with HIFI reveal 
    that the profiles do not change significantly with excitation up to $E_{\rm up}$ = 250 K
     (Mottram et al. in 2014); $J-$type shock components typically contribute $<$ 10\% of
      the emission. It is unclear if the trend of line profiles not changing with excitation
       continues to higher upper-level energies, in particular all the way up to 
       $E_{\rm up}$ = 1070 K ($J$ = 8$_{18}$—$7_{07}$ at 63.32 $\mu$m). OH and [\ion{O}{i}],
        on the other hand, almost certainly trace dissociative $J-$type shocks 
        \citep[e.g.,][]{vK10,Wa13} but a full analysis of 
        their emission will be presented in a forthcoming paper. Thus, in the following the 
        focus remains on comparing emission to models of $C-$type shocks.

Figure~\ref{summary} summarizes the different line ratios as a
function of pre-shock density discussed in the previous
sections. General agreement is found between the observations and
models when line ratios of different transitions of {\it the same
  species} are used (see top row for H$_\mathrm{2}$O, CO, and OH
examples), indicating that the excitation of individual species is
reproduced well by the models.
The H$_\mathrm{2}$O line ratios are a sensitive tracer of the pre-shock gas density 
since they vary less with shock velocity than those of
CO. The $C$ shock models from KN96 with pre-shock gas densities in
the range of 10$^{4}$-10$^{5}$ cm$^{-3}$ are a best match to the
observed ratios, consistent with values of 10$^{5}$ cm$^{-3}$ from the
$C$ shock models of F+PdF10. For the considered range of shock
velocities, the compression factor in those shocks, defined as the
ratio of the post-shock and pre-shock gas densities, varies from about
10 to 30 \citep{ND89,Dr93,Ka13}. The resulting values of post-shock
densities, traced by the observed molecules, are therefore expected to be 
$\geq10^{5}$-$10^{6}$ cm$^{-3}$.  

The CO line ratios, on the other hand, are not only sensitive to
density, but also to the shock velocities, due to their connection to
the peak temperature attained in the shock. In the pre-shock density
range of $\geq10^{4}$-$10^{5}$ cm$^{-3}$, indicated by the
H$_\mathrm{2}$O line ratios, shocks with velocities above 20 km\,s$^{-1}$ 
best agree with the CO observations. Within this range of
densities, the predictions from both the KN96 and F+PdF10 $C$ shock models 
show a very good agreement with each other.

The ratio of two OH lines from the KN96 models compared with the
observations suggest higher pre-shock densities above $10^{5}$
cm$^{-3}$, but this ratio may be affected by infrared pumping
\citep{Wa13}. Also, some OH emission traces (dissociative) $J-$shocks,
based on its spatial connection and flux correlations to [\ion{O}{I}]
emission \citep{Wa10,Wa13,Ka13}. The single spectrally-resolved OH spectrum towards 
Ser SMM1 \citep[Fig.~3,][]{Kr13} suggests that the contribution of the 
dissociative and non-dissociative shocks is comparable. 
Thus, observed CO/OH and H$_2$O/OH line ratios are only affected at the factor $\sim$2 level 
and the discrepancy in the bottom row of Fig.~\ref{summary} remains.

Overall, the observed CO and H$_2$O
line ratios are best fit with $C-$shock models with pre-shock densities
of $\sim$10$^{5}$\,cm$^{-3}$ and velocities $\gtrsim$20 km\,s$^{-1}$,
with higher velocities needed for the excitation of the highest$-J$ CO lines.

The shock conditions inferred here can be compared to the temperatures
and densities found from single-point non-LTE excitation and
radiative-transfer models, e.g., \textsc{radex} (van der Tak et al. 2007) and from non-LTE radiative
transfer analysis of line intensity ratios \citep[][Mottram et al., in
  prep.]{Kr13} toward  various sources. Typically, densities $\gtrsim10^6$\,cm$^{-3}$ and temperatures 
of $\sim300$ K and $\gtrsim700$ K are required to account for the line emission
\citep{He12,Go12,Sa12,Va12,Ka13,Sa13}. 
Within this range of densities, the predictions from both the KN96 and F+PdF10 
$C-$shock models reproduce CO observations. However, the disconnect
between predicted pre-shock conditions required to reproduce H$_2$O 
and CO is puzzling (see below).

A small number of individual sources have been compared directly
to shock models (Lee et al. 2013, Dionatos et al. 2013) and the
conclusions are similar to what is reported here: pre-shock conditions of
typically $10^{4}-10^{5}$ cm$^{-3}$, and emission originating in 
$C-$type shocks. None of the sources analyzed previously were
therefore special or atypical, rather these shock conditions appear to
exist toward every embedded protostar. 

At shock positions away from the protostar,
dissociative or non-dissociative $J-$type shocks at the same pre-shock
densities are typically invoked to explain the FIR line emission \citep{Be12,Sa12,Bu14}.
Differences between the protostar position and the distant shock positions are revealed 
primarily by our line ratios using the
low-excitation H$_\mathrm{2}$O 2$_{12}$-1$_{01}$ (Table 3) and can be
ascribed to the differences in the filling factors and column densities between the
immediate surrounding of the protostar and the more distant shock positions (Mottram et al. 2014).

\subsection{Abundances and need for UV radiation}

\begin{table}
\centering
\caption{Fraction of H$_2$O and OH emission with respect to total far-IR molecular emission}
\label{tab:cool}
\renewcommand{\footnoterule}{}  
\begin{tabular}{lcrcccccc}
\hline
log $n_\mathrm{H}$(cm $^{-3}$) & Obs. (\%) & \multicolumn{3}{c}{KN96 models (\%)} \\
~ & ~ & \multicolumn{3}{c}{$\varv$ (km s$^{-1}$):} \\\cline{3-5}
~ & ~ & 20 & 30 & 40 \\
\hline
\multicolumn{5}{c}{H$_2$O / (CO+H$_2$O+OH)} \\
4    & 29 & 95.5 & 92.6 & 90.0 \\
4.5  & 29 & 91.6 & 86.3 & 82.8 \\
5    & 29 & 84.6 & 78.2 & 75.3 \\
5.5  & 29 & 77.3 & 71.2 & 70.4 \\
6    & 29 & 71.3 & 68.5 & 70.7 \\
6.5  & 29 & 68.3 & 69.6 & 73.3\\
\hline
\multicolumn{5}{c}{OH / (CO+H$_2$O+OH)} \\
4    & 25 & 0.7 & 0.3 & 0.2 \\
4.5  & 25 & 0.7 & 0.3 & 0.3 \\
5    & 25 & 0.7 & 0.3 & 0.4 \\
5.5  & 25 & 0.9 & 0.4 & 0.6 \\
6    & 25 & 1.3 & 0.5 & 1.1 \\
6.5  & 25 & 1.9 & 0.8 & 2.4 \\
\hline
\end{tabular}
\tablefoot{In this analysis, the following lines are used: H$_\mathrm{2}$O lines at 179.5 $\mu$m, 
125.4 $\mu$m, 108.1 $\mu$m, and 90.0 $\mu$m; CO lines at 162.8 $\mu$m, 124.2 $\mu$m, 
108.7 $\mu$m, and 90.2 $\mu$m; OH lines at 79 $\mu$m and 84 $\mu$m. Median values of the fractions  
calculated from 18 sources with line detections are adopted in case of observations (column Obs.).}
\end{table}

In contrast with the ratios of two H$_\mathrm{2}$O or CO lines, the
ratios calculated using {\it different} species do not agree with the
shock models (Figure~\ref{summary}, bottom row). The ratios of H$_\mathrm{2}$O-to-CO lines are
overproduced by the $C$ shock models from both the KN96 and F+PdF10
grids by at least an order of magnitude, irrespective of the assumed
shock velocity. Although there are a few exceptions (e.g. the ratio of
H$_\mathrm{2}$O 2$_{12}$-1$_{01}$ and CO 16-15), the majority of the
investigated sets of H$_\mathrm{2}$O and CO lines follow the same
trend. Observations agree only with slow, $<$20 km\,s$^{-1}$, $J$ shock
models, but as shown above, those models do not seem to reproduce the
excitation properly (Fig.~\ref{allvel}).

The discrepancy between the models and observations is even larger in
the case of the H$_\mathrm{2}$O-to-OH line ratios, as illustrated in
Fig.~\ref{summary}. The two orders of magnitude disagreement with
the $C$ shock models cannot be accounted by any excitation effects for
any realistic shock parameters. Additional comparison to $J$ shock
models is not possible due to a lack of OH predictions for $J$ shocks in
the F+PdF10 models.
 
The CO-to-OH ratios are overproduced by about an order of magnitude in
the $C$ shock models, similar to the H$_\mathrm{2}$O-to-CO ratios. The
agreement improves for fast ($\varv$ = 40 km s$^{-1}$) shocks in high density pre-shock medium
($\sim10^{6.5}$ cm$^{-3}$), but those parameters are not consistent
with the line ratios from the same species.

An additional test of the disagreement between models and observations
is provided by calculating the fraction of each species with regard to
the sum of CO, H$_\mathrm{2}$O and OH emission.  For that purpose,
only the strongest lines observed in our program are used.  As seen in
Table \ref{tab:cool}, the observed percentage (median) of
H$_\mathrm{2}$O is about 30 \% and OH is about 25 \%. In contrast,
KN96 models predict typically 70-90 \% of flux in the chosen
H$_\mathrm{2}$O lines and only up to 2\% in the OH lines.
 
The only possible way to reconcile the models with the observations, after
concluding that the excitation is treated properly in the models, is
to reconsider the assumed abundances.  The fact that the
H$_\mathrm{2}$O-to-CO and CO-to-OH ratios are simultaneously
overestimated suggests a problem with the abundances of
H$_\mathrm{2}$O and OH, rather than that of CO.  The scenario with the
overestimated H$_\mathrm{2}$O abundances and underestimated OH
abundances would translate into a too large H$_\mathrm{2}$O-to-OH
abundance ratio in the models. A possible and likely solution is
photodissocation of H$_2$O to OH and subsequently to atomic
oxygen. As noted above, some OH also comes from the dissociative shock 
seen in [\ion{O}{I}]. 

A significant shortcoming of all these shock models lies in their inability
to account for grain-grain interactions, which has been shown to significantly
alter the structure of the shocks propagating in dense media 
\citep[$n_{\rm H} > 10^5$~cm$^{-3}$,][]{Gu07,Gui09,Gu11}.
These grain-grain interactions mostly consist of coagulation, vaporization,
and shattering effects affecting the grains. Their inclusion in shock models
necessitates a sophisticated treatment of the grains, especially following
their charge and size distribution \citep{Gu07}. Most remarkably,
such interactions eventually result in the creation of small grain fragments
in large numbers, which increases the total dust grain surface area and
thereby changes the coupling between the neutral and the charged fluids
within the shock layer. The net effect is that the shock layer is 
significantly hotter and thinner \citep{Gu11}, which in turn
affects the chemistry and emission of molecules \citep{Gui09}.

Unfortunately, at the moment these models are computationally expensive
and are not well-suited for a grid analysis; moreover, the solutions
found by Guillet et al. do not converge for preshock densities of
10$^6$~cm$^{-3}$ or higher. A recent study by \citet{An13}
shows that it is possible to approximate these effects in a computationally
efficient way, and subsequently evaluated the line intensities of CO, OH
and H$_2$O on a small grid of models. When including grain-grain interactions,
CO lines were found to be significantly less emitting than in \lq simpler' models,
while a smaller decrease was found for H$_2$O lines, and a small increase for OH.
These trends probably still need to be systematically investigated on larger grids
of models before they can be applied to our present comparison efforts.

Regardless of the effect of grain-grain interactions, 
\citet{Sn05} invoked several scenarios to reconcile high absolute
H$_\mathrm{2}$O fluxes with the shock models for the case of supernova
remnants. These include (i) the high ratio of atomic to molecular hydrogen, 
which drives H$_\mathrm{2}$O back to OH and O, (ii) freeze-out of H$_\mathrm{2}$O in the post-shock gas, (iii)
freeze-out of H$_\mathrm{2}$O in the pre-shock gas, and (iv)
photodissociation of H$_\mathrm{2}$O in the pre- and post-shock
gas. Due to the high activation barrier of the H$_\mathrm{2}$O + H $\to$
OH + H$_\mathrm{2}$ reaction ($\sim10^4$ K), the first scenario is not
viable. The freeze-out in the post-shock gas (ii) is not effective in
the low density regions considered in \citet{Sn05}, but can play a
role in the vicinity of protostars, where densities above
$\sim10^6$ cm$^{-3}$ are found \citep[e.g.][this work]{Kr12}.
However, this mechanism alone would not explain the bright OH and
high$-J$ H$_\mathrm{2}$O lines seen toward many deeply-embedded
sources \citep[e.g.][]{Ka13,Wa13}. A similar problem is related to the
freeze-out in the pre-shock gas (iii), which decreases the amount of
e.g. O, OH, and H$_\mathrm{2}$O in the gas phase for shock velocities
below 15 km\,s$^{-1}$.
 
Therefore, the most likely reason for the overproduction of
H$_\mathrm{2}$O in the current generation of shock models, at the
expense of OH, is the omission of the effects of ultraviolet
irradiation (scenario iv) of the shocked material. The presence of UV
radiation is directly seen in Ly-$\alpha$ emission both in the
outflow-envelope shocks \citep{Cu95,Wa03} and at the protostar
position \citep{Va00,Yan12}. Additionally, UV radiation on scales of a few 1000
AU has been inferred from the narrow profiles of $^{13}$CO 6-5
observed from the ground toward a few low-mass protostars
\citep{Sp95,vK09,Yi12}. H$_\mathrm{2}$O can be photodissociated into
OH over a broad range of far-UV wavelengths, including by Ly-$\alpha$,
and this would provide an explanation for the disagreement between our
observations and the models. Photodissociation of CO is less likely,
given the fact that it cannot be dissociated by Ly-$\alpha$ and only
by very hard UV photons with wavelengths $<$ 1000 \AA. The lack of CO
photodissociation is consistent with weak [\ion{C}{i}] and
[\ion{C}{ii}] emission observed toward low-mass YSOs
\citep{Yi12,Go12,Ka13}. At the positions away from the protostars, on
the other hand, the bow-shocks at the tip of the protostellar jets can
produce significant emission in the [\ion{C}{i}] \citep{vK09}.
Therefore, it is unlikely that lower line excitation at those
positions is due to the weaker UV. The differences seen in the
resolved line profiles \citep[e.g.][Mottram et al. in
  prep.]{Sa12,Va12} indicate that the lower column densities involved
are the more likely reason for differences in the excitation.

\citet{Vi12} proposed a scenario in which the lower-lying CO
transitions observed with PACS ($14<J_\mathrm{u}<23$) originate in
UV--heated gas and higher$-J$ transitions ($J_\mathrm{u}$ $>$ 24) in
shocked material. All water emission would be associated with the same
shocks as responsible for the higher-$J$ CO emission with less than 1\% of
the water emission coming from the PDR layer. Although not
modeled explicitly, the UV irradiation from the star-disk boundary
impinging on the shocks naturally accounts for the lower
H$_\mathrm{2}$O abundance and exceeds by at least two orders of magnitude the 
H$_\mathrm{2}$O destruction rate by He$^{+}$ and H$_{\mathrm{3}}^{+}$, assuming 
a normal interstellar radiation field (G$_\mathrm{0}=1$).
The authors predict that while the dynamics of the hot layers where
both shocks and UV irradiation play a role will be dominated by the
shocks, only the UV photons penetrate further into the envelope, where
the dynamics would resemble the quiescent envelope. The
lower-temperature UV-heated gas has indeed been observed to be
quiescent on the spatial scales of the outflow cavity through
observations of medium-$J$ $^{13}$CO lines (Y{\i}ld{\i}z et al. 2012,
subm.).

\citet{FP13} proposed a model where all emission originates in a
non-stationary shock wave, where a $J-$type shock is embedded in a
$C-$type shock. Without a detailed modeling of individual sources based
on different source parameters it is not possible to rule out any of
these solutions. However, the trends reported here suggest that it is
possible to find a pure shock solution, in agreement with
\citet{FP13}, as long as UV photons are incorporated to provide
dissociation of H$_\mathrm{2}$O. Complementary observations,
preferably at higher angular resolution, are required to break the
solution degeneracy and determine the relative role the shocks and UV photons
play on the spatial scales of the thickness of the cavity wall. Models
whose results depend sensitively on a single parameter such as time,
are ruled out by the fact that the observed line ratios are so similar
across sources.
  
\section{Conclusions}

We have compared the line ratios of the main molecular cooling lines
detected in 22 low-mass protostars using \textit{Herschel}/PACS with
publicly available one-dimensional shock models. Our conclusions are
the following:
\begin{itemize}
\item Line ratios of various species and transitions are remarkably similar for all observed 
sources. No correlation is found with source physical parameters. 

\item Line ratios observed toward the protostellar position are consistent with the values 
reported for the positions away from the protostar, except for some ratios 
involving the low-excitation 
H$_\mathrm{2}$O 2$_{12}$-1$_{01}$ line. 
Coupled with the larger absolute fluxes of highly-excited H$_\mathrm{2}$O and CO lines 
at the protostellar positions, this indicates that lines at distant off-source shock positions are less excited. 

\item General agreement is found between the observed line ratios of the same species (H$_\mathrm{2}$O, 
CO, and OH) and the $C$ shock models from Kaufman \& Neufeld (1996) and Flower \& Pineau des For\^ets (2010). 
Ratios of H$_\mathrm{2}$O are particularly good tracers of the density of the ambient material 
and indicate pre-shock densities of order $\geq10^{5}$ cm$^{-3}$ and thus post-shock densities of order $10^6$ cm$^{-3}$.
 Ratios of CO lines are more sensitive to the shock velocities and, for the derived range of pre-shock densities, indicate
shock velocities above 20 km\,s$^{-1}$. 

\item Ratios of CO lines located in the \lq warm' component of CO ladders 
(with $T_\mathrm{rot}\sim$300 K) are reproduced with shock velocities of 20-30 km\,s$^{-1}$
and pre-shock densities of $10^{5}$ cm$^{-3}$. The CO ratios using the lines from the 
 \lq hot' component ($T_\mathrm{rot}\gtrsim$700 K) are better reproduced by models with shock velocities above 
25 km s$^{-1}$.

\item A lack of agreement is found between models and the observed line ratios of different species.
The H$_\mathrm{2}$O-to-CO, H$_\mathrm{2}$O-to-OH, and CO-to-OH line ratios are all overproduced by 
the models by 1-2 orders of magnitude for the majority of the considered sets of transitions.

\item Since the observed molecular excitation is properly reproduced
  in the $C$ shock models, the most likely reason for disagreement with
  observations is the abundances in the shock models, which are too
  high in case of H$_\mathrm{2}$O and too low in case of OH. Invoking
  UV irradiation of the shocked material, together with a dissociative
  $J$ shock contribution to OH and [\ion{O}{i}], would lower the
  H$_2$O abundance and reconcile the models and observations.

\end{itemize}

New UV-irradiated shock models will allow us to constrain the UV field needed
 to reconcile the shock models with observations (M. Kaufman, priv. comm.) 
 Those models should also account for the grain-grain processing, which 
 affects significantly the shocks structure at densities $\sim10^6$ cm$^{-3}$ \citep{Gu11}.
 The effects of shock irradiation as a function of the distance from a protostar 
 will help to understand the differences in the observed spectrally-resolved 
 lines from HIFI at \lq on source' and distant shock-spot positions.

\begin{acknowledgements}
Herschel is an ESA space observatory with science instruments provided
by European-led Principal Investigator consortia and with important
participation from NASA. AK acknowledges support from  
the Polish National Science Center grant 2013/11/N/ST9/00400. 
Astrochemistry in Leiden is supported by the Netherlands Research
School for Astronomy (NOVA), by a Royal Netherlands Academy of Arts
and Sciences (KNAW) professor prize, by a Spinoza grant and grant
614.001.008 from the Netherlands Organisation for Scientific Research
(NWO), and by the European Community's Seventh Framework Programme
FP7/2007-2013 under grant agreement 238258 (LASSIE). NJE was supported by NASA through
an award issued by the Jet Propulsion Laboratory, California Institute of Technology.
\end{acknowledgements}   
   
\bibliographystyle{aa}
\bibliography{biblio14}

\begin{thebibliography}{120}
\expandafter\ifx\csname natexlab\endcsname\relax\def\natexlab#1{#1}\fi

\bibitem[{{Anderl} {et~al.}(2013){Anderl}, {Guillet}, {Pineau des For{\^e}ts},
  \& {Flower}}]{An13}
{Anderl}, S., {Guillet}, V., {Pineau des For{\^e}ts}, G., \& {Flower}, D.~R.
  2013, \aap, 556, A69

\bibitem[{{Andr{\'e}} {et~al.}(2010){Andr{\'e}}, {Men'shchikov}, {Bontemps},
  {K{\"o}nyves}, {Motte}, {Schneider}, {Didelon}, {Minier}, {Saraceno},
  {Ward-Thompson}, {di Francesco}, {White}, {Molinari}, {Testi}, {Abergel},
  {Griffin}, {Henning}, {Royer}, {Mer{\'{\i}}n}, {Vavrek}, {Attard},
  {Arzoumanian}, {Wilson}, {Ade}, {Aussel}, {Baluteau}, {Benedettini},
  {Bernard}, {Blommaert}, {Cambr{\'e}sy}, {Cox}, {di Giorgio}, {Hargrave},
  {Hennemann}, {Huang}, {Kirk}, {Krause}, {Launhardt}, {Leeks}, {Le Pennec},
  {Li}, {Martin}, {Maury}, {Olofsson}, {Omont}, {Peretto}, {Pezzuto}, {Prusti},
  {Roussel}, {Russeil}, {Sauvage}, {Sibthorpe}, {Sicilia-Aguilar}, {Spinoglio},
  {Waelkens}, {Woodcraft}, \& {Zavagno}}]{An10}
{Andr{\'e}}, P., {Men'shchikov}, A., {Bontemps}, S., {et~al.} 2010, \aap, 518,
  L102

\bibitem[{{Arce} {et~al.}(2010){Arce}, {Borkin}, {Goodman}, {Pineda}, \&
  {Halle}}]{Ar10}
{Arce}, H.~G., {Borkin}, M.~A., {Goodman}, A.~A., {Pineda}, J.~E., \& {Halle},
  M.~W. 2010, \apj, 715, 1170

\bibitem[{{Arce} {et~al.}(2007){Arce}, {Shepherd}, {Gueth}, {Lee}, {Bachiller},
  {Rosen}, \& {Beuther}}]{Ar07}
{Arce}, H.~G., {Shepherd}, D., {Gueth}, F., {et~al.} 2007, Protostars and
  Planets V, University of Arizona Press (2006), eds. B. Reipurth, D. Jewitt,
  and K. Keil, 245

\bibitem[{{Baulch} {et~al.}(1992){Baulch}, {Cobos}, {Cox}, {Esser}, {Frank},
  {Just}, {Kerr}, {Pilling}, {Troe}, {Walker}, \& {Warnatz}}]{Ba12}
{Baulch}, D., {Cobos}, C., {Cox}, R., {et~al.} 1992, J. Phys. Chem. Ref. Data,
  21, 411

\bibitem[{{Benedettini} {et~al.}(2012){Benedettini}, {Busquet}, {Lefloch},
  {Codella}, {Cabrit}, {Ceccarelli}, {Giannini}, {Nisini}, {Vasta},
  {Cernicharo}, {Lorenzani}, \& {di Giorgio}}]{Be12}
{Benedettini}, M., {Busquet}, G., {Lefloch}, B., {et~al.} 2012, \aap, 539, L3

\bibitem[{{Bergin} {et~al.}(1998){Bergin}, {Neufeld}, \& {Melnick}}]{Be98}
{Bergin}, E.~A., {Neufeld}, D.~A., \& {Melnick}, G.~J. 1998, \apj, 499, 777

\bibitem[{{Brinch} \& {Hogerheijde}(2010)}]{Br10}
{Brinch}, C. \& {Hogerheijde}, M.~R. 2010, \aap, 523, A25

\bibitem[{{Busquet} {et~al.}(2014){Busquet}, {Lefloch}, {Benedettini},
  {Ceccarelli}, {Codella}, {Cabrit}, {Nisini}, {Viti}, {G{\'o}mez-Ruiz},
  {Gusdorf}, {di Giorgio}, \& {Wiesenfeld}}]{Bu14}
{Busquet}, G., {Lefloch}, B., {Benedettini}, M., {et~al.} 2014, \aap, 561, A120

\bibitem[{{Ceccarelli} {et~al.}(2002){Ceccarelli}, {Boogert}, {Tielens},
  {Caux}, {Hogerheijde}, \& {Parise}}]{Ce02}
{Ceccarelli}, C., {Boogert}, A.~C.~A., {Tielens}, A.~G.~G.~M., {et~al.} 2002,
  \aap, 395, 863

\bibitem[{{Chieze} {et~al.}(1998){Chieze}, {Pineau des Forets}, \&
  {Flower}}]{Ch98}
{Chieze}, J.-P., {Pineau des Forets}, G., \& {Flower}, D.~R. 1998, \mnras, 295,
  672

\bibitem[{{Clegg} {et~al.}(1996){Clegg}, {Ade}, {Armand}, {Baluteau}, {Barlow},
  {Buckley}, {Berges}, {Burgdorf}, {Caux}, {Ceccarelli}, {Cerulli}, {Church},
  {Cotin}, {Cox}, {Cruvellier}, {Culhane}, {Davis}, {di Giorgio}, {Diplock},
  {Drummond}, {Emery}, {Ewart}, {Fischer}, {Furniss}, {Glencross},
  {Greenhouse}, {Griffin}, {Gry}, {Harwood}, {Hazell}, {Joubert}, {King},
  {Lim}, {Liseau}, {Long}, {Lorenzetti}, {Molinari}, {Murray}, {Naylor},
  {Nisini}, {Norman}, {Omont}, {Orfei}, {Patrick}, {Pequignot}, {Pouliquen},
  {Price}, {Nguyen-Q-Rieu}, {Rogers}, {Robinson}, {Saisse}, {Saraceno},
  {Serra}, {Sidher}, {Smith}, {Smith}, {Spinoglio}, {Swinyard}, {Texier},
  {Towlson}, {Trams}, {Unger}, \& {White}}]{LWS}
{Clegg}, P.~E., {Ade}, P.~A.~R., {Armand}, C., {et~al.} 1996, \aap, 315, L38

\bibitem[{{Codella} {et~al.}(2012){Codella}, {Ceccarelli}, {Lefloch},
  {Fontani}, {Busquet}, {Caselli}, {Kahane}, {Lis}, {Taquet}, {Vasta}, {Viti},
  \& {Wiesenfeld}}]{Cod12}
{Codella}, C., {Ceccarelli}, C., {Lefloch}, B., {et~al.} 2012, \apjl, 757, L9

\bibitem[{{Curiel} {et~al.}(1995){Curiel}, {Raymond}, {Wolfire}, {Hartigan},
  {Morse}, {Schwartz}, \& {Nisenson}}]{Cu95}
{Curiel}, S., {Raymond}, J.~C., {Wolfire}, M., {et~al.} 1995, \apj, 453, 322

\bibitem[{{Davis} {et~al.}(2008){Davis}, {Scholz}, {Lucas}, {Smith}, \&
  {Adamson}}]{Da08}
{Davis}, C.~J., {Scholz}, P., {Lucas}, P., {Smith}, M.~D., \& {Adamson}, A.
  2008, \mnras, 387, 954

\bibitem[{{de Graauw} {et~al.}(2010){de Graauw}, {Helmich}, {Phillips},
  {Stutzki}, {Caux}, {Whyborn}, {Dieleman}, {Roelfsema}, {Aarts}, {Assendorp},
  {Bachiller}, {Baechtold}, {Barcia}, {Beintema}, {Belitsky}, {Benz}, {Bieber},
  {Boogert}, {Borys}, {Bumble}, {Ca{\"i}s}, {Caris}, {Cerulli-Irelli},
  {Chattopadhyay}, {Cherednichenko}, {Ciechanowicz}, {Coeur-Joly}, {Comito},
  {Cros}, {de Jonge}, {de Lange}, {Delforges}, {Delorme}, {den Boggende},
  {Desbat}, {Diez-Gonz{\'a}lez}, {di Giorgio}, {Dubbeldam}, {Edwards},
  {Eggens}, {Erickson}, {Evers}, {Fich}, {Finn}, {Franke}, {Gaier}, {Gal},
  {Gao}, {Gallego}, {Gauffre}, {Gill}, {Glenz}, {Golstein}, {Goulooze},
  {Gunsing}, {G{\"u}sten}, {Hartogh}, {Hatch}, {Higgins}, {Honingh}, {Huisman},
  {Jackson}, {Jacobs}, {Jacobs}, {Jarchow}, {Javadi}, {Jellema}, {Justen},
  {Karpov}, {Kasemann}, {Kawamura}, {Keizer}, {Kester}, {Klapwijk}, {Klein},
  {Kollberg}, {Kooi}, {Kooiman}, {Kopf}, {Krause}, {Krieg}, {Kramer},
  {Kruizenga}, {Kuhn}, {Laauwen}, {Lai}, {Larsson}, {Leduc}, {Leinz}, {Lin},
  {Liseau}, {Liu}, {Loose}, {L{\'o}pez-Fernandez}, {Lord}, {Luinge}, {Marston},
  {Mart{\'{\i}}n-Pintado}, {Maestrini}, {Maiwald}, {McCoey}, {Mehdi}, {Megej},
  {Melchior}, {Meinsma}, {Merkel}, {Michalska}, {Monstein}, {Moratschke},
  {Morris}, {Muller}, {Murphy}, {Naber}, {Natale}, {Nowosielski}, {Nuzzolo},
  {Olberg}, {Olbrich}, {Orfei}, {Orleanski}, {Ossenkopf}, {Peacock}, {Pearson},
  {Peron}, {Phillip-May}, {Piazzo}, {Planesas}, {Rataj}, {Ravera}, {Risacher},
  {Salez}, {Samoska}, {Saraceno}, {Schieder}, {Schlecht}, {Schl{\"o}der},
  {Schm{\"u}lling}, {Schultz}, {Schuster}, {Siebertz}, {Smit}, {Szczerba},
  {Shipman}, {Steinmetz}, {Stern}, {Stokroos}, {Teipen}, {Teyssier}, {Tils},
  {Trappe}, {van Baaren}, {van Leeuwen}, {van de Stadt}, {Visser}, {Wildeman},
  {Wafelbakker}, {Ward}, {Wesselius}, {Wild}, {Wulff}, {Wunsch}, {Tielens},
  {Zaal}, {Zirath}, {Zmuidzinas}, \& {Zwart}}]{dG10}
{de Graauw}, T., {Helmich}, F.~P., {Phillips}, T.~G., {et~al.} 2010, \aap, 518,
  L6

\bibitem[{{Dionatos} {et~al.}(2013){Dionatos}, {J{\o}rgensen}, {Green},
  {Herczeg}, {Evans}, {Kristensen}, {Lindberg}, \& {van Dishoeck}}]{Di13}
{Dionatos}, O., {J{\o}rgensen}, J.~K., {Green}, J.~D., {et~al.} 2013, \aap,
  558, A88

\bibitem[{{Draine}(1980)}]{Dr80}
{Draine}, B.~T. 1980, \apj, 241, 1021

\bibitem[{{Draine} \& {McKee}(1993)}]{Dr93}
{Draine}, B.~T. \& {McKee}, C.~F. 1993, \araa, 31, 373

\bibitem[{{Draine} {et~al.}(1983){Draine}, {Roberge}, \& {Dalgarno}}]{Dr83}
{Draine}, B.~T., {Roberge}, W.~G., \& {Dalgarno}, A. 1983, \apj, 264, 485

\bibitem[{{Elitzur} \& {de Jong}(1978)}]{EJ78b}
{Elitzur}, M. \& {de Jong}, T. 1978, \aap, 67, 323

\bibitem[{{Elitzur} \& {Watson}(1978)}]{EJ78}
{Elitzur}, M. \& {Watson}, W.~D. 1978, \aap, 70, 443

\bibitem[{{Enoch} {et~al.}(2009){Enoch}, {Evans}, {Sargent}, \& {Glenn}}]{En09}
{Enoch}, M.~L., {Evans}, II, N.~J., {Sargent}, A.~I., \& {Glenn}, J. 2009,
  \apj, 692, 973

\bibitem[{{Enoch} {et~al.}(2006){Enoch}, {Young}, {Glenn}, {Evans}, {Golwala},
  {Sargent}, {Harvey}, {Aguirre}, {Goldin}, {Haig}, {Huard}, {Lange},
  {Laurent}, {Maloney}, {Mauskopf}, {Rossinot}, \& {Sayers}}]{En06}
{Enoch}, M.~L., {Young}, K.~E., {Glenn}, J., {et~al.} 2006, \apj, 638, 293

\bibitem[{{Evans} {et~al.}(2009){Evans}, {Dunham}, {J{\o}rgensen}, {Enoch},
  {Mer{\'{\i}}n}, {van Dishoeck}, {Alcal{\'a}}, {Myers}, {Stapelfeldt},
  {Huard}, {Allen}, {Harvey}, {van Kempen}, {Blake}, {Koerner}, {Mundy},
  {Padgett}, \& {Sargent}}]{Ev09}
{Evans}, II, N.~J., {Dunham}, M.~M., {J{\o}rgensen}, J.~K., {et~al.} 2009,
  \apjs, 181, 321

\bibitem[{{Faure} {et~al.}(2007){Faure}, {Crimier}, {Ceccarelli}, {Valiron},
  {Wiesenfeld}, \& {Dubernet}}]{Fa07}
{Faure}, A., {Crimier}, N., {Ceccarelli}, C., {et~al.} 2007, \aap, 472, 1029

\bibitem[{{Flower} \& {Gusdorf}(2009)}]{FG09}
{Flower}, D.~R. \& {Gusdorf}, A. 2009, \mnras, 395, 234

\bibitem[{{Flower} \& {Pineau des For{\^e}ts}(2003)}]{FP03}
{Flower}, D.~R. \& {Pineau des For{\^e}ts}, G. 2003, \mnras, 343, 390

\bibitem[{{Flower} \& {Pineau des For{\^e}ts}(2010)}]{FP10}
{Flower}, D.~R. \& {Pineau des For{\^e}ts}, G. 2010, \mnras, 406, 1745

\bibitem[{{Flower} \& {Pineau des For{\^e}ts}(2012)}]{FP12}
{Flower}, D.~R. \& {Pineau des For{\^e}ts}, G. 2012, \mnras, 421, 2786

\bibitem[{{Flower} \& {Pineau des For{\^e}ts}(2013)}]{FP13}
{Flower}, D.~R. \& {Pineau des For{\^e}ts}, G. 2013, \mnras, 436, 2143

\bibitem[{{Frank} {et~al.}(2014){Frank}, {Ray}, {Cabrit}, {Hartigan}, {Arce},
  {Bacciotti}, {Bally}, {Benisty}, {Eisl{\"o}ffel}, {G{\"u}del}, {Lebedev},
  {Nisini}, \& {Raga}}]{Fr14}
{Frank}, A., {Ray}, T.~P., {Cabrit}, S., {et~al.} 2014, Protostars and Planets
  VI, University of Arizona Press (2014), eds. H. Beuther, R. Klessen, C.
  Dullemond, Th. Henning, in press

\bibitem[{{Giannini} {et~al.}(2001){Giannini}, {Nisini}, \&
  {Lorenzetti}}]{Gi01}
{Giannini}, T., {Nisini}, B., \& {Lorenzetti}, D. 2001, \apj, 555, 40

\bibitem[{{Giannini} {et~al.}(2011){Giannini}, {Nisini}, {Neufeld}, {Yuan},
  {Antoniucci}, \& {Gusdorf}}]{Gi11}
{Giannini}, T., {Nisini}, B., {Neufeld}, D., {et~al.} 2011, \apj, 738, 80

\bibitem[{{Goicoechea} {et~al.}(2012){Goicoechea}, {Cernicharo}, {Karska},
  {Herczeg}, {Polehampton}, {Wampfler}, {Kristensen}, {van Dishoeck},
  {Etxaluze}, {Bern{\'e}}, \& {Visser}}]{Go12}
{Goicoechea}, J.~R., {Cernicharo}, J., {Karska}, A., {et~al.} 2012, \aap, 548,
  A77

\bibitem[{{Goldsmith} \& {Langer}(1978)}]{GL78}
{Goldsmith}, P.~F. \& {Langer}, W.~D. 1978, \apj, 222, 881

\bibitem[{{Green} {et~al.}(2013){Green}, {Evans}, {J{\o}rgensen}, {Herczeg},
  {Kristensen}, {Lee}, {Dionatos}, {Yildiz}, {Salyk}, {Meeus}, {Bouwman},
  {Visser}, {Bergin}, {van Dishoeck}, {Rascati}, {Karska}, {van Kempen},
  {Dunham}, {Lindberg}, {Fedele}, \& {DIGIT Team}}]{Gr13}
{Green}, J.~D., {Evans}, II, N.~J., {J{\o}rgensen}, J.~K., {et~al.} 2013, \apj,
  770, 123

\bibitem[{{Green} {et~al.}(1993){Green}, {Maluendes}, \& {McLean}}]{Gr93}
{Green}, S., {Maluendes}, S., \& {McLean}, A.~D. 1993, \apjs, 85, 181

\bibitem[{{Gueth} \& {Guilloteau}(1999)}]{Gu99}
{Gueth}, F. \& {Guilloteau}, S. 1999, \aap, 343, 571

\bibitem[{{Guillet} {et~al.}(2009){Guillet}, {Jones}, \& {Pineau Des
  For{\^e}ts}}]{Gui09}
{Guillet}, V., {Jones}, A.~P., \& {Pineau Des For{\^e}ts}, G. 2009, \aap, 497,
  145

\bibitem[{{Guillet} {et~al.}(2007){Guillet}, {Pineau Des For{\^e}ts}, \&
  {Jones}}]{Gu07}
{Guillet}, V., {Pineau Des For{\^e}ts}, G., \& {Jones}, A.~P. 2007, \aap, 476,
  263

\bibitem[{{Guillet} {et~al.}(2011){Guillet}, {Pineau Des For{\^e}ts}, \&
  {Jones}}]{Gu11}
{Guillet}, V., {Pineau Des For{\^e}ts}, G., \& {Jones}, A.~P. 2011, \aap, 527,
  A123

\bibitem[{{Gusdorf} {et~al.}(2011){Gusdorf}, {Giannini}, {Flower}, {Parise},
  {G{\"u}sten}, \& {Kristensen}}]{Gus11}
{Gusdorf}, A., {Giannini}, T., {Flower}, D.~R., {et~al.} 2011, \aap, 532, A53

\bibitem[{{Gusdorf} {et~al.}(2008){Gusdorf}, {Pineau Des For{\^e}ts}, {Cabrit},
  \& {Flower}}]{Gus08}
{Gusdorf}, A., {Pineau Des For{\^e}ts}, G., {Cabrit}, S., \& {Flower}, D.~R.
  2008, \aap, 490, 695

\bibitem[{{Gutermuth} {et~al.}(2009){Gutermuth}, {Megeath}, {Myers}, {Allen},
  {Pipher}, \& {Fazio}}]{Gu09}
{Gutermuth}, R.~A., {Megeath}, S.~T., {Myers}, P.~C., {et~al.} 2009, \apjs,
  184, 18

\bibitem[{{Gutermuth} {et~al.}(2010){Gutermuth}, {Megeath}, {Myers}, {Allen},
  {Pipher}, \& {Fazio}}]{Gu10}
{Gutermuth}, R.~A., {Megeath}, S.~T., {Myers}, P.~C., {et~al.} 2010, \apjs,
  189, 352

\bibitem[{{Hatchell} {et~al.}(2007{\natexlab{a}}){Hatchell}, {Fuller}, \&
  {Richer}}]{Ha07b}
{Hatchell}, J., {Fuller}, G.~A., \& {Richer}, J.~S. 2007{\natexlab{a}}, \aap,
  472, 187

\bibitem[{{Hatchell} {et~al.}(2007{\natexlab{b}}){Hatchell}, {Fuller},
  {Richer}, {Harries}, \& {Ladd}}]{Ha07a}
{Hatchell}, J., {Fuller}, G.~A., {Richer}, J.~S., {Harries}, T.~J., \& {Ladd},
  E.~F. 2007{\natexlab{b}}, \aap, 468, 1009

\bibitem[{{Herczeg} {et~al.}(2012){Herczeg}, {Karska}, {Bruderer},
  {Kristensen}, {van Dishoeck}, {J{\o}rgensen}, {Visser}, {Wampfler}, {Bergin},
  {Y{\i}ld{\i}z}, {Pontoppidan}, \& {Gracia-Carpio}}]{He12}
{Herczeg}, G.~J., {Karska}, A., {Bruderer}, S., {et~al.} 2012, \aap, 540, A84

\bibitem[{{Hirano} {et~al.}(2010){Hirano}, {Ho}, {Liu}, {Shang}, {Lee}, \&
  {Bourke}}]{Hi10}
{Hirano}, N., {Ho}, P.~P.~T., {Liu}, S.-Y., {et~al.} 2010, \apj, 717, 58

\bibitem[{{Hirota} {et~al.}(2008){Hirota}, {Bushimata}, {Choi}, {Honma},
  {Imai}, {Iwadate}, {Jike}, {Kameya}, {Kamohara}, {Kan-Ya}, {Kawaguchi},
  {Kijima}, {Kobayashi}, {Kuji}, {Kurayama}, {Manabe}, {Miyaji}, {Nagayama},
  {Nakagawa}, {Oh}, {Omodaka}, {Oyama}, {Sakai}, {Sasao}, {Sato}, {Shibata},
  {Tamura}, \& {Yamashita}}]{Hir08}
{Hirota}, T., {Bushimata}, T., {Choi}, Y.~K., {et~al.} 2008, \pasj, 60, 37

\bibitem[{{Hollenbach}(1997)}]{Hol97}
{Hollenbach}, D. 1997, in IAU Symposium, Vol. 182, Herbig-Haro Flows and the
  Birth of Stars, ed. B.~{Reipurth} \& C.~{Bertout}, 181--198

\bibitem[{{Hollenbach} {et~al.}(1989){Hollenbach}, {Chernoff}, \&
  {McKee}}]{Hol89}
{Hollenbach}, D.~J., {Chernoff}, D.~F., \& {McKee}, C.~F. 1989, in ESA Special
  Publication, Vol. 290, Infrared Spectroscopy in Astronomy, ed.
  E.~{B{\"o}hm-Vitense}, 245--258

\bibitem[{{J{\o}rgensen} {et~al.}(2006){J{\o}rgensen}, {Harvey}, {Evans},
  {Huard}, {Allen}, {Porras}, {Blake}, {Bourke}, {Chapman}, {Cieza}, {Koerner},
  {Lai}, {Mundy}, {Myers}, {Padgett}, {Rebull}, {Sargent}, {Spiesman},
  {Stapelfeldt}, {van Dishoeck}, {Wahhaj}, \& {Young}}]{Jo06}
{J{\o}rgensen}, J.~K., {Harvey}, P.~M., {Evans}, II, N.~J., {et~al.} 2006,
  \apj, 645, 1246

\bibitem[{{J{\o}rgensen} {et~al.}(2007){J{\o}rgensen}, {Johnstone}, {Kirk}, \&
  {Myers}}]{Jo07b}
{J{\o}rgensen}, J.~K., {Johnstone}, D., {Kirk}, H., \& {Myers}, P.~C. 2007,
  \apj, 656, 293

\bibitem[{{Karska} {et~al.}(2013){Karska}, {Herczeg}, {van Dishoeck},
  {Wampfler}, {Kristensen}, {Goicoechea}, {Visser}, {Nisini}, {San
  Jos{\'e}-Garc{\'{\i}}a}, {Bruderer}, {{\'S}niady}, {Doty}, {Fedele},
  {Y{\i}ld{\i}z}, {Benz}, {Bergin}, {Caselli}, {Herpin}, {Hogerheijde},
  {Johnstone}, {J{\o}rgensen}, {Liseau}, {Tafalla}, {van der Tak}, \&
  {Wyrowski}}]{Ka13}
{Karska}, A., {Herczeg}, G.~J., {van Dishoeck}, E.~F., {et~al.} 2013, \aap,
  552, A141

\bibitem[{{Kaufman} \& {Neufeld}(1996)}]{KN96}
{Kaufman}, M.~J. \& {Neufeld}, D.~A. 1996, \apj, 456, 611

\bibitem[{{Kessler} {et~al.}(1996){Kessler}, {Steinz}, {Anderegg}, {Clavel},
  {Drechsel}, {Estaria}, {Faelker}, {Riedinger}, {Robson}, {Taylor}, \&
  {Xim{\'e}nez de Ferr{\'a}n}}]{ISO}
{Kessler}, M.~F., {Steinz}, J.~A., {Anderegg}, M.~E., {et~al.} 1996, \aap, 315,
  L27

\bibitem[{{Knee} \& {Sandell}(2000)}]{KS00}
{Knee}, L.~B.~G. \& {Sandell}, G. 2000, \aap, 361, 671

\bibitem[{{Kristensen} {et~al.}(2007){Kristensen}, {Ravkilde}, {Field},
  {Lemaire}, \& {Pineau Des For{\^e}ts}}]{Kr07}
{Kristensen}, L.~E., {Ravkilde}, T.~L., {Field}, D., {Lemaire}, J.~L., \&
  {Pineau Des For{\^e}ts}, G. 2007, \aap, 469, 561

\bibitem[{{Kristensen} {et~al.}(2013){Kristensen}, {van Dishoeck}, {Benz},
  {Bruderer}, {Visser}, \& {Wampfler}}]{Kr13}
{Kristensen}, L.~E., {van Dishoeck}, E.~F., {Benz}, A.~O., {et~al.} 2013, \aap,
  557, A23

\bibitem[{{Kristensen} {et~al.}(2012){Kristensen}, {van Dishoeck}, {Bergin},
  {Visser}, {Y{\i}ld{\i}z}, {San Jose-Garcia}, {J{\o}rgensen}, {Herczeg},
  {Johnstone}, {Wampfler}, {Benz}, {Bruderer}, {Cabrit}, {Caselli}, {Doty},
  {Harsono}, {Herpin}, {Hogerheijde}, {Karska}, {van Kempen}, {Liseau},
  {Nisini}, {Tafalla}, {van der Tak}, \& {Wyrowski}}]{Kr12}
{Kristensen}, L.~E., {van Dishoeck}, E.~F., {Bergin}, E.~A., {et~al.} 2012,
  \aap, 542, A8

\bibitem[{{Kristensen} {et~al.}(2010){Kristensen}, {Visser}, {van Dishoeck},
  {Y{\i}ld{\i}z}, {Doty}, {Herczeg}, {Liu}, {Parise}, {J{\o}rgensen}, {van
  Kempen}, {Brinch}, {Wampfler}, {Bruderer}, {Benz}, {Hogerheijde}, {Deul},
  {Bachiller}, {Baudry}, {Benedettini}, {Bergin}, {Bjerkeli}, {Blake},
  {Bontemps}, {Braine}, {Caselli}, {Cernicharo}, {Codella}, {Daniel}, {de
  Graauw}, {di Giorgio}, {Dominik}, {Encrenaz}, {Fich}, {Fuente}, {Giannini},
  {Goicoechea}, {Helmich}, {Herpin}, {Jacq}, {Johnstone}, {Kaufman}, {Larsson},
  {Lis}, {Liseau}, {Marseille}, {McCoey}, {Melnick}, {Neufeld}, {Nisini},
  {Olberg}, {Pearson}, {Plume}, {Risacher}, {Santiago-Garc{\'{\i}}a},
  {Saraceno}, {Shipman}, {Tafalla}, {Tielens}, {van der Tak}, {Wyrowski},
  {Beintema}, {de Jonge}, {Dieleman}, {Ossenkopf}, {Roelfsema}, {Stutzki}, \&
  {Whyborn}}]{Kr10}
{Kristensen}, L.~E., {Visser}, R., {van Dishoeck}, E.~F., {et~al.} 2010, \aap,
  521, L30

\bibitem[{{Kwon} {et~al.}(2006){Kwon}, {Looney}, {Crutcher}, \& {Kirk}}]{Kw06}
{Kwon}, W., {Looney}, L.~W., {Crutcher}, R.~M., \& {Kirk}, J.~M. 2006, \apj,
  653, 1358

\bibitem[{{Lee} {et~al.}(2013){Lee}, {Lee}, {Lee}, {Green}, {Evans}, {Choi},
  {Kristensen}, {Dionatos}, {J{\o}rgensen}, \& {the DIGIT team}}]{Lee13}
{Lee}, J., {Lee}, J.-E., {Lee}, S., {et~al.} 2013, \apjs, 209, 4

\bibitem[{{Lefloch} {et~al.}(2012{\natexlab{a}}){Lefloch}, {Cabrit}, {Busquet},
  {Codella}, {Ceccarelli}, {Cernicharo}, {Pardo}, {Benedettini}, {Lis}, \&
  {Nisini}}]{Lef12}
{Lefloch}, B., {Cabrit}, S., {Busquet}, G., {et~al.} 2012{\natexlab{a}}, \apjl,
  757, L25

\bibitem[{{Lefloch} {et~al.}(2012{\natexlab{b}}){Lefloch}, {Cabrit}, {Busquet},
  {Codella}, {Ceccarelli}, {Cernicharo}, {Pardo}, {Benedettini}, {Lis}, \&
  {Nisini}}]{Le12}
{Lefloch}, B., {Cabrit}, S., {Busquet}, G., {et~al.} 2012{\natexlab{b}}, \apjl,
  757, L25

\bibitem[{{Lesaffre} {et~al.}(2004{\natexlab{a}}){Lesaffre}, {Chi{\`e}ze},
  {Cabrit}, \& {Pineau des For{\^e}ts}}]{Le04b}
{Lesaffre}, P., {Chi{\`e}ze}, J.-P., {Cabrit}, S., \& {Pineau des For{\^e}ts},
  G. 2004{\natexlab{a}}, \aap, 427, 147

\bibitem[{{Lesaffre} {et~al.}(2004{\natexlab{b}}){Lesaffre}, {Chi{\`e}ze},
  {Cabrit}, \& {Pineau des For{\^e}ts}}]{Le04a}
{Lesaffre}, P., {Chi{\`e}ze}, J.-P., {Cabrit}, S., \& {Pineau des For{\^e}ts},
  G. 2004{\natexlab{b}}, \aap, 427, 157

\bibitem[{{Lindberg} {et~al.}(2014){Lindberg}, {J{\o}rgensen}, {Green},
  {Herczeg}, {Dionatos}, {Evans}, {Karska}, \& {Wampfler}}]{Li14}
{Lindberg}, J.~E., {J{\o}rgensen}, J.~K., {Green}, J.~D., {et~al.} 2014, \aap,
  565, A29

\bibitem[{{Looney} {et~al.}(2000){Looney}, {Mundy}, \& {Welch}}]{Lo00}
{Looney}, L.~W., {Mundy}, L.~G., \& {Welch}, W.~J. 2000, \apj, 529, 477

\bibitem[{{Manoj} {et~al.}(2013){Manoj}, {Watson}, {Neufeld}, {Megeath},
  {Vavrek}, {Yu}, {Visser}, {Bergin}, {Fischer}, {Tobin}, {Stutz}, {Ali},
  {Wilson}, {Di Francesco}, {Osorio}, {Maret}, \& {Poteet}}]{Ma12}
{Manoj}, P., {Watson}, D.~M., {Neufeld}, D.~A., {et~al.} 2013, \apj, 763, 83

\bibitem[{{Maret} {et~al.}(2009){Maret}, {Bergin}, {Neufeld}, {Green},
  {Watson}, {Harwit}, {Kristensen}, {Melnick}, {Sonnentrucker}, {Tolls},
  {Werner}, {Willacy}, \& {Yuan}}]{Ma09}
{Maret}, S., {Bergin}, E.~A., {Neufeld}, D.~A., {et~al.} 2009, \apj, 698, 1244

\bibitem[{{Mathis} {et~al.}(1977){Mathis}, {Rumpl}, \& {Nordsieck}}]{Ma77}
{Mathis}, J.~S., {Rumpl}, W., \& {Nordsieck}, K.~H. 1977, \apj, 217, 425

\bibitem[{{McElroy} {et~al.}(2013){McElroy}, {Walsh}, {Markwick}, {Cordiner},
  {Smith}, \& {Millar}}]{UMIST}
{McElroy}, D., {Walsh}, C., {Markwick}, A.~J., {et~al.} 2013, \aap, 550, A36

\bibitem[{{Melnick} {et~al.}(2000){Melnick}, {Stauffer}, {Ashby}, {Bergin},
  {Chin}, {Erickson}, {Goldsmith}, {Harwit}, {Howe}, {Kleiner}, {Koch},
  {Neufeld}, {Patten}, {Plume}, {Schieder}, {Snell}, {Tolls}, {Wang},
  {Winnewisser}, \& {Zhang}}]{SWAS}
{Melnick}, G.~J., {Stauffer}, J.~R., {Ashby}, M.~L.~N., {et~al.} 2000, \apjl,
  539, L77

\bibitem[{{Mottram} {et~al.}(2014){Mottram}, {Kristensen}, \& {van
  Dishoeck}}]{Mo14}
{Mottram}, J., {Kristensen}, L., \& {van Dishoeck}, E. a.~a. 2014, A\&A, in
  press

\bibitem[{{M{\"u}ller} {et~al.}(2005){M{\"u}ller}, {Schl{\"o}der}, {Stutzki},
  \& {Winnewisser}}]{CDMS}
{M{\"u}ller}, H.~S.~P., {Schl{\"o}der}, F., {Stutzki}, J., \& {Winnewisser}, G.
  2005, Journal of Molecular Structure, 742, 215

\bibitem[{{M{\"u}ller} {et~al.}(2001){M{\"u}ller}, {Thorwirth}, {Roth}, \&
  {Winnewisser}}]{CDMS2}
{M{\"u}ller}, H.~S.~P., {Thorwirth}, S., {Roth}, D.~A., \& {Winnewisser}, G.
  2001, \aap, 370, L49

\bibitem[{{Neufeld}(2012)}]{Ne12}
{Neufeld}, D.~A. 2012, \apj, 749, 125

\bibitem[{{Neufeld} \& {Dalgarno}(1989)}]{ND89}
{Neufeld}, D.~A. \& {Dalgarno}, A. 1989, \apj, 344, 251

\bibitem[{{Nisini} {et~al.}(2010{\natexlab{a}}){Nisini}, {Benedettini},
  {Codella}, {Giannini}, {Liseau}, {Neufeld}, {Tafalla}, {van Dishoeck},
  {Bachiller}, {Baudry}, {Benz}, {Bergin}, {Bjerkeli}, {Blake}, {Bontemps},
  {Braine}, {Bruderer}, {Caselli}, {Cernicharo}, {Daniel}, {Encrenaz}, {di
  Giorgio}, {Dominik}, {Doty}, {Fich}, {Fuente}, {Goicoechea}, {de Graauw},
  {Helmich}, {Herczeg}, {Herpin}, {Hogerheijde}, {Jacq}, {Johnstone},
  {J{\o}rgensen}, {Kaufman}, {Kristensen}, {Larsson}, {Lis}, {Marseille},
  {McCoey}, {Melnick}, {Olberg}, {Parise}, {Pearson}, {Plume}, {Risacher},
  {Santiago}, {Saraceno}, {Shipman}, {van Kempen}, {Visser}, {Viti},
  {Wampfler}, {Wyrowski}, {van der Tak}, {Y{\i}ld{\i}z}, {Delforge}, {Desbat},
  {Hatch}, {P{\'e}ron}, {Schieder}, {Stern}, {Teyssier}, \& {Whyborn}}]{Ni10}
{Nisini}, B., {Benedettini}, M., {Codella}, C., {et~al.} 2010{\natexlab{a}},
  \aap, 518, L120

\bibitem[{{Nisini} {et~al.}(1999){Nisini}, {Benedettini}, {Giannini}, {Caux},
  {di Giorgio}, {Liseau}, {Lorenzetti}, {Molinari}, {Saraceno}, {Smith},
  {Spinoglio}, \& {White}}]{Ni99}
{Nisini}, B., {Benedettini}, M., {Giannini}, T., {et~al.} 1999, \aap, 350, 529

\bibitem[{{Nisini} {et~al.}(2000){Nisini}, {Benedettini}, {Giannini},
  {Codella}, {Lorenzetti}, {di Giorgio}, \& {Richer}}]{Ni00}
{Nisini}, B., {Benedettini}, M., {Giannini}, T., {et~al.} 2000, \aap, 360, 297

\bibitem[{{Nisini} {et~al.}(2002){Nisini}, {Giannini}, \& {Lorenzetti}}]{Ni02}
{Nisini}, B., {Giannini}, T., \& {Lorenzetti}, D. 2002, \apj, 574, 246

\bibitem[{{Nisini} {et~al.}(2010{\natexlab{b}}){Nisini}, {Giannini}, {Neufeld},
  {Yuan}, {Antoniucci}, {Bergin}, \& {Melnick}}]{Ni10b}
{Nisini}, B., {Giannini}, T., {Neufeld}, D.~A., {et~al.} 2010{\natexlab{b}},
  \apj, 724, 69

\bibitem[{{Nisini} {et~al.}(2013){Nisini}, {Santangelo}, {Antoniucci},
  {Benedettini}, {Codella}, {Giannini}, {Lorenzani}, {Liseau}, {Tafalla},
  {Bjerkeli}, {Cabrit}, {Caselli}, {Kristensen}, {Neufeld}, {Melnick}, \& {van
  Dishoeck}}]{Ni13}
{Nisini}, B., {Santangelo}, G., {Antoniucci}, S., {et~al.} 2013, \aap, 549, A16

\bibitem[{{Ott}(2010)}]{Ot10}
{Ott}, S. 2010, in Astronomical Society of the Pacific Conference Series, Vol.
  434, Astronomical Data Analysis Software and Systems XIX, ed. Y.~{Mizumoto},
  K.-I. {Morita}, \& M.~{Ohishi}, 139

\bibitem[{{Pickett} {et~al.}(1998){Pickett}, {Poynter}, {Cohen}, {Delitsky},
  {Pearson}, \& {M{\"u}ller}}]{JPL}
{Pickett}, H.~M., {Poynter}, R.~L., {Cohen}, E.~A., {et~al.} 1998, \jqsrt, 60,
  883

\bibitem[{{Poglitsch} {et~al.}(2010){Poglitsch}, {Waelkens}, {Geis},
  {Feuchtgruber}, {Vandenbussche}, {Rodriguez}, {Krause}, {Renotte}, {van
  Hoof}, {Saraceno}, {Cepa}, {Kerschbaum}, {Agn{\`e}se}, {Ali}, {Altieri},
  {Andreani}, {Augueres}, {Balog}, {Barl}, {Bauer}, {Belbachir}, {Benedettini},
  {Billot}, {Boulade}, {Bischof}, {Blommaert}, {Callut}, {Cara}, {Cerulli},
  {Cesarsky}, {Contursi}, {Creten}, {De Meester}, {Doublier}, {Doumayrou},
  {Duband}, {Exter}, {Genzel}, {Gillis}, {Gr{\"o}zinger}, {Henning},
  {Herreros}, {Huygen}, {Inguscio}, {Jakob}, {Jamar}, {Jean}, {de Jong},
  {Katterloher}, {Kiss}, {Klaas}, {Lemke}, {Lutz}, {Madden}, {Marquet},
  {Martignac}, {Mazy}, {Merken}, {Montfort}, {Morbidelli}, {M{\"u}ller},
  {Nielbock}, {Okumura}, {Orfei}, {Ottensamer}, {Pezzuto}, {Popesso},
  {Putzeys}, {Regibo}, {Reveret}, {Royer}, {Sauvage}, {Schreiber}, {Stegmaier},
  {Schmitt}, {Schubert}, {Sturm}, {Thiel}, {Tofani}, {Vavrek}, {Wetzstein},
  {Wieprecht}, \& {Wiezorrek}}]{Po10}
{Poglitsch}, A., {Waelkens}, C., {Geis}, N., {et~al.} 2010, \aap, 518, L2

\bibitem[{{Rebull} {et~al.}(2007){Rebull}, {Stapelfeldt}, {Evans},
  {J{\o}rgensen}, {Harvey}, {Brooke}, {Bourke}, {Padgett}, {Chapman}, {Lai},
  {Spiesman}, {Noriega-Crespo}, {Mer{\'{\i}}n}, {Huard}, {Allen}, {Blake},
  {Jarrett}, {Koerner}, {Mundy}, {Myers}, {Sargent}, {van Dishoeck}, {Wahhaj},
  \& {Young}}]{Re07}
{Rebull}, L.~M., {Stapelfeldt}, K.~R., {Evans}, II, N.~J., {et~al.} 2007,
  \apjs, 171, 447

\bibitem[{{Robitaille} {et~al.}(2007){Robitaille}, {Whitney}, {Indebetouw}, \&
  {Wood}}]{Ro07}
{Robitaille}, T.~P., {Whitney}, B.~A., {Indebetouw}, R., \& {Wood}, K. 2007,
  \apjs, 169, 328

\bibitem[{{Robitaille} {et~al.}(2006){Robitaille}, {Whitney}, {Indebetouw},
  {Wood}, \& {Denzmore}}]{Ro06}
{Robitaille}, T.~P., {Whitney}, B.~A., {Indebetouw}, R., {Wood}, K., \&
  {Denzmore}, P. 2006, \apjs, 167, 256

\bibitem[{{Sadavoy} {et~al.}(2014){Sadavoy}, {Di Francesco}, {Andr{\'e}},
  {Pezzuto}, {Bernard}, {Maury}, {Men'shchikov}, {Motte},
  {Nguy{\tilde}{\^e}n-Lu'o'ng}, {Schneider}, {Arzoumanian}, {Benedettini},
  {Bontemps}, {Elia}, {Hennemann}, {Hill}, {K{\"o}nyves}, {Louvet}, {Peretto},
  {Roy}, \& {White}}]{Sad14}
{Sadavoy}, S.~I., {Di Francesco}, J., {Andr{\'e}}, P., {et~al.} 2014, \apjl,
  787, L18

\bibitem[{{San Jos{\'e}-Garc{\'{\i}}a} {et~al.}(2013){San
  Jos{\'e}-Garc{\'{\i}}a}, {Mottram}, {Kristensen}, {van Dishoeck},
  {Y{\i}ld{\i}z}, {van der Tak}, {Herpin}, {Visser}, {McCoey}, {Wyrowski},
  {Braine}, \& {Johnstone}}]{IreneCO}
{San Jos{\'e}-Garc{\'{\i}}a}, I., {Mottram}, J.~C., {Kristensen}, L.~E.,
  {et~al.} 2013, \aap, 553, A125

\bibitem[{{Santangelo} {et~al.}(2013){Santangelo}, {Nisini}, {Antoniucci},
  {Codella}, {Cabrit}, {Giannini}, {Herczeg}, {Liseau}, {Tafalla}, \& {van
  Dishoeck}}]{Sa13}
{Santangelo}, G., {Nisini}, B., {Antoniucci}, S., {et~al.} 2013, \aap, 557, A22

\bibitem[{{Santangelo} {et~al.}(2014){Santangelo}, {Nisini}, {Codella},
  {Lorenzani}, {Yildiz}, {Antoniucci}, {Bjerkeli}, {Cabrit}, {Giannini},
  {Kristensen}, {Liseau}, {Mottram}, {Tafalla}, \& {van Dishoeck}}]{Sa14}
{Santangelo}, G., {Nisini}, B., {Codella}, C., {et~al.} 2014, ArXiv No.
  1406.6302

\bibitem[{{Santangelo} {et~al.}(2012){Santangelo}, {Nisini}, {Giannini},
  {Antoniucci}, {Vasta}, {Codella}, {Lorenzani}, {Tafalla}, {Liseau}, {van
  Dishoeck}, \& {Kristensen}}]{Sa12}
{Santangelo}, G., {Nisini}, B., {Giannini}, T., {et~al.} 2012, \aap, 538, A45

\bibitem[{{Snell} {et~al.}(2005){Snell}, {Hollenbach}, {Howe}, {Neufeld},
  {Kaufman}, {Melnick}, {Bergin}, \& {Wang}}]{Sn05}
{Snell}, R.~L., {Hollenbach}, D., {Howe}, J.~E., {et~al.} 2005, \apj, 620, 758

\bibitem[{{Spaans} {et~al.}(1995){Spaans}, {Hogerheijde}, {Mundy}, \& {van
  Dishoeck}}]{Sp95}
{Spaans}, M., {Hogerheijde}, M.~R., {Mundy}, L.~G., \& {van Dishoeck}, E.~F.
  1995, \apjl, 455, L167

\bibitem[{{Suutarinen} {et~al.}(2014){Suutarinen}, {Kristensen}, {Mottram},
  {Fraser}, \& {van Dishoeck}}]{Su14}
{Suutarinen}, A.~N., {Kristensen}, L.~E., {Mottram}, J.~C., {Fraser}, H.~J., \&
  {van Dishoeck}, E.~F. 2014, \mnras, 440, 1844

\bibitem[{{Tafalla} {et~al.}(2013){Tafalla}, {Liseau}, {Nisini}, {Bachiller},
  {Santiago-Garc{\'{\i}}a}, {van Dishoeck}, {Kristensen}, {Herczeg}, \&
  {Y{\i}ld{\i}z}}]{Ta13}
{Tafalla}, M., {Liseau}, R., {Nisini}, B., {et~al.} 2013, \aap, 551, A116

\bibitem[{{Valenti} {et~al.}(2000){Valenti}, {Johns-Krull}, \& {Linsky}}]{Va00}
{Valenti}, J.~A., {Johns-Krull}, C.~M., \& {Linsky}, J.~L. 2000, \apjs, 129,
  399

\bibitem[{{van Dishoeck}(2004)}]{EvDISO}
{van Dishoeck}, E.~F. 2004, \araa, 42, 119

\bibitem[{{van Dishoeck} {et~al.}(2013){van Dishoeck}, {Herbst}, \&
  {Neufeld}}]{EvD13}
{van Dishoeck}, E.~F., {Herbst}, E., \& {Neufeld}, D.~A. 2013, Chemical
  Reviews, 113, 9043

\bibitem[{{van Dishoeck} {et~al.}(2011){van Dishoeck}, {Kristensen}, {Benz},
  {Bergin}, {Caselli}, {Cernicharo}, {Herpin}, {Hogerheijde}, {Johnstone},
  {Liseau}, {Nisini}, {Shipman}, {Tafalla}, {van der Tak}, {Wyrowski},
  {Aikawa}, {Bachiller}, {Baudry}, {Benedettini}, {Bjerkeli}, {Blake},
  {Bontemps}, {Braine}, {Brinch}, {Bruderer}, {Chavarr{\'{\i}}a}, {Codella},
  {Daniel}, {de Graauw}, {Deul}, {di Giorgio}, {Dominik}, {Doty}, {Dubernet},
  {Encrenaz}, {Feuchtgruber}, {Fich}, {Frieswijk}, {Fuente}, {Giannini},
  {Goicoechea}, {Helmich}, {Herczeg}, {Jacq}, {J{\o}rgensen}, {Karska},
  {Kaufman}, {Keto}, {Larsson}, {Lefloch}, {Lis}, {Marseille}, {McCoey},
  {Melnick}, {Neufeld}, {Olberg}, {Pagani}, {Pani{\'c}}, {Parise}, {Pearson},
  {Plume}, {Risacher}, {Salter}, {Santiago-Garc{\'{\i}}a}, {Saraceno},
  {St{\"a}uber}, {van Kempen}, {Visser}, {Viti}, {Walmsley}, {Wampfler}, \&
  {Y{\i}ld{\i}z}}]{WISH}
{van Dishoeck}, E.~F., {Kristensen}, L.~E., {Benz}, A.~O., {et~al.} 2011,
  \pasp, 123, 138

\bibitem[{{van Kempen} {et~al.}(2010){van Kempen}, {Kristensen}, {Herczeg},
  {Visser}, {van Dishoeck}, {Wampfler}, {Bruderer}, {Benz}, {Doty}, {Brinch},
  {Hogerheijde}, {J{\o}rgensen}, {Tafalla}, {Neufeld}, {Bachiller}, {Baudry},
  {Benedettini}, {Bergin}, {Bjerkeli}, {Blake}, {Bontemps}, {Braine},
  {Caselli}, {Cernicharo}, {Codella}, {Daniel}, {di Giorgio}, {Dominik},
  {Encrenaz}, {Fich}, {Fuente}, {Giannini}, {Goicoechea}, {de Graauw},
  {Helmich}, {Herpin}, {Jacq}, {Johnstone}, {Kaufman}, {Larsson}, {Lis},
  {Liseau}, {Marseille}, {McCoey}, {Melnick}, {Nisini}, {Olberg}, {Parise},
  {Pearson}, {Plume}, {Risacher}, {Santiago-Garc{\'{\i}}a}, {Saraceno},
  {Shipman}, {van der Tak}, {Wyrowski}, {Y{\i}ld{\i}z}, {Ciechanowicz},
  {Dubbeldam}, {Glenz}, {Huisman}, {Lin}, {Morris}, {Murphy}, \&
  {Trappe}}]{vK10}
{van Kempen}, T.~A., {Kristensen}, L.~E., {Herczeg}, G.~J., {et~al.} 2010,
  \aap, 518, L121

\bibitem[{{van Kempen} {et~al.}(2009){van Kempen}, {van Dishoeck},
  {G{\"u}sten}, {Kristensen}, {Schilke}, {Hogerheijde}, {Boland}, {Nefs},
  {Menten}, {Baryshev}, \& {Wyrowski}}]{vK09}
{van Kempen}, T.~A., {van Dishoeck}, E.~F., {G{\"u}sten}, R., {et~al.} 2009,
  \aap, 501, 633

\bibitem[{{Van Loo} {et~al.}(2013){Van Loo}, {Ashmore}, {Caselli}, {Falle}, \&
  {Hartquist}}]{vL13}
{Van Loo}, S., {Ashmore}, I., {Caselli}, P., {Falle}, S.~A.~E.~G., \&
  {Hartquist}, T.~W. 2013, \mnras, 428, 381

\bibitem[{{Vasta} {et~al.}(2012){Vasta}, {Codella}, {Lorenzani}, {Santangelo},
  {Nisini}, {Giannini}, {Tafalla}, {Liseau}, {van Dishoeck}, \&
  {Kristensen}}]{Va12}
{Vasta}, M., {Codella}, C., {Lorenzani}, A., {et~al.} 2012, \aap, 537, A98

\bibitem[{{Velusamy} {et~al.}(2014){Velusamy}, {Langer}, \& {Thompson}}]{Ve14}
{Velusamy}, T., {Langer}, W.~D., \& {Thompson}, T. 2014, \apj, 783, 6

\bibitem[{{Visser} {et~al.}(2012){Visser}, {Kristensen}, {Bruderer}, {van
  Dishoeck}, {Herczeg}, {Brinch}, {Doty}, {Harsono}, \& {Wolfire}}]{Vi12}
{Visser}, R., {Kristensen}, L.~E., {Bruderer}, S., {et~al.} 2012, \aap, 537,
  A55

\bibitem[{{Wagner} \& {Graff}(1987)}]{Wa87}
{Wagner}, A.~F. \& {Graff}, M.~M. 1987, \apj, 317, 423

\bibitem[{{Walter} {et~al.}(2003){Walter}, {Herczeg}, {Brown}, {Ardila},
  {Gahm}, {Johns-Krull}, {Lissauer}, {Simon}, \& {Valenti}}]{Wa03}
{Walter}, F.~M., {Herczeg}, G., {Brown}, A., {et~al.} 2003, \aj, 126, 3076

\bibitem[{{Wampfler} {et~al.}(2013){Wampfler}, {Bruderer}, {Karska}, {Herczeg},
  {van Dishoeck}, {Kristensen}, {Goicoechea}, {Benz}, {Doty}, {McCoey},
  {Baudry}, {Giannini}, \& {Larsson}}]{Wa13}
{Wampfler}, S.~F., {Bruderer}, S., {Karska}, A., {et~al.} 2013, \aap, 552, A56

\bibitem[{{Wampfler} {et~al.}(2010){Wampfler}, {Herczeg}, {Bruderer}, {Benz},
  {van Dishoeck}, {Kristensen}, {Visser}, {Doty}, {Melchior}, {van Kempen},
  {Y{\i}ld{\i}z}, {Dedes}, {Goicoechea}, {Baudry}, {Melnick}, {Bachiller},
  {Benedettini}, {Bergin}, {Bjerkeli}, {Blake}, {Bontemps}, {Braine},
  {Caselli}, {Cernicharo}, {Codella}, {Daniel}, {di Giorgio}, {Dominik},
  {Encrenaz}, {Fich}, {Fuente}, {Giannini}, {de Graauw}, {Helmich}, {Herpin},
  {Hogerheijde}, {Jacq}, {Johnstone}, {J{\o}rgensen}, {Larsson}, {Lis},
  {Liseau}, {Marseille}, {McCoey}, {Neufeld}, {Nisini}, {Olberg}, {Parise},
  {Pearson}, {Plume}, {Risacher}, {Santiago-Garc{\'{\i}}a}, {Saraceno},
  {Shipman}, {Tafalla}, {van der Tak}, {Wyrowski}, {Roelfsema}, {Jellema},
  {Dieleman}, {Caux}, \& {Stutzki}}]{Wa10}
{Wampfler}, S.~F., {Herczeg}, G.~J., {Bruderer}, S., {et~al.} 2010, \aap, 521,
  L36

\bibitem[{{Yang} {et~al.}(2010){Yang}, {Stancil}, {Balakrishnan}, \&
  {Forrey}}]{Ya10}
{Yang}, B., {Stancil}, P.~C., {Balakrishnan}, N., \& {Forrey}, R.~C. 2010,
  \apj, 718, 1062

\bibitem[{{Yang} {et~al.}(2012){Yang}, {Herczeg}, {Linsky}, {Brown},
  {Johns-Krull}, {Ingleby}, {Calvet}, {Bergin}, \& {Valenti}}]{Yan12}
{Yang}, H., {Herczeg}, G.~J., {Linsky}, J.~L., {et~al.} 2012, \apj, 744, 121

\bibitem[{{Y{\i}ld{\i}z} {et~al.}(2012){Y{\i}ld{\i}z}, {Kristensen}, {van
  Dishoeck}, {Belloche}, {van Kempen}, {Hogerheijde}, {G{\"u}sten}, \& {van der
  Marel}}]{Yi12}
{Y{\i}ld{\i}z}, U.~A., {Kristensen}, L.~E., {van Dishoeck}, E.~F., {et~al.}
  2012, \aap, 542, A86

\bibitem[{{Y{\i}ld{\i}z} {et~al.}(2013){Y{\i}ld{\i}z}, {Kristensen}, {van
  Dishoeck}, {San Jos{\'e}-Garc{\'{\i}}a}, {Karska}, {Harsono}, {Tafalla},
  {Fuente}, {Visser}, {J{\o}rgensen}, \& {Hogerheijde}}]{Yi13}
{Y{\i}ld{\i}z}, U.~A., {Kristensen}, L.~E., {van Dishoeck}, E.~F., {et~al.}
  2013, \aap, 556, A89

\end{thebibliography}

\newpage
\appendix

\section{Supplementary material}
Table \ref{lines} provides molecular / atomic information about the lines observed in the WILL program.
\begin{table}
\centering 
\caption{\label{lines} Atomic and molecular data\tablefootmark{a} for lines observed in the WILL program}              
\renewcommand{\footnoterule}{}  
\begin{tabular}{l l c c c c c c}     
\hline\hline       
Species & Transition & Wave. & Freq.\tablefootmark{b} & $E_\mathrm{u}$/$k_\mathrm{B}$ & $A_\mathrm{ul}$\tablefootmark{c} \\
~ & ~ & ($\mu$m) & (GHz) &  (K) & (s$^{-1}$) \\
\hline    
H$_2$O & 2$_{21}$-2$_{12}$ & 180.488 & 1661.0 & 194.1 & 3.1(-2)  \\       
H$_2$O & 2$_{12}$-1$_{01}$ & 179.527 & 1669.9 & 114.4 & 5.6(-2)  \\
OH & $\nicefrac{3}{2}$,$\nicefrac{1}{2}$-$\nicefrac{1}{2}$,$\nicefrac{1}{2}$ & 163.398 & 1834.7 & 269.8 & 2.1(-2) \\
OH & $\nicefrac{3}{2}$,$\nicefrac{1}{2}$-$\nicefrac{1}{2}$,$\nicefrac{1}{2}$ & 163.131 & 1837.7 & 270.1 & 2.1(-2) \\
CO & 16-15 & 162.812 & 1841.3 & 751.7 & 4.1(-4) \\
{[\ion{C}{ii}]} & $^{2}P_{\nicefrac{3}{2}}-^{2}P_{\nicefrac{1}{2}}$ &  157.74 & 2060.0 & 326.6 & 1.8(-5) \\
H$_2$O & 4$_{04}$-3$_{13}$ & 125.354 & 2391.6 & 319.5 & 1.7(-1) \\
CO & 21-20 & 124.193 & 2413.9 & 1276.1 & 8.8(-4) \\
CO & 24-23 & 108.763 & 2756.4 & 1656.5 & 1.3(-3)  \\
H$_2$O & 2$_{21}$-1$_{10}$ & 108.073 & 2774.0 & 194.1 & 2.6(-1) \\
CO & 29-28 & 90.163 & 3325.0 & 2399.8 & 2.1(-3) \\
H$_2$O & 3$_{22}$-2$_{11}$ & 89.988 & 3331.5 & 296.8 & 3.5(-1)  \\
H$_2$O & 7$_{16}$-7$_{07}$ & 84.767 & 3536.7 & 1013.2 & 2.1(-1) \\
OH & $\nicefrac{7}{2}$,$\nicefrac{3}{2}$-$\nicefrac{5}{2}$,$\nicefrac{3}{2}$ & 84.596 & 3543.8 & 290.5 & 4.9(-1) \\
OH & $\nicefrac{7}{2}$,$\nicefrac{3}{2}$-$\nicefrac{5}{2}$,$\nicefrac{3}{2}$ & 84.420 & 3551.2 & 291.2 & 2.5(-2) \\
CO & 31-30 & 84.411 & 3551.6 & 2735.3 & 2.5(-3) \\
H$_2$O & 6$_{16}$-5$_{05}$ & 82.032 & 3654.6 & 643.5 & 7.5(-1) \\ 
CO & 32-31 & 81.806 & 3664.7 & 2911.2 & 2.7(-3)  \\
CO & 33-32 & 79.360 & 3777.6 & 3092.5 & 3.0(-3)  \\
OH & $\nicefrac{1}{2}$,$\nicefrac{1}{2}$-$\nicefrac{3}{2}$,$\nicefrac{3}{2}$ & 79.182 & 3786.1 & 181.7 & 2.9(-2) \\
OH & $\nicefrac{1}{2}$,$\nicefrac{1}{2}$-$\nicefrac{3}{2}$,$\nicefrac{3}{2}$ & 79.116 & 3789.3 & 181.9 & 5.8(-3) \\
H$_2$O & 6$_{15}$-5$_{24}$ & 78.928 & 3798.3 & 781.1 & 4.6(-1) \\
H$_2$O & 4$_{23}$-3$_{12}$ & 78.742 & 3807.3 & 432.2 & 4.9(-1) \\
H$_2$O & 8$_{18}$-7$_{07}$ & 63.324 & 4734.3 & 1070.7 & 1.8 \\
{[\ion{O}{i}]} & $^{3}P_{1}-^{3}P_{2}$ & 63.184 & 4744.8 & 227.7 & 8.9(-5) \\
\hline                  
\end{tabular}
\tablefoot{
\tablefoottext{a}{Compiled using the CDMS \citep{CDMS2,CDMS} and JPL \citep{JPL} databases.}
\tablefoottext{b}{Frequencies are rest frequencies.}
\tablefoottext{c}{ $A(B)\equiv A\times10^{B}$}}
\end{table}

\begin{table*}
\caption{\label{log} Log of PACS observations}             
\centering     
\renewcommand{\footnoterule}{}  
\begin{tabular}{lccccccccccccc}     
\hline\hline       
Source & OBSID & OD & Date & Total time & RA & DEC \\
~ & ~ & ~ & ~ & (s) & ($^\mathrm{h}$ $^\mathrm{m}$ $^\mathrm{s}$) & ($^{\mathrm{o}}$ $\mathrm{'}$ $\mathrm{''}$) & ~ \\   
\hline
Per01   	& 1342263508 & 1370 & 2013-02-12 & 851   & 3 25 22.32 & +30 45 13.9   \\
~		 	& 1342263509 & 1370 & 2013-02-12 & 1986  & 3 25 22.32 & +30 45 13.9   \\
Per02		& 1342263506 & 1370 & 2013-02-12 & 851   & 3 25 36.49 & +30 45 22.2  \\
~			& 1342263507 & 1370 & 2013-02-12 & 1986  & 3 25 36.49 & +30 45 22.2  \\
Per03		& 1342263510 & 1370 & 2013-02-12 & 851   & 3 25 39.12 &	+30 43 58.2  \\
~			& 1342263511 & 1370 & 2013-02-12 & 1986  & 3 25 39.12 &	+30 43 58.2 \\
Per04		& 1342264250 & 1383 & 2013-02-25 & 851   & 3 26 37.47 & +30 15 28.1  \\
~			& 1342264251 & 1383 & 2013-02-25 & 1986  & 3 26 37.47 &	+30 15 28.1  \\
Per05		& 1342264248 & 1383 & 2013-02-25 & 851   & 3 28 37.09 &	+31 13 30.8  \\
~			& 1342264249 & 1383 & 2013-02-25 & 1986  & 3 28 37.09 &	+31 13 30.8  \\
Per06		& 1342264247 & 1383 & 2013-02-25 & 1986  & 3 28 57.36 &	+31 14 15.9  \\
~			& 1342264246 & 1383 & 2013-02-25 & 851   & 3 28 57.36 &	+31 14 15.9  \\
Per07		& 1342264244 & 1383 & 2013-02-25 & 1986  & 3 29 00.55 & +31 12 00.8  \\
~			& 1342264245 & 1383 & 2013-02-25 & 851   & 3 29 00.55 &	+31 12 00.8  \\
Per08		& 1342264242 & 1383 & 2013-02-25 & 1986  & 3 29 01.56 &	+31 20 20.6  \\
~			& 1342264243 & 1383 & 2013-02-25 & 851   & 3 29 01.56 &	+31 20 20.6  \\
Per09		& 1342267611 & 1401 & 2013-03-15 & 1986  & 3 29 07.78 &	+31 21 57.3  \\
~			& 1342267612 & 1401 & 2013-03-15 & 851   & 3 29 07.78 &	+31 21 57.3  \\
Per10		& 1342267615 & 1401 & 2013-03-15 & 1986  & 3 29 10.68 &	+31 18 20.6  \\
~			& 1342267616 & 1401 & 2013-03-15 & 851   & 3 29 10.68 &	+31 18 20.6  \\
Per11		& 1342267607 & 1401 & 2013-03-15 & 1986  & 3 29 12.06 & +31 13 01.7  \\
~			& 1342267608 & 1401 & 2013-03-15 & 851   & 3 29 12.06 & +31 13 01.7   \\
Per12		& 1342267609 & 1401 & 2013-03-15 & 1986  & 3 29 13.54 &	+31 13 58.2  \\
~			& 1342267610 & 1401 & 2013-03-15 & 851   & 3 29 13.54 &	+31 13 58.2  \\
Per13		& 1342267613 & 1401 & 2013-03-15 & 1986  & 3 29 51.82 &	+31 39 06.0  \\
~			& 1342267614 & 1401 & 2013-03-15 & 851   & 3 29 51.82 &	+31 39 06.0  \\
Per14		& 1342263512 & 1370 & 2013-02-12 & 1986  & 3 30 15.14 & +30 23 49.4  \\
~			& 1342263513 & 1370 & 2013-02-12 & 851   & 3 30 15.14 & +30 23 49.4  \\
Per15		& 1342263514 & 1370 & 2013-02-12 & 1986  & 3 31 20.98 & +30 45 30.1  \\
~			& 1342263515 & 1370 & 2013-02-12 & 851   & 3 31 20.98 & +30 45 30.1  \\
Per16		& 1342265447 & 1374 & 2013-02-16 & 1986  & 3 32 17.96 &	+30 49 47.5  \\
~			& 1342265448 & 1374 & 2013-02-16 & 851   & 3 32 17.96 &	+30 49 47.5  \\
Per17		& 1342263486 & 1369 & 2013-02-11 & 1986 & 3 33 14.38 &	+31 07 10.9  \\
~			& 1342263487 & 1369 & 2013-02-11 & 851 & 3 33 14.38 &	+31 07 10.9  \\
Per18		& 1342265449 & 1374 & 2013-02-16 & 1986 & 3 33 16.44 &	+31 06 52.5  \\
~			& 1342265450 & 1374 & 2013-02-16 & 851 & 3 33 16.44 &	+31 06 52.5  \\
Per19		& 1342265451 & 1374 & 2013-02-16 & 1986 & 3 33 27.29 &	+31 07 10.2  \\
~			& 1342265452 & 1374 & 2013-02-16 & 851 & 3 33 27.29 &	+31 07 10.2  \\
Per20		& 1342265453 & 1374 & 2013-02-16 & 1986 & 3 43 56.52 &	+32 00 52.8  \\
~			& 1342265454 & 1374 & 2013-02-16 & 851 & 3 43 56.52 &	+32 00 52.8  \\
Per21		& 1342265455 & 1374 & 2013-02-16 & 1986 & 3 43 56.84 &	+32 03 04.7  \\
~			& 1342265456 & 1374 & 2013-02-16 & 851 & 3 43 56.84 &	+32 03 04.7  \\
Per22		& 1342265701 & 1381 & 2013-02-23 & 1986 & 3 44 43.96 &	+32 01 36.2  \\
~			& 1342265702 & 1381 & 2013-02-23 & 851 & 3 44 43.96 &	+32 01 36.2  \\
\hline
\end{tabular}
\end{table*}

Table \ref{log} shows the observing log of PACS observations including
observations identifications (OBSID), observation day (OD), date of observation, total integration time, 
 and pointed coordinates (RA, DEC). 
 
Table \ref{det} informs about which lines are detected toward  the Perseus sources. The full list of line 
fluxes for all WILL sources including Perseus will be tabulated in the forthcoming paper (Karska et al. in prep.).

\begin{figure*}[!tb]
  \begin{minipage}[t]{.3\textwidth}
    \hspace{+0.2cm}
      \includegraphics[angle=90,height=6.3cm]{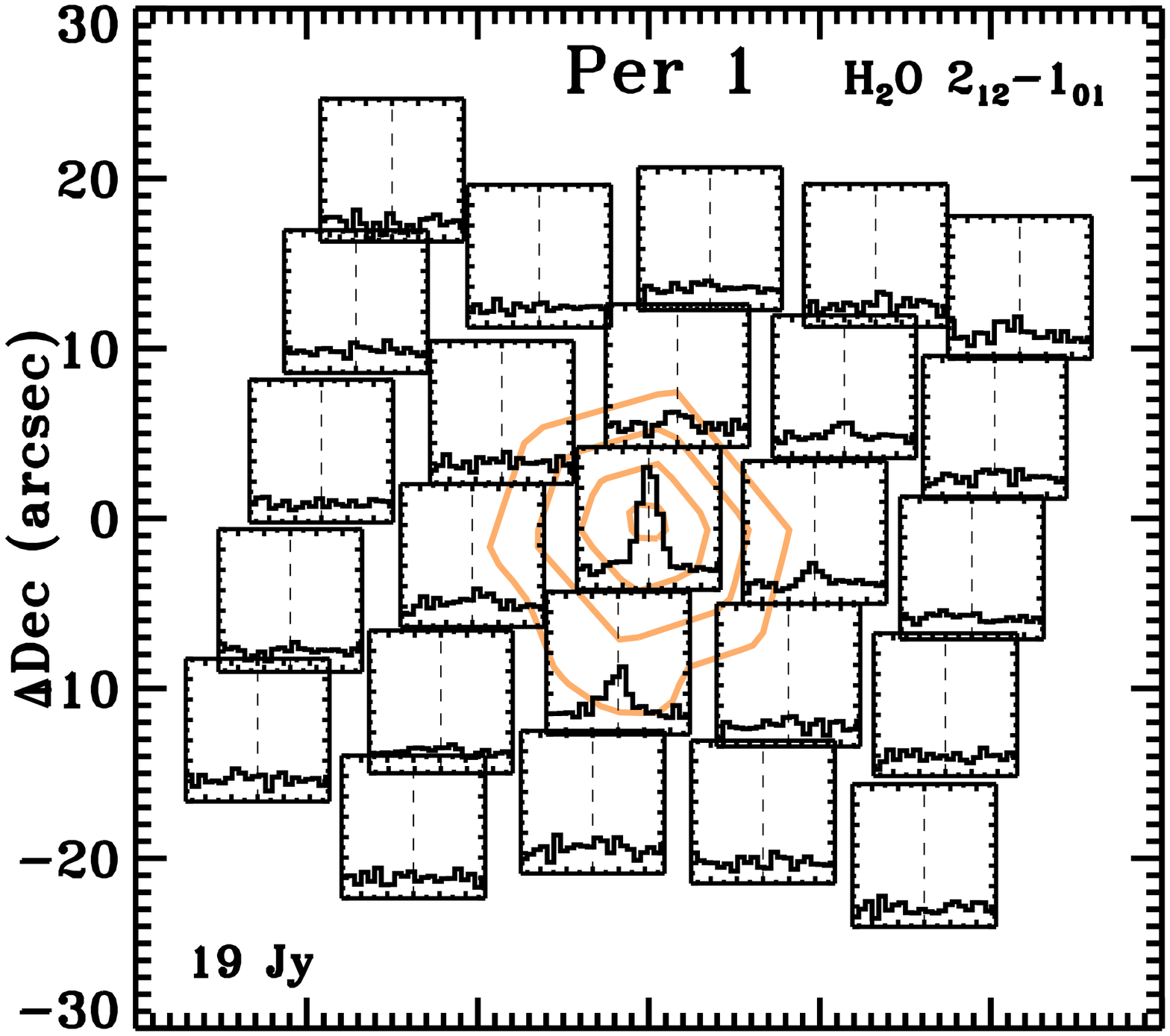}
     \vspace{-1cm}
     
     \hspace{+0.2cm}
       \includegraphics[angle=90,height=6.3cm]{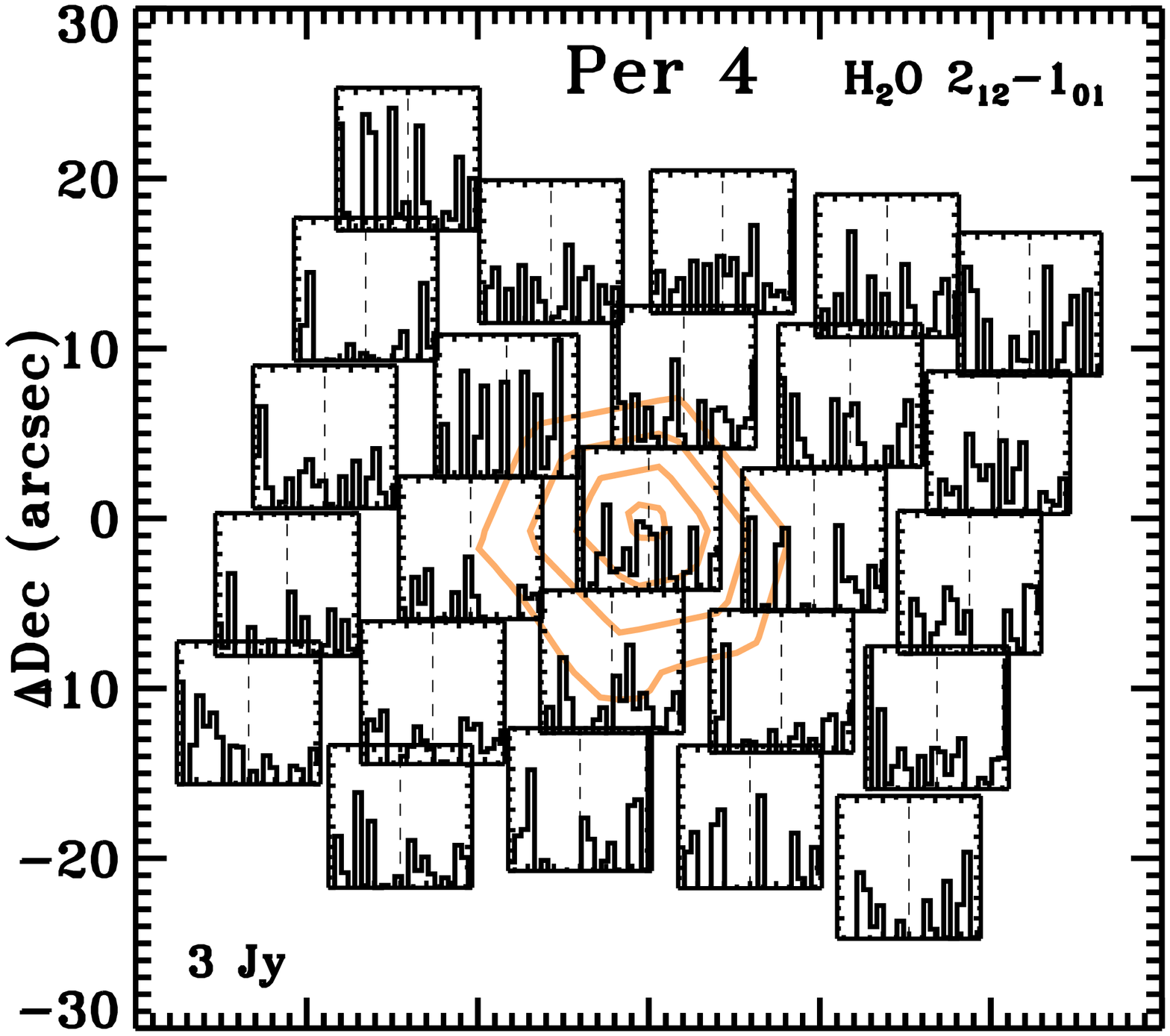}
       \vspace{-1cm}
       
       \hspace{+0.2cm}
       \includegraphics[angle=90,height=6.3cm]{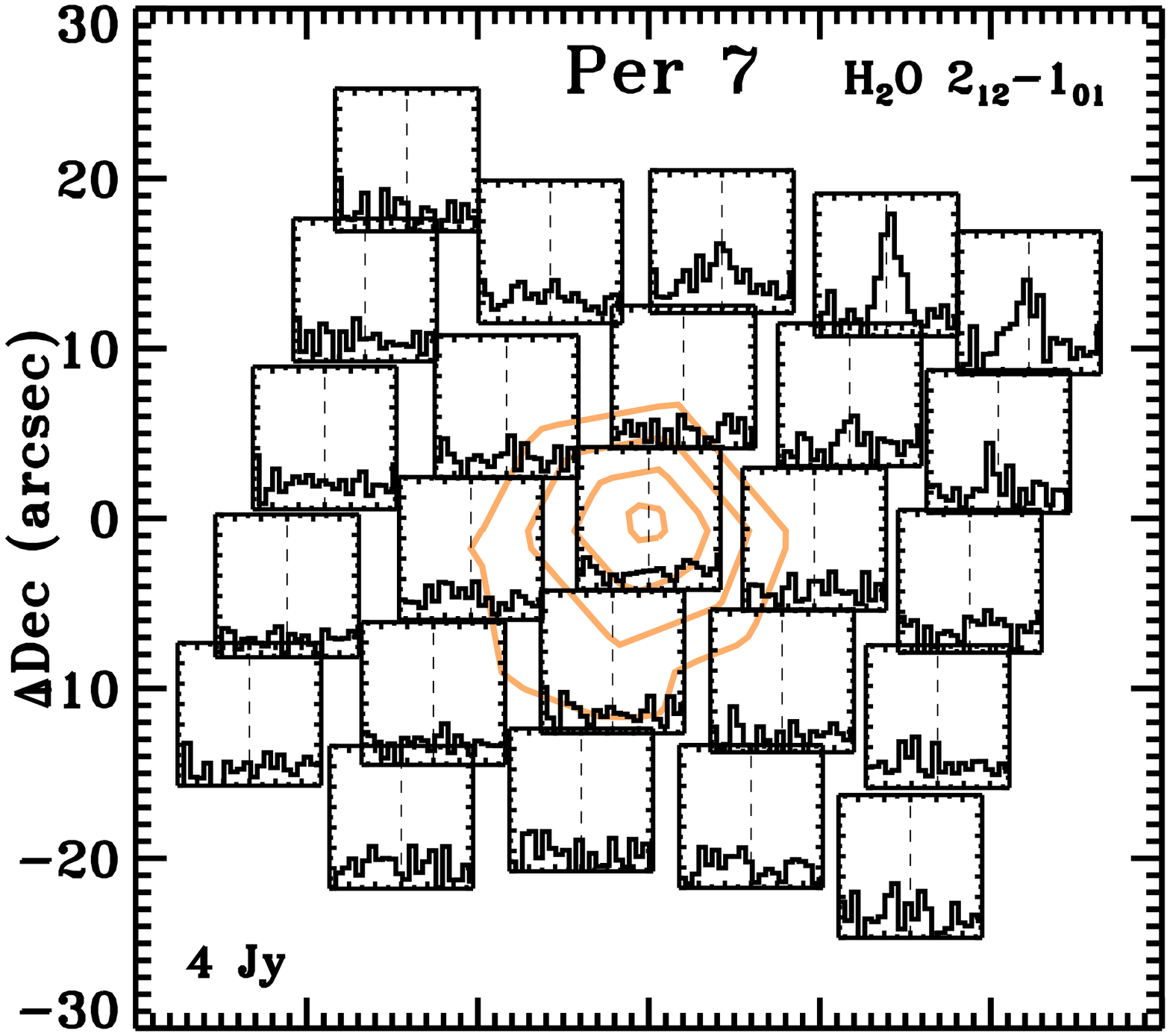}
              \vspace{-1cm}
              
          \hspace{+0.2cm}    
       \includegraphics[angle=90,height=6.3cm]{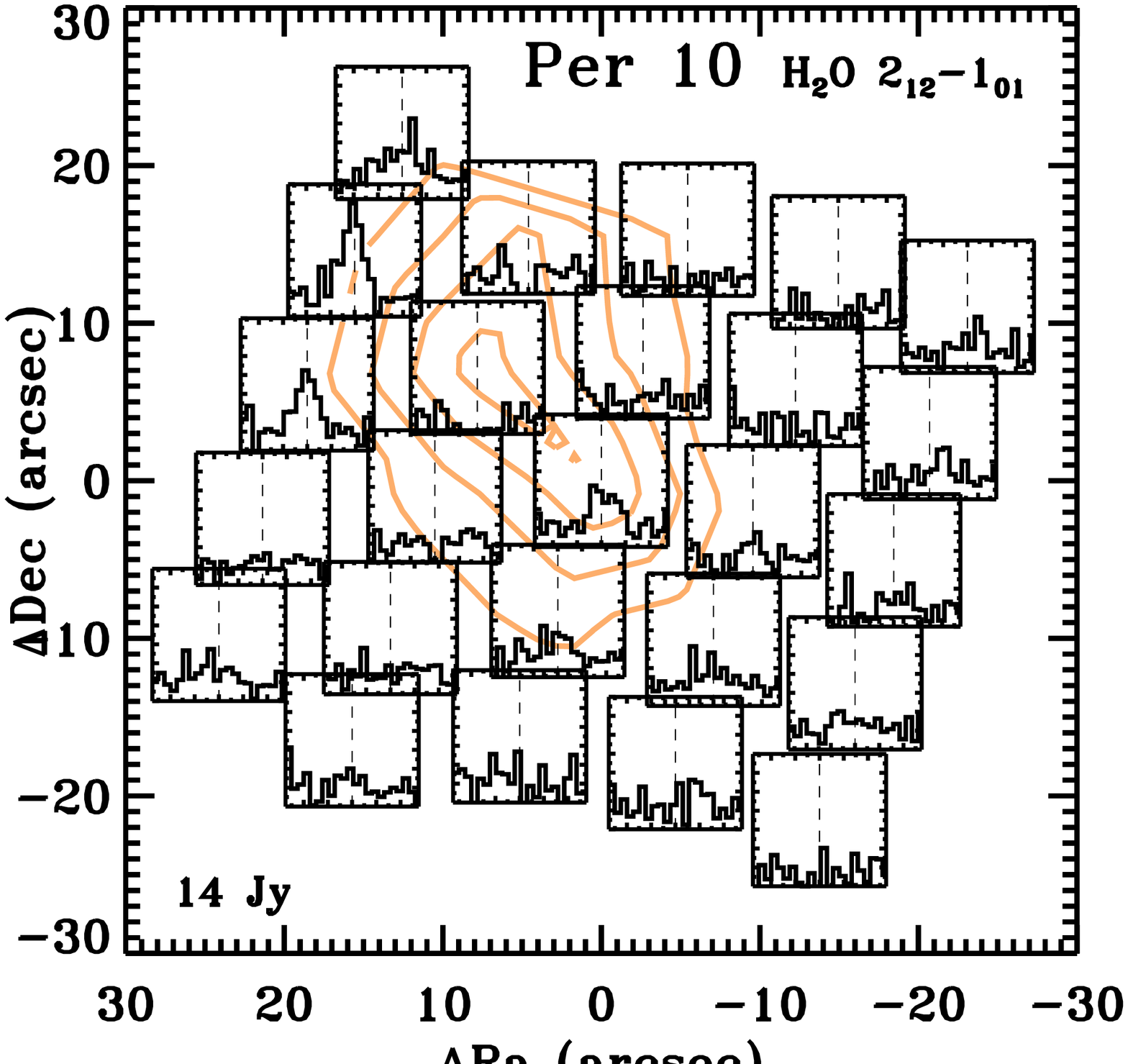}
                     \vspace{+0.2cm}
  \end{minipage}
  \hfill
  \begin{minipage}[t]{.3\textwidth}
 \hspace{-0.2cm} 
    \includegraphics[angle=90,height=6.3cm]{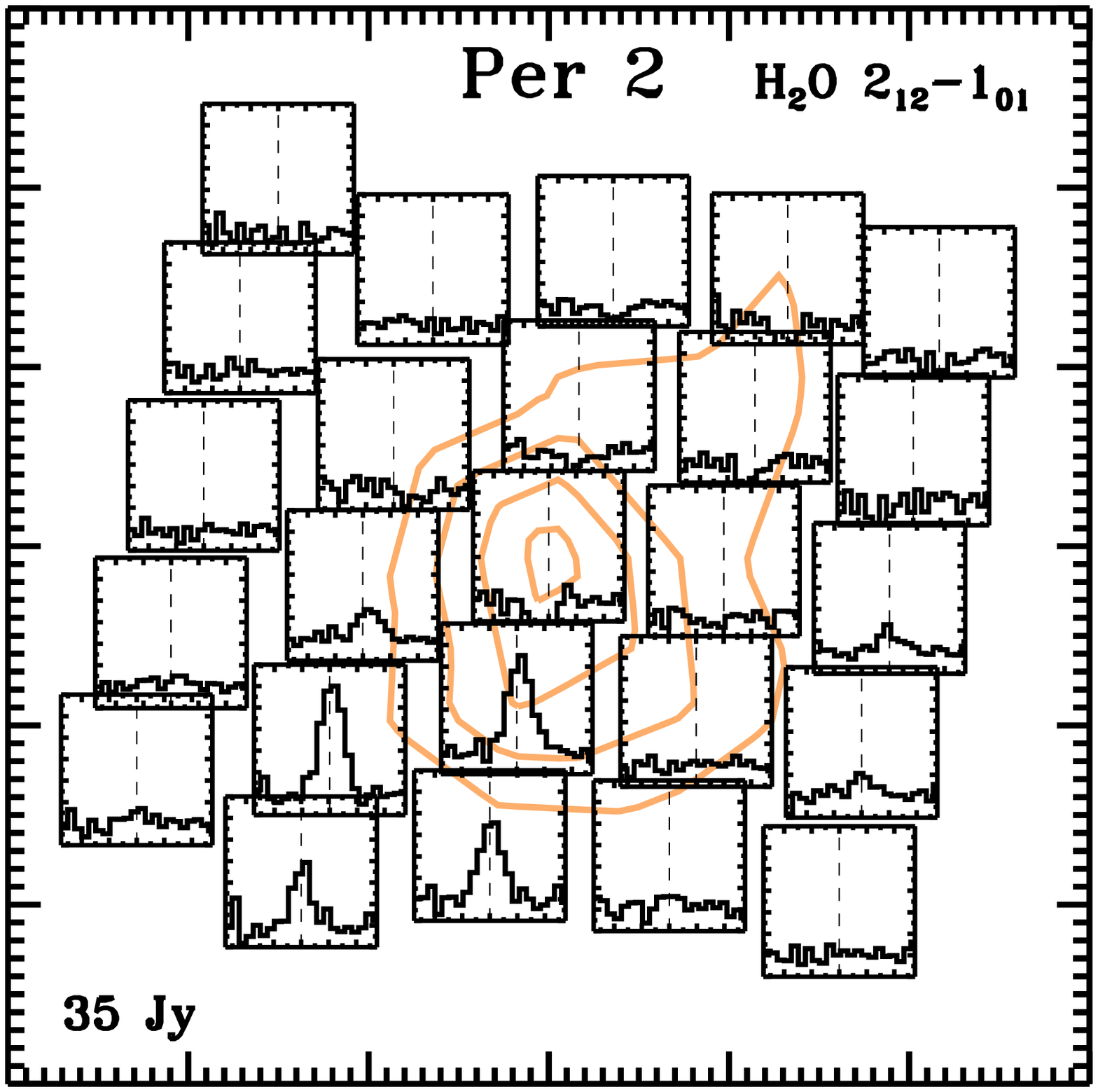} 
         \vspace{-1cm}
         
          \hspace{-0.2cm} 
     \includegraphics[angle=90,height=6.3cm]{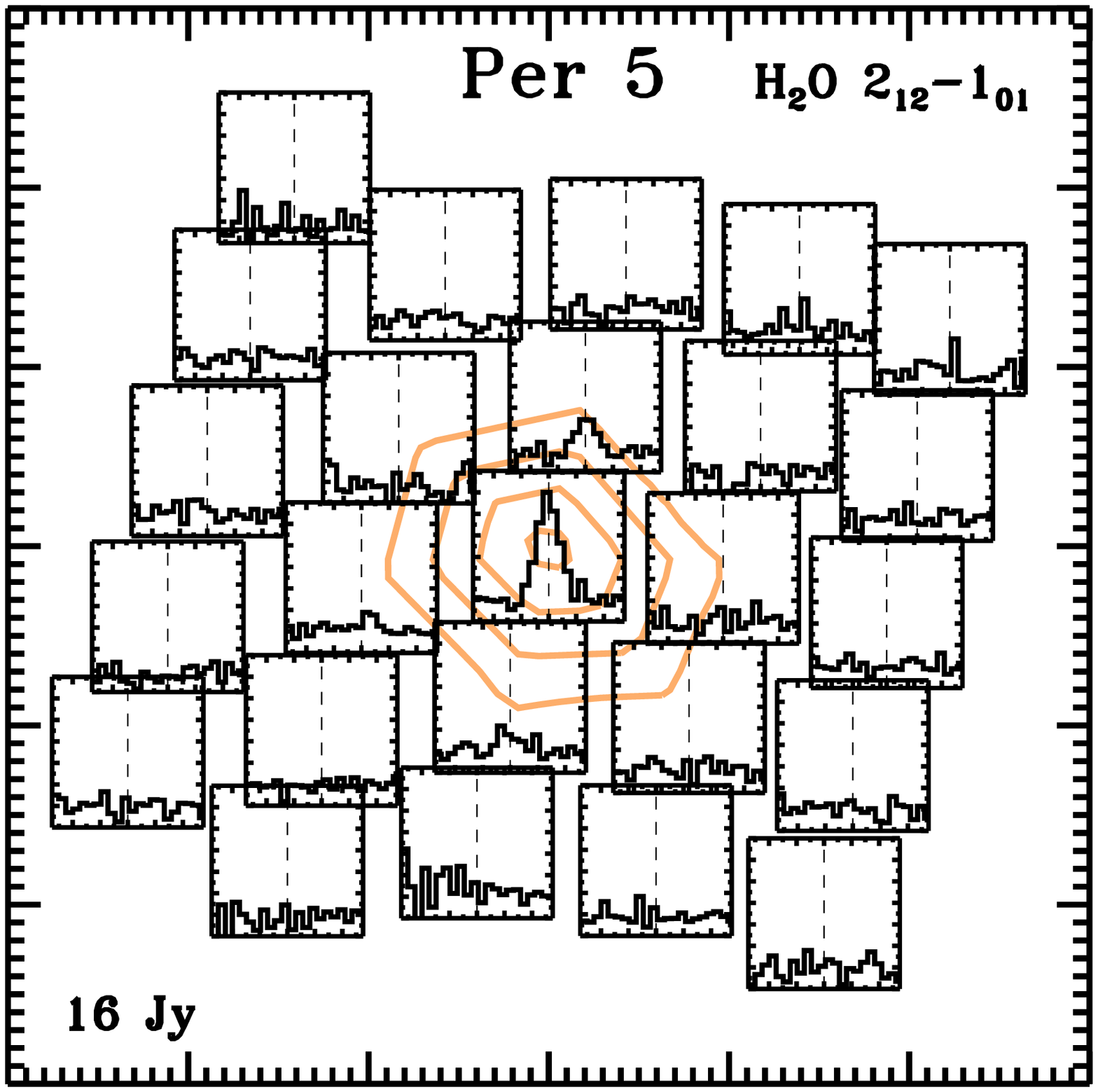}
              \vspace{-1cm}
              
         \hspace{-0.2cm}       
     \includegraphics[angle=90,height=6.3cm]{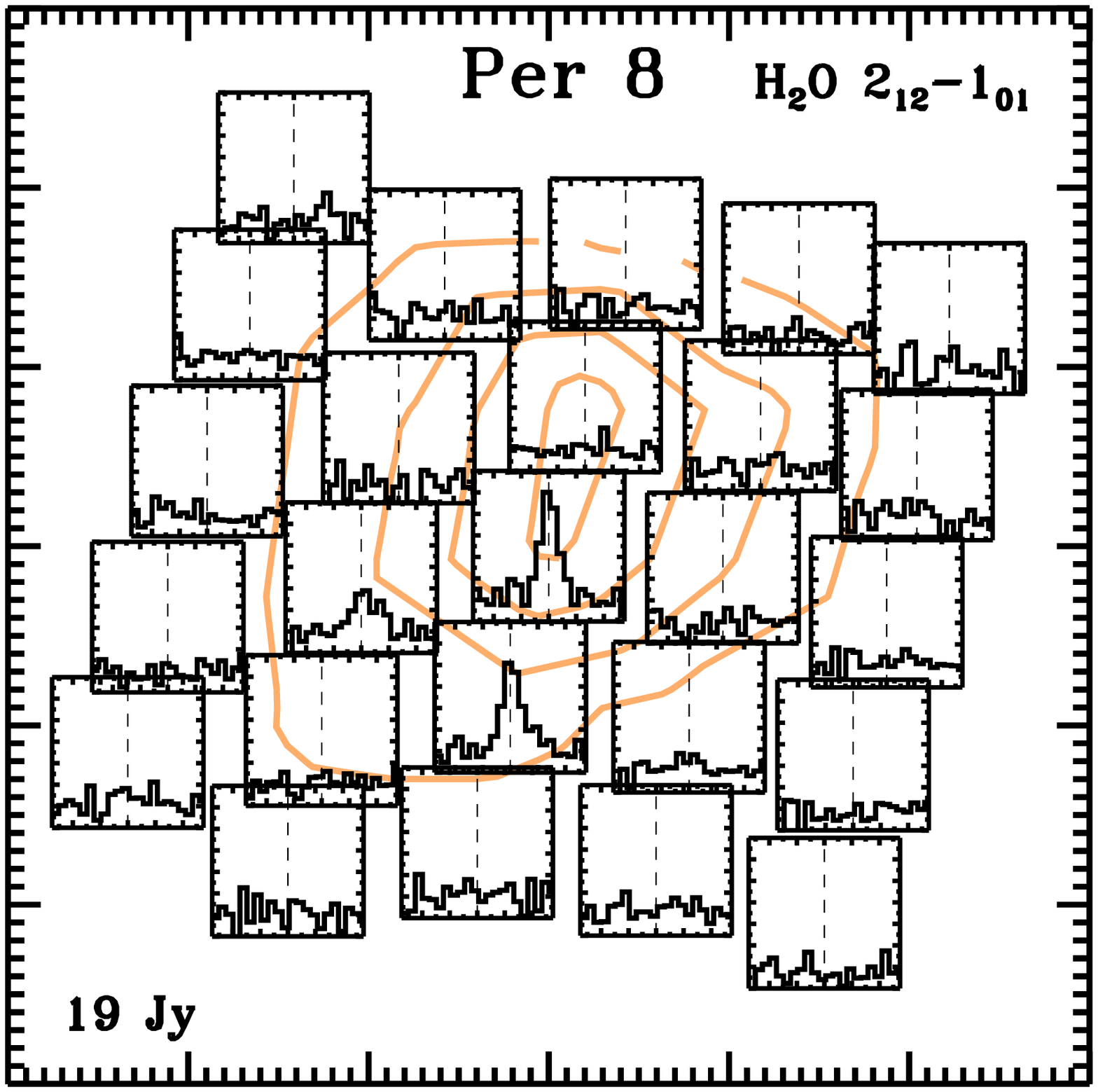}           
        \vspace{-1cm}
        
         \hspace{-0.2cm} 
       \includegraphics[angle=90,height=6.3cm]{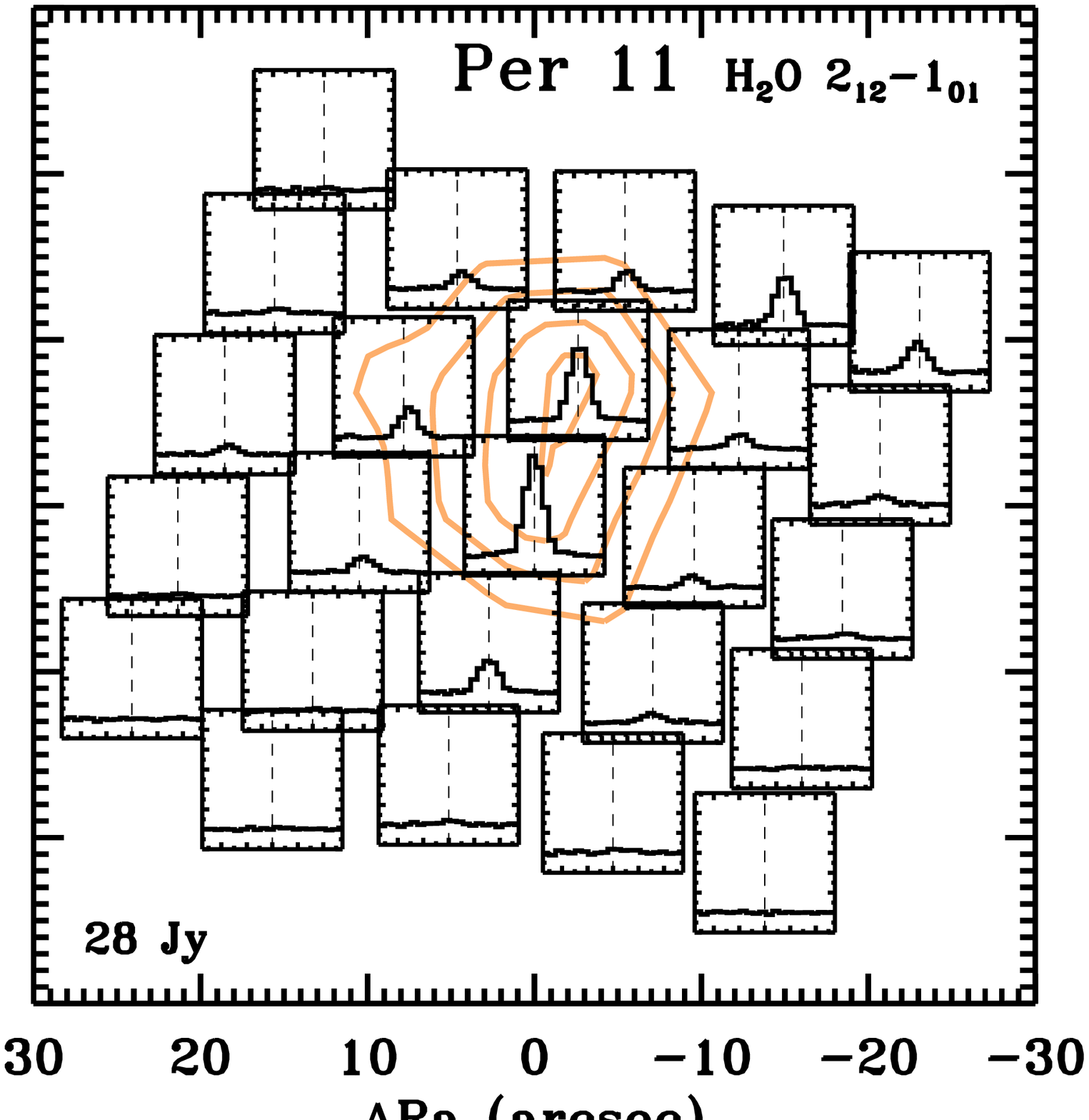}
                     \vspace{+0.2cm}
  \end{minipage}
    \hfill
  \begin{minipage}[t]{.3\textwidth}
  \hspace{-0.6cm}
     \includegraphics[angle=90,height=6.3cm]{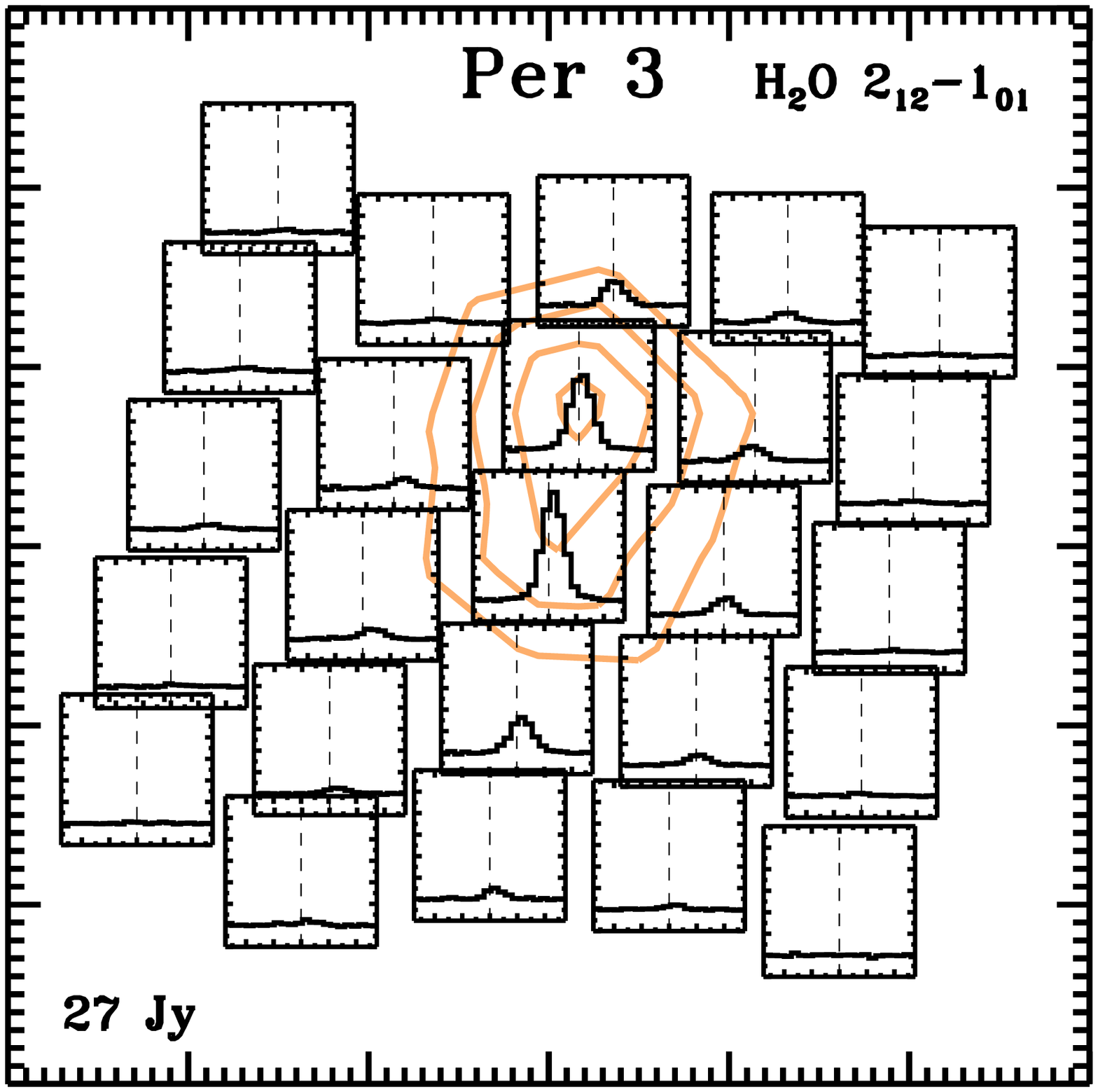}
          \vspace{-0.65cm}
          
          \hspace{-0.6cm}
     \includegraphics[angle=90,height=6.3cm]{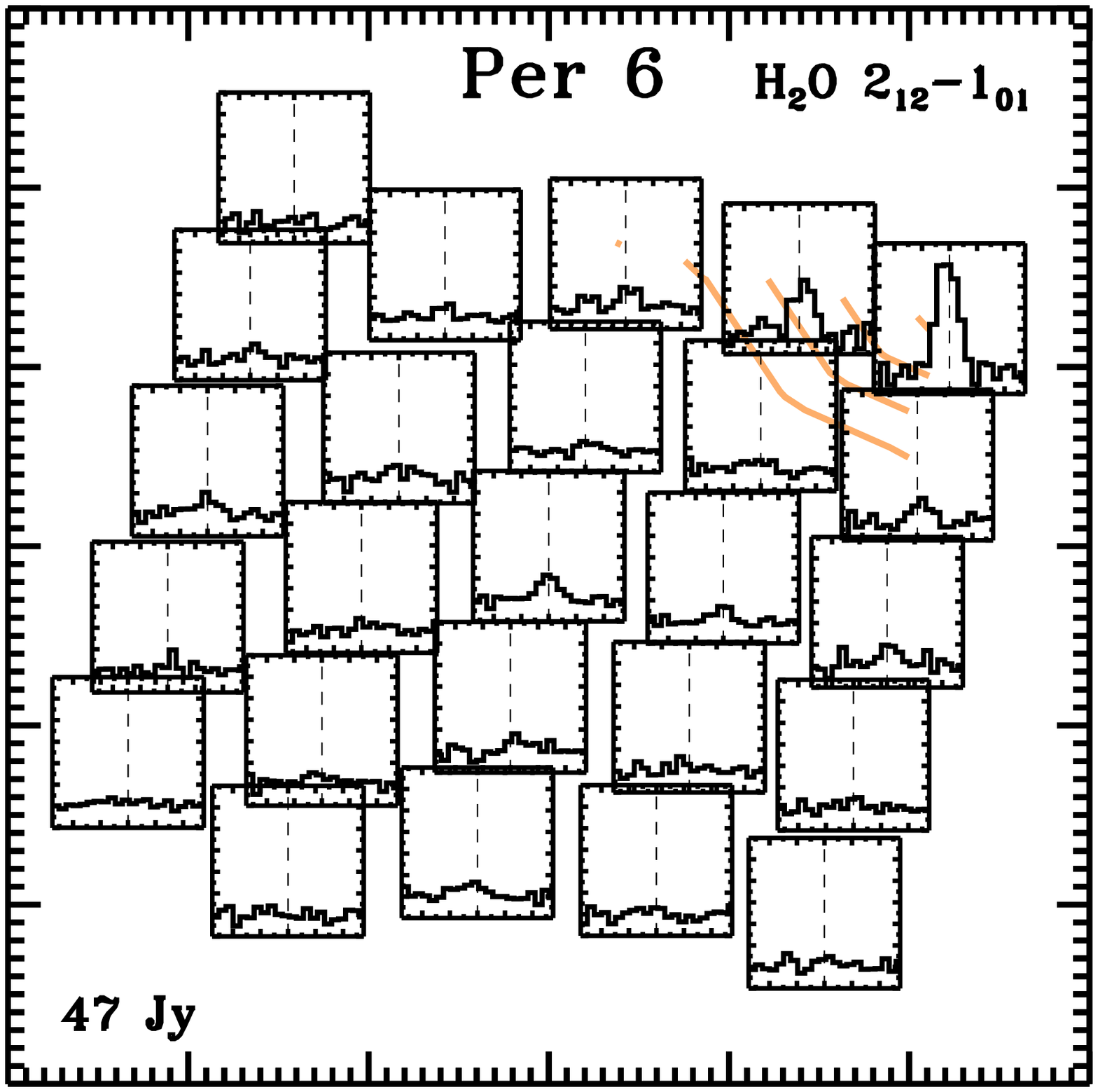}
               \vspace{-0.65cm}
               
       \hspace{-0.6cm}
     \includegraphics[angle=90,height=6.3cm]{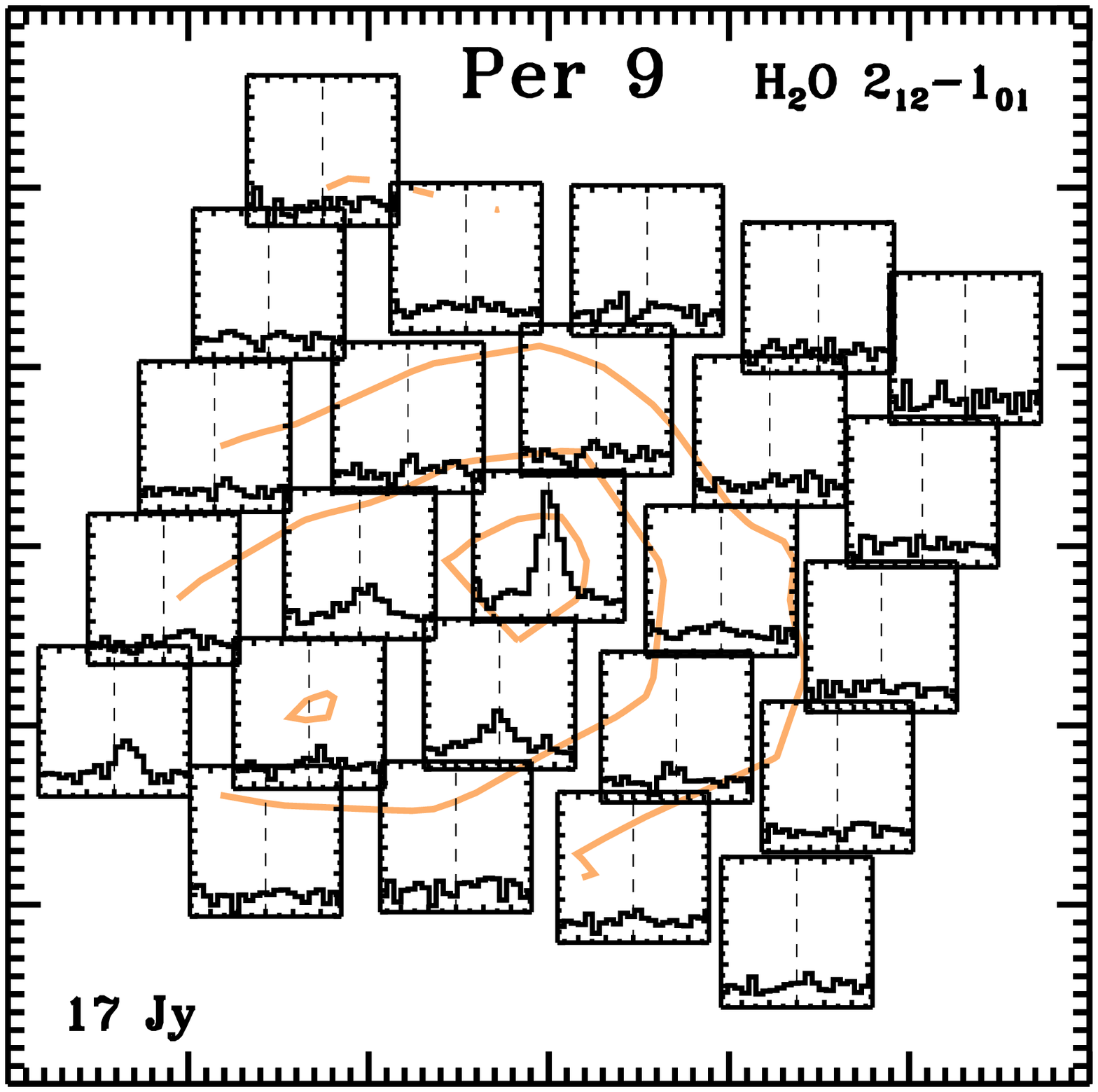}
                   \vspace{-0.65cm}
                   
     \hspace{-0.6cm}
       \includegraphics[angle=90,height=6.3cm]{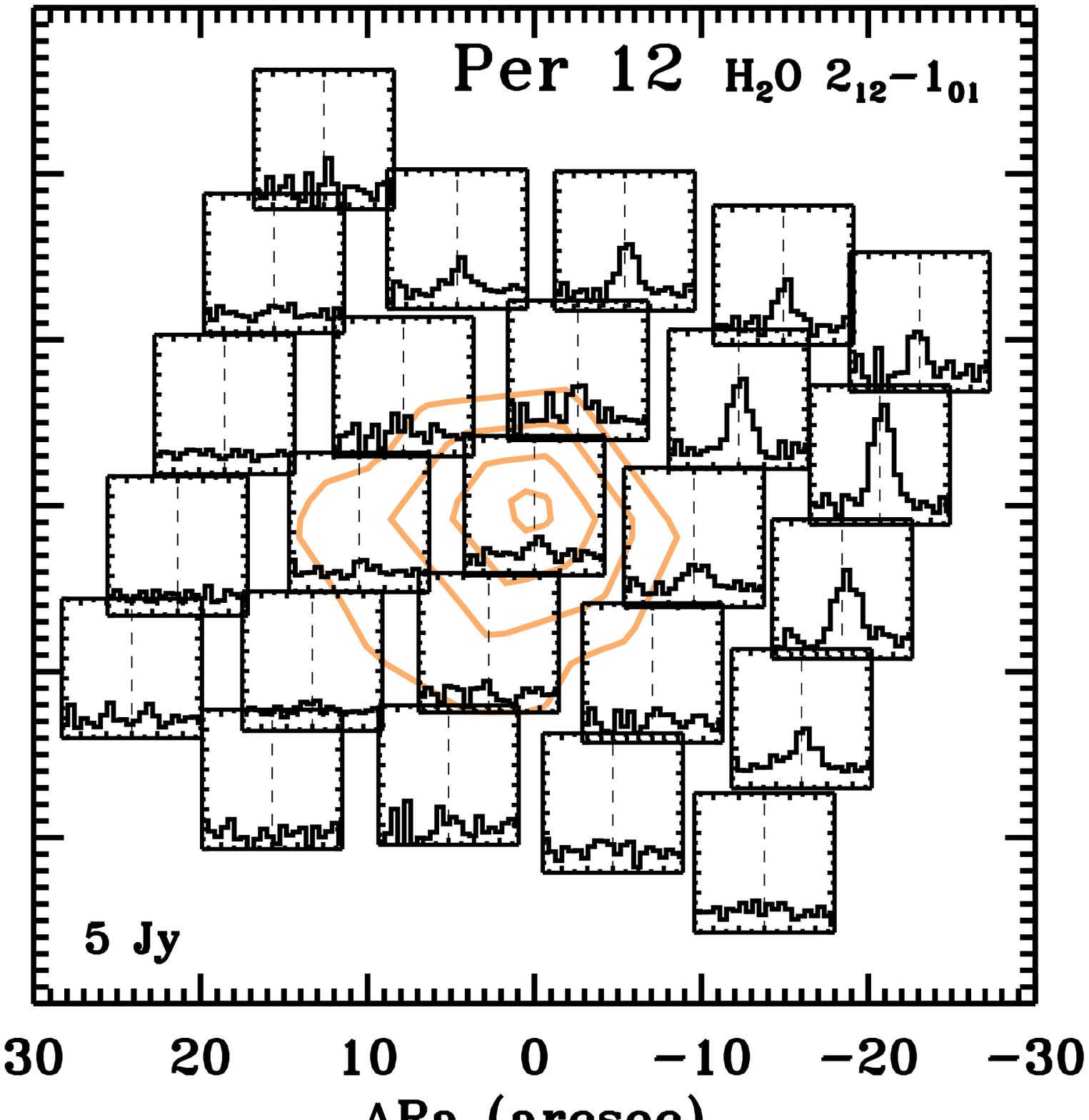}
                     \vspace{+0.2cm}
  \end{minipage} 
        \caption{\label{specmap1}PACS spectral maps in the H$_2$O 2$_{12}$-1$_{01}$ 
        line at 179.527 $\mu$m. Wavelengths in  microns are translated to the velocity scale on the X-axis using laboratory wavelengths of the
species and cover the range from -600 to 600 km\,s$^{-1}$. The Y-axis shows fluxes normalized to the 
    spaxel with the brightest line on the map in a range -0.2 to 1.2. The orange contours show continuum emission 
        at 30\%, 50\%, 70\%, and 90\% of the peak value written in the bottom left corner of each map. }
\end{figure*}

\begin{figure*}[!tb]
  \begin{minipage}[t]{.3\textwidth}
    \hspace{+0.2cm}
      \includegraphics[angle=90,height=6.3cm]{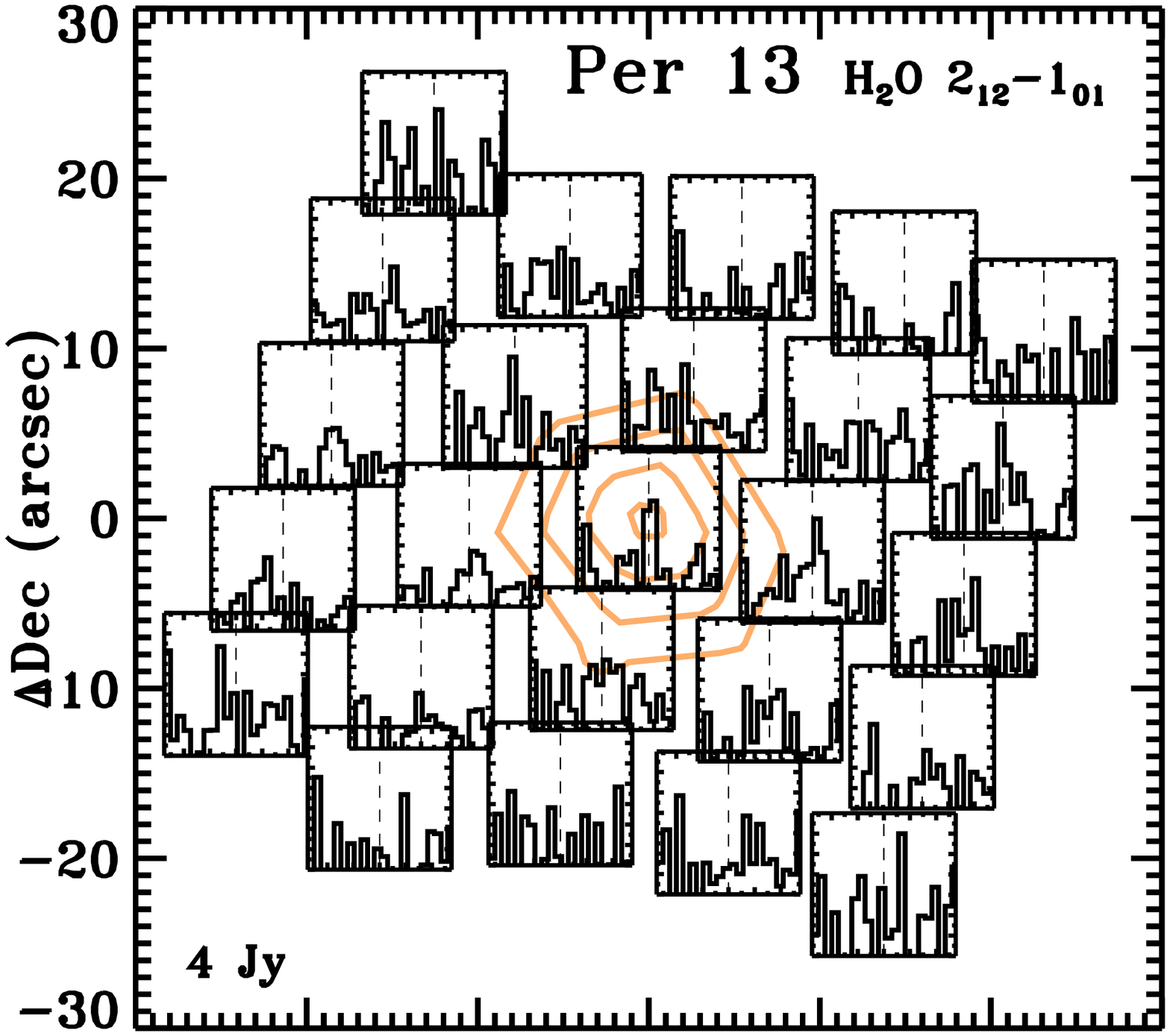}
     \vspace{-1cm}
     
         \hspace{+0.2cm}
       \includegraphics[angle=90,height=6.3cm]{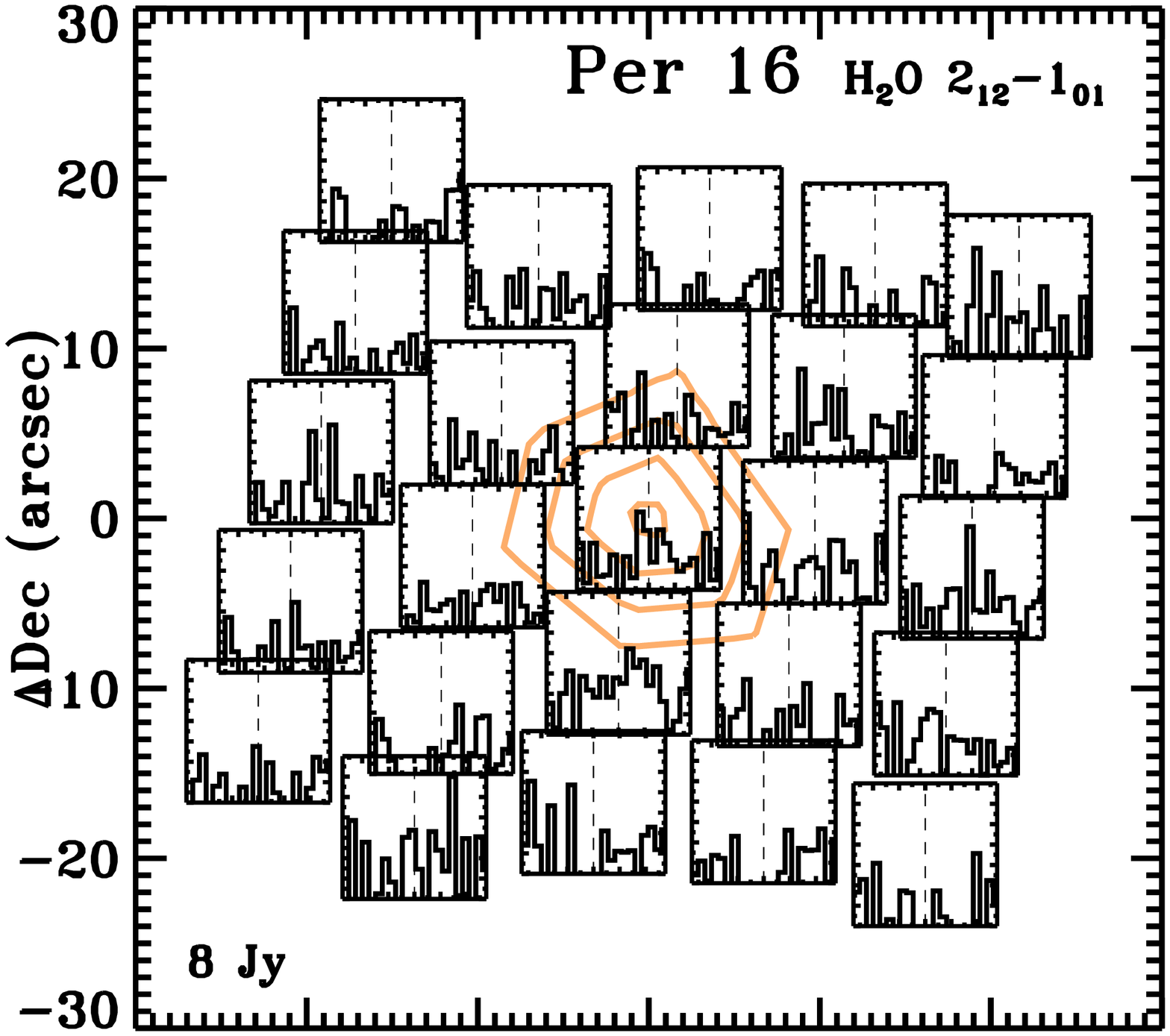}
       \vspace{-1cm}
     
         \hspace{+0.2cm}
       \includegraphics[angle=90,height=6.3cm]{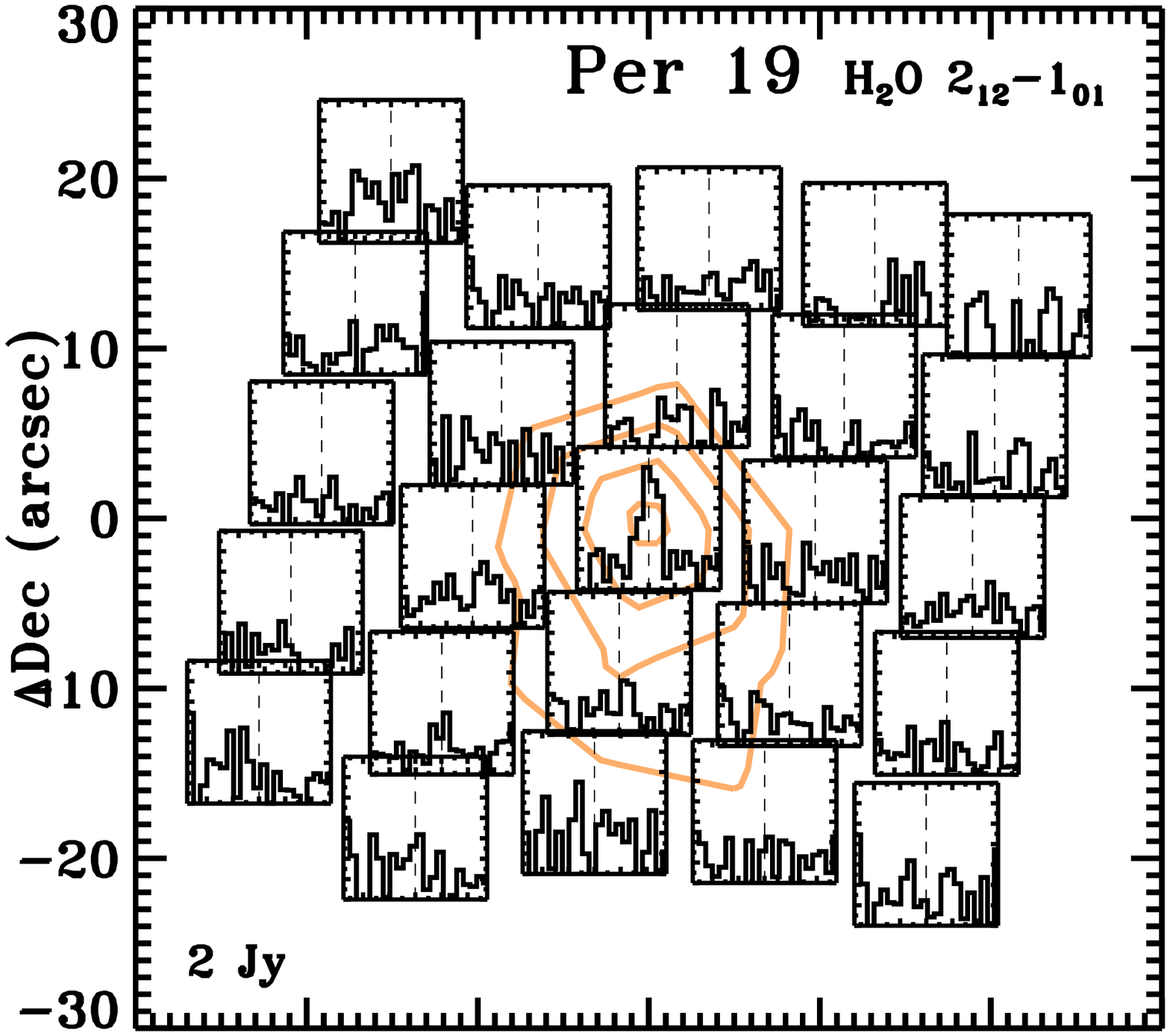}
              \vspace{-1cm}
     
     \hspace{+0.2cm}
       \includegraphics[angle=90,height=6.3cm]{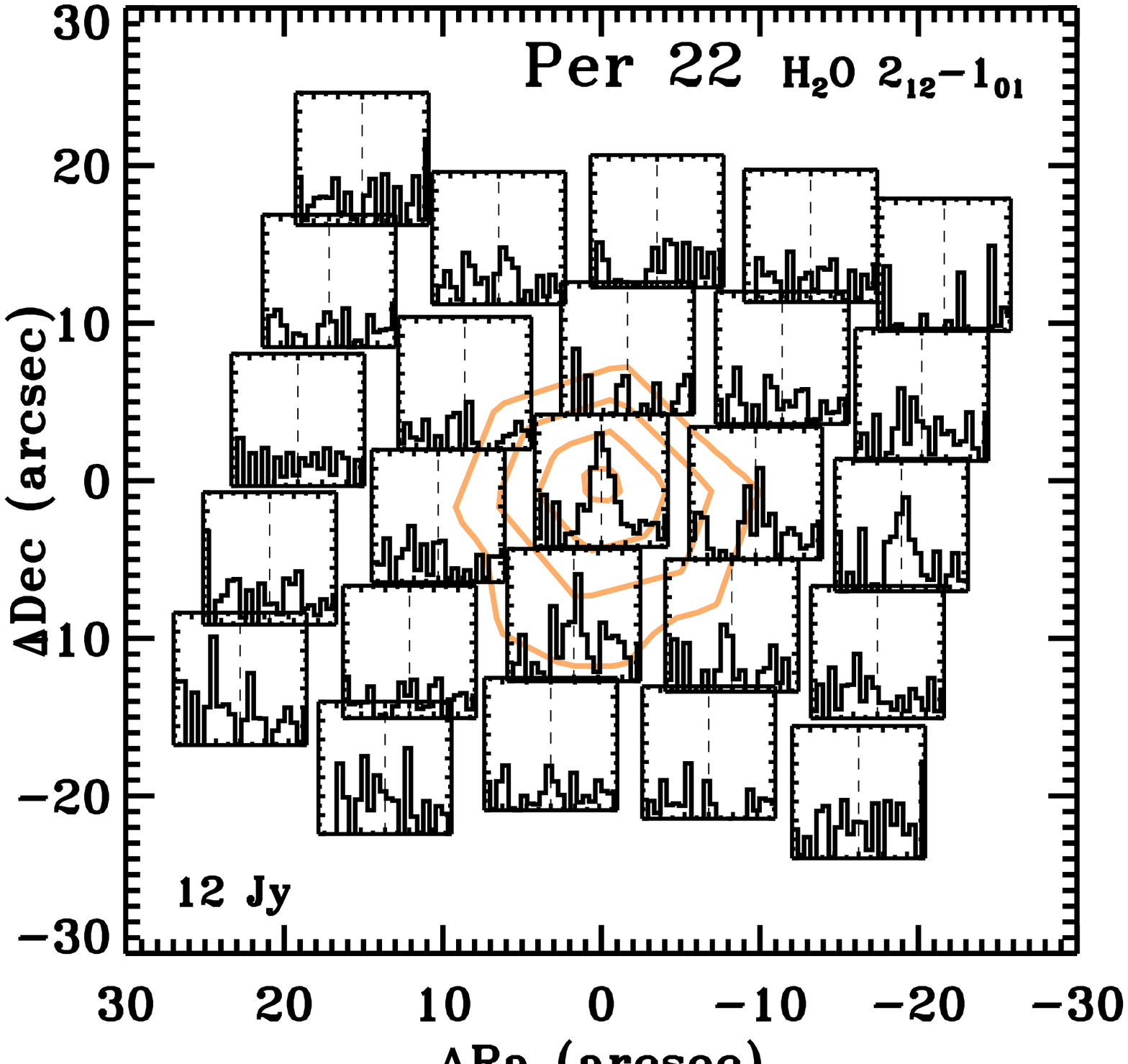}
                     \vspace{+0.2cm}
  \end{minipage}
  \hfill
  \begin{minipage}[t]{.3\textwidth}
     \hspace{-0.2cm} 
    \includegraphics[angle=90,height=6.3cm]{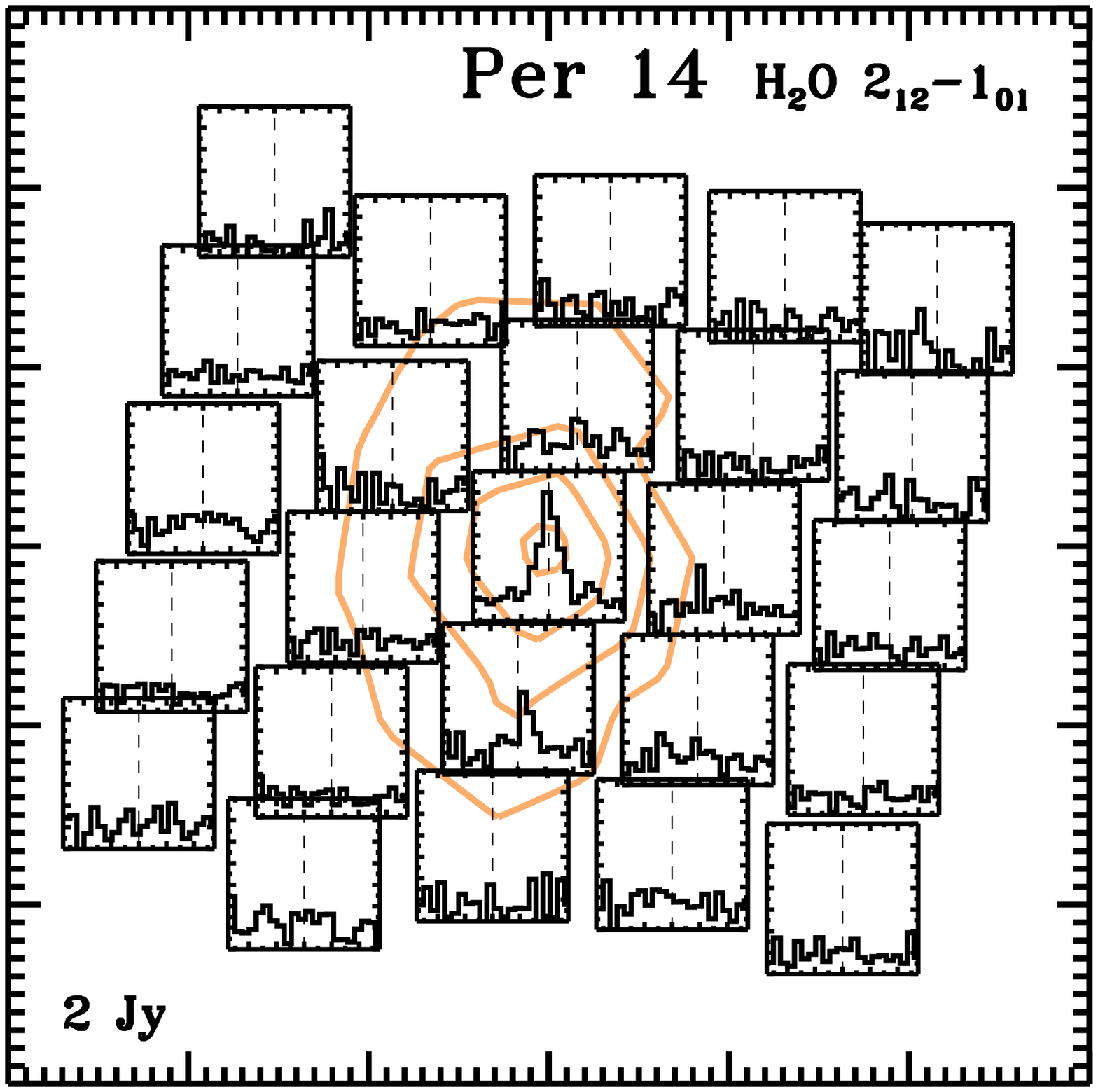} 
         \vspace{-1cm}
       
        \hspace{-0.2cm} 
     \includegraphics[angle=90,height=6.3cm]{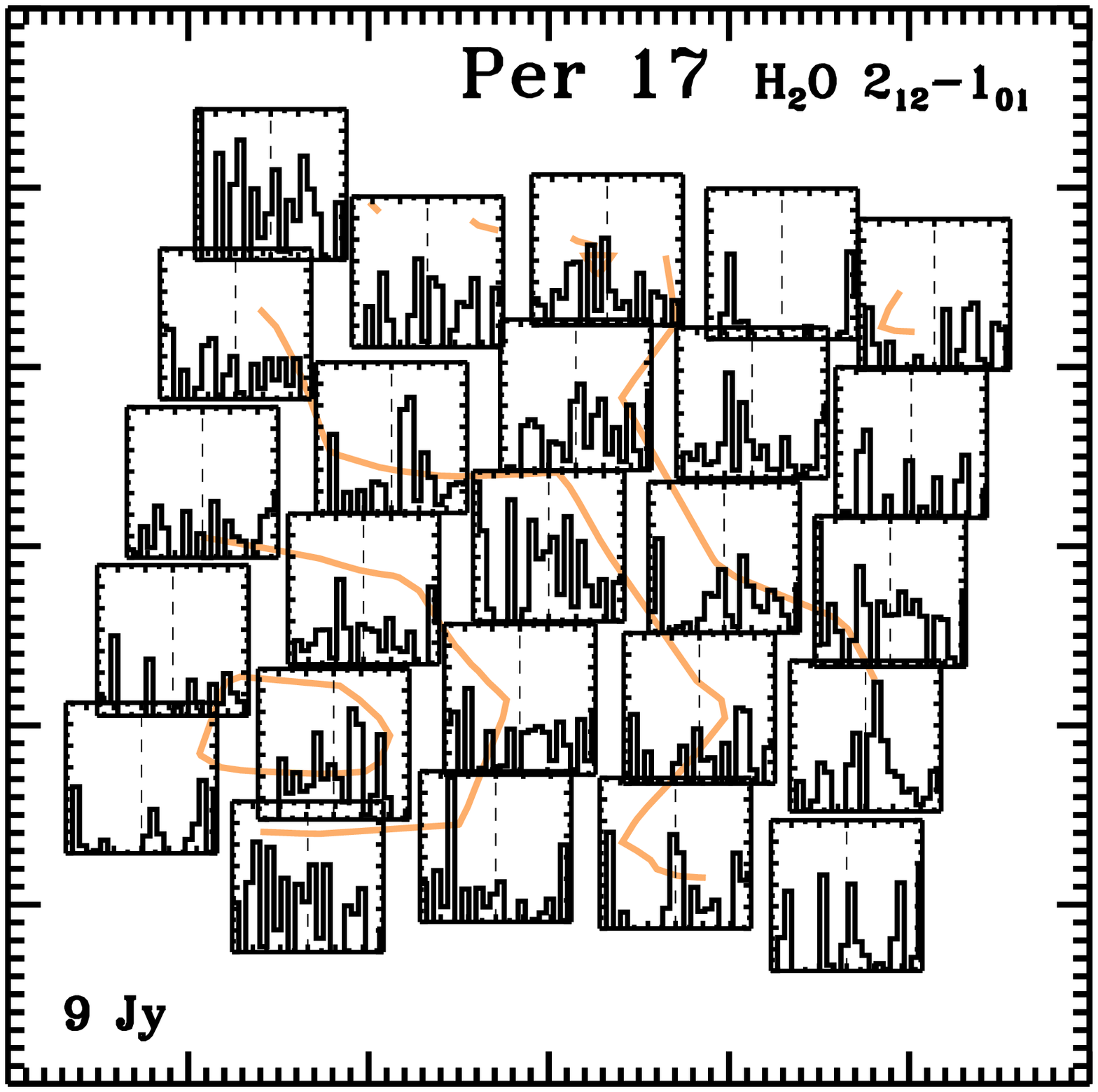}
              \vspace{-1cm}
       
        \hspace{-0.2cm} 
     \includegraphics[angle=90,height=6.3cm]{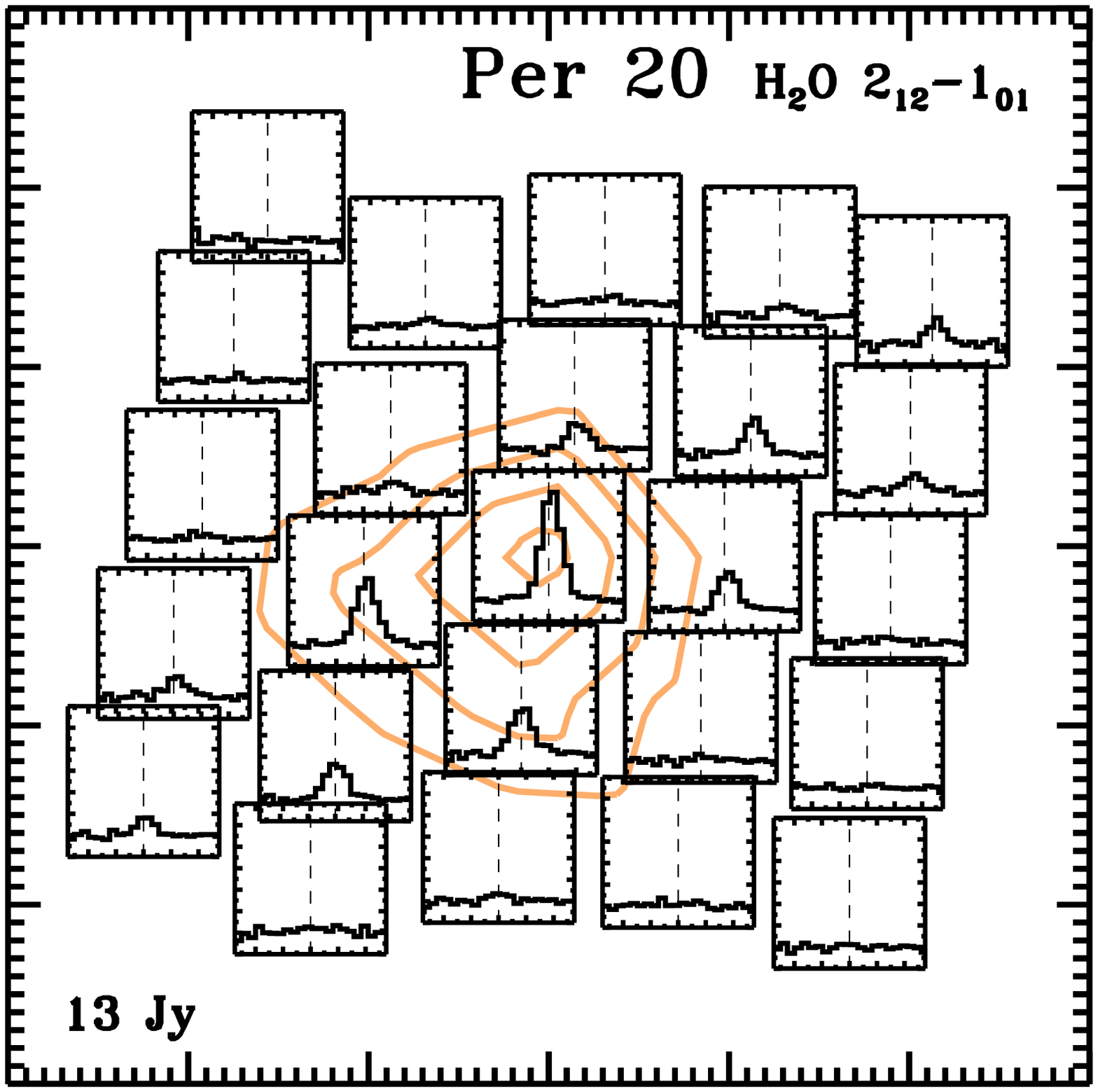}           
        \vspace{+0.2cm}
  \end{minipage}
    \hfill
  \begin{minipage}[t]{.3\textwidth}
 \hspace{-0.6cm}
     \includegraphics[angle=90,height=6.3cm]{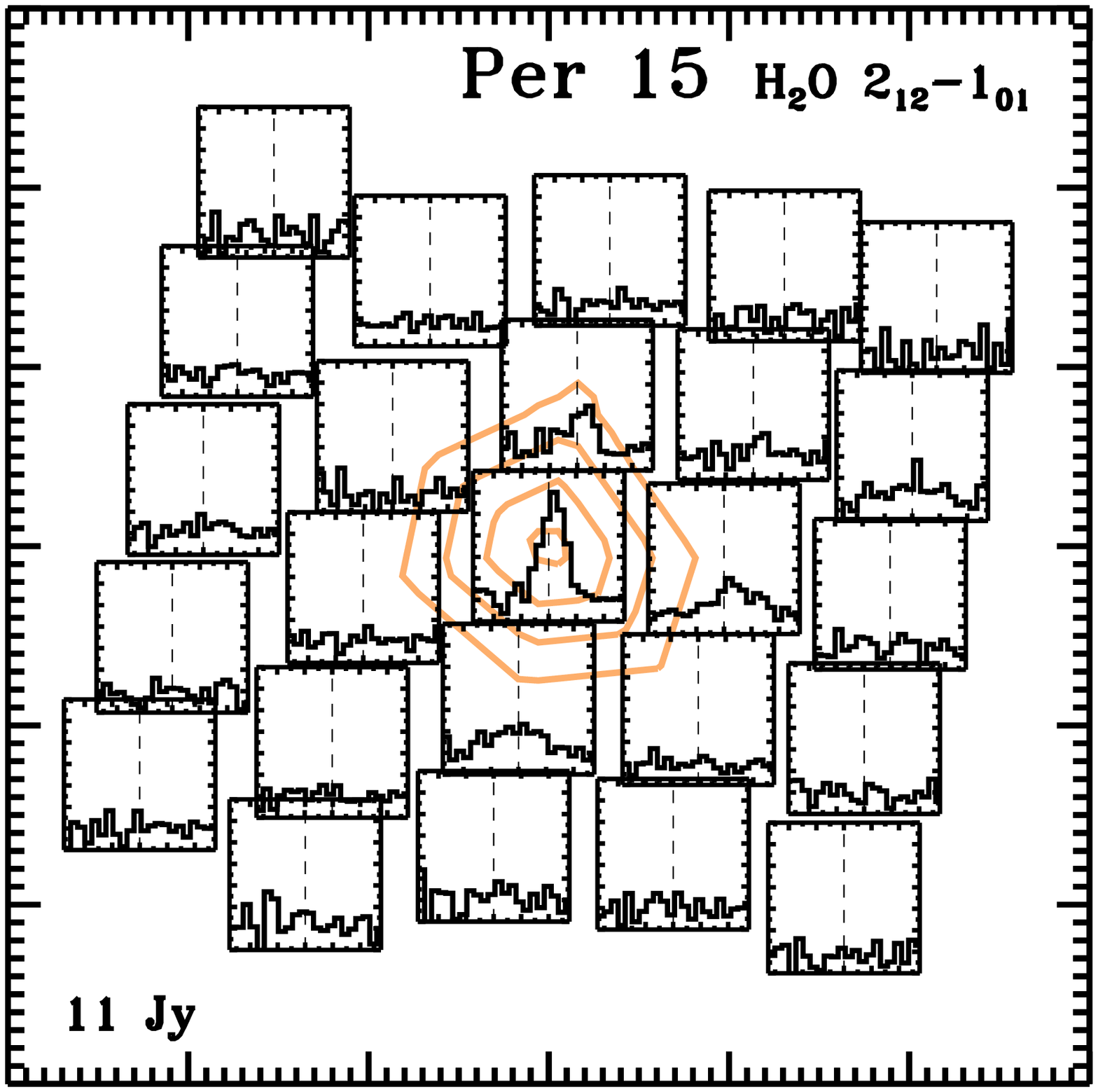}
          \vspace{-0.65cm}
       
       \hspace{-0.6cm}
     \includegraphics[angle=90,height=6.3cm]{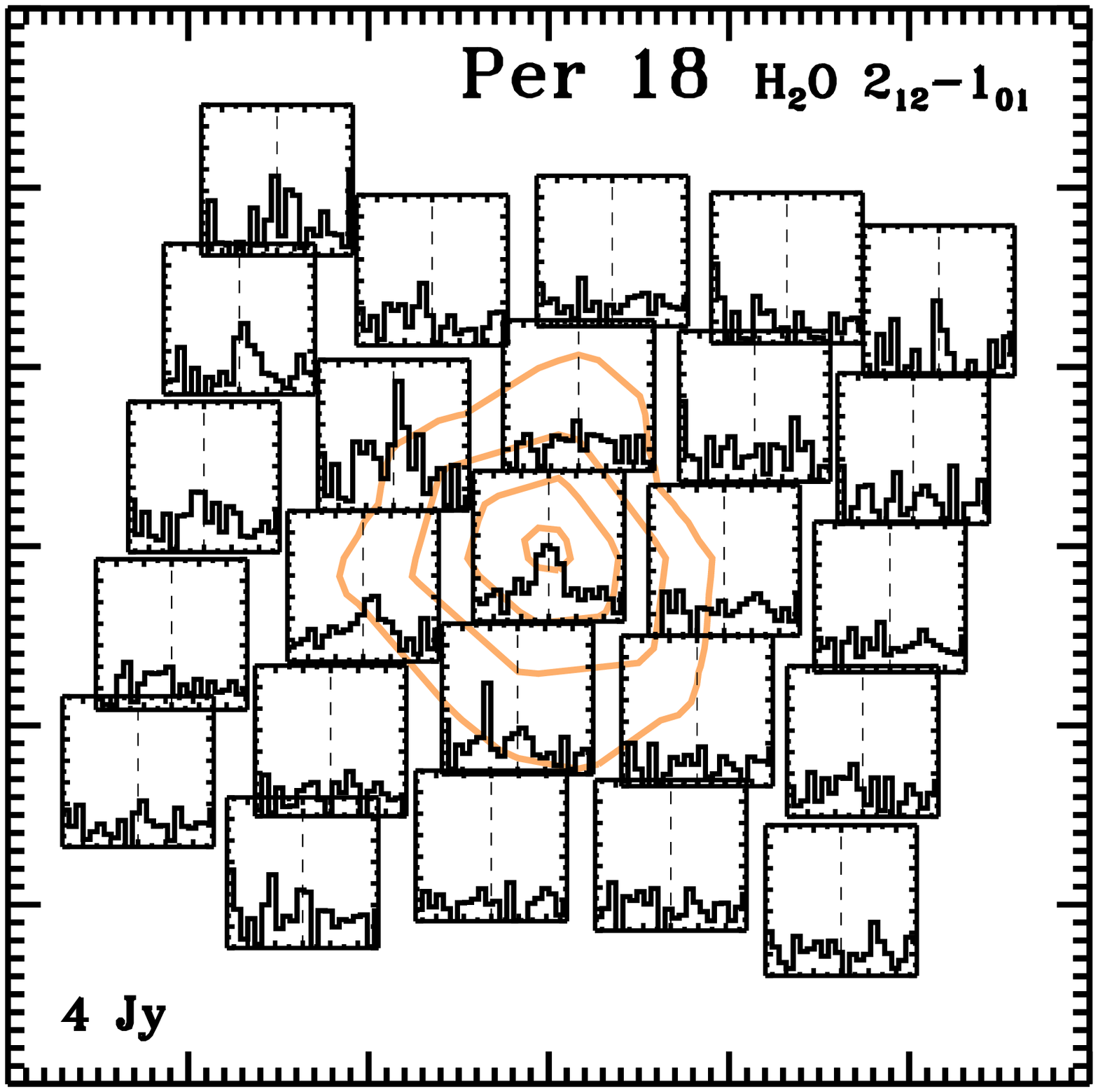}
             \vspace{-0.65cm}
       
       \hspace{-0.6cm}
     \includegraphics[angle=90,height=6.3cm]{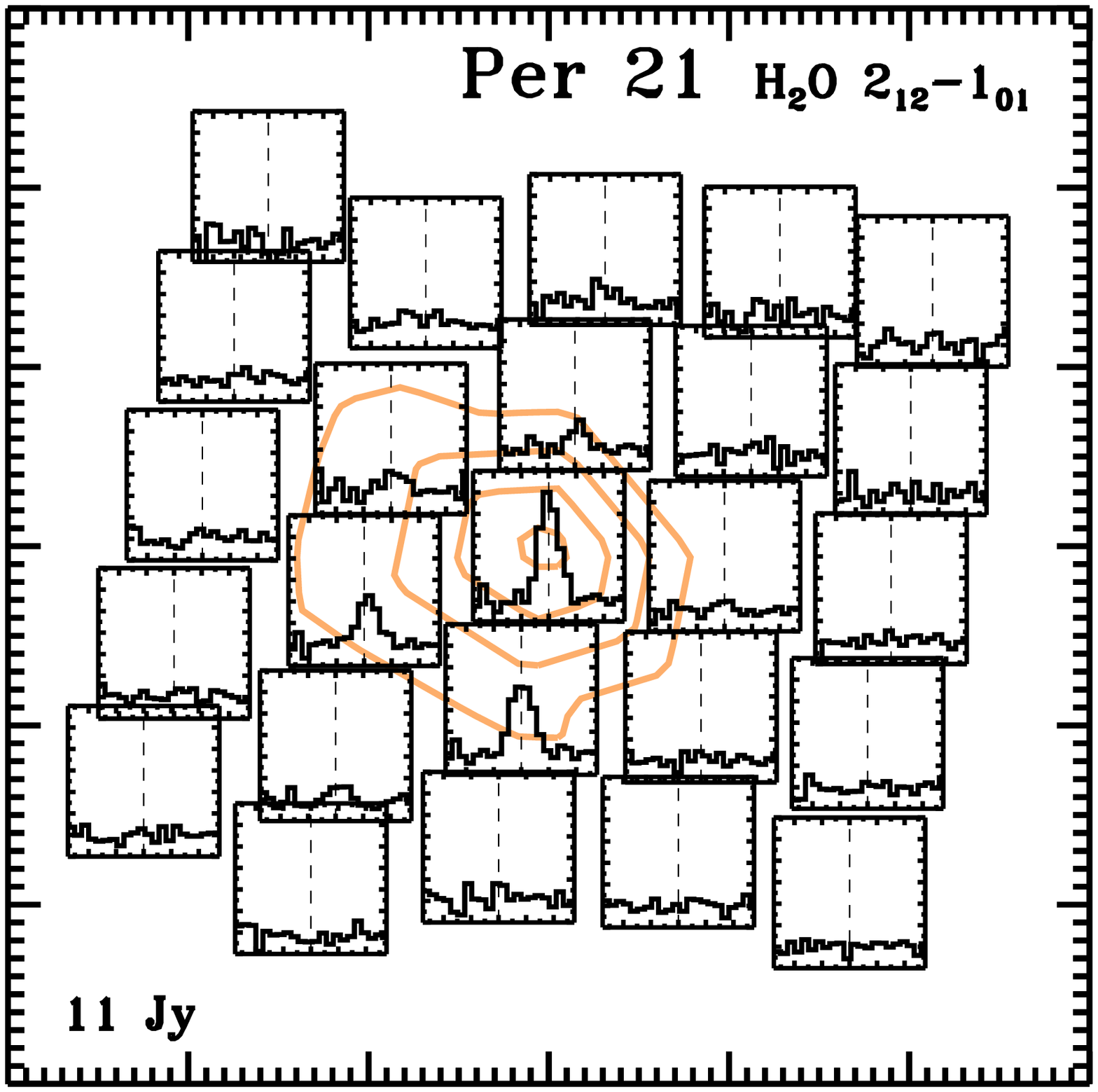}
               \vspace{-0.65cm}
  \end{minipage}
        \caption{\label{specmap2}The same as Fig.~\ref{specmap1} but for the remaining 
        sources.}
\end{figure*}

Figures \ref{specmap1} and \ref{specmap2} show line and continuum maps around 179.5 $\mu$m for all the Perseus sources
in the WILL program.

\begin{figure*}[!tb]
  \begin{minipage}[t]{.3\textwidth}
    \hspace{+0.2cm}
      \includegraphics[angle=90,height=6.3cm]{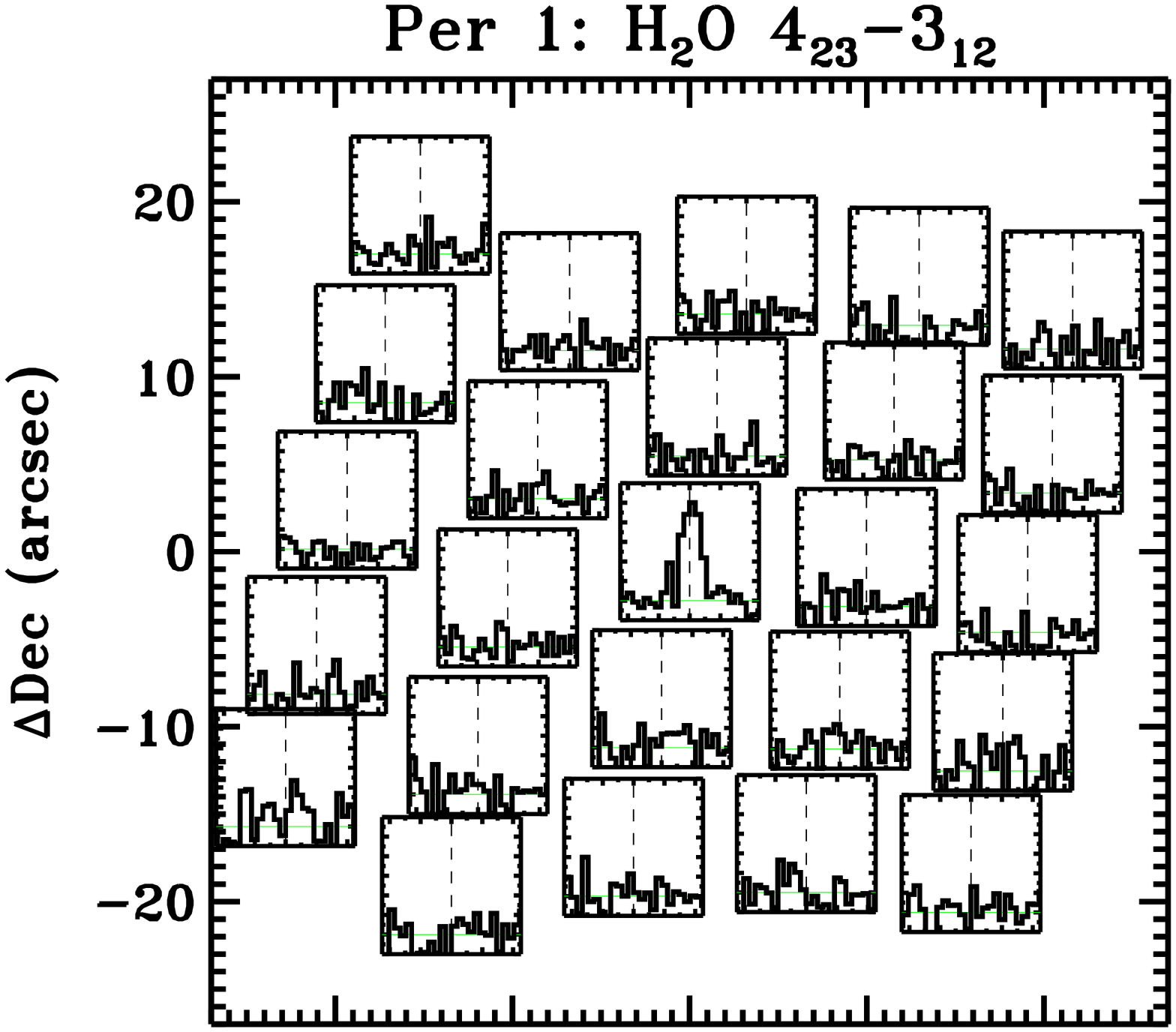}
     \vspace{-1cm}
     
     \hspace{+0.2cm}
       \includegraphics[angle=90,height=6.3cm]{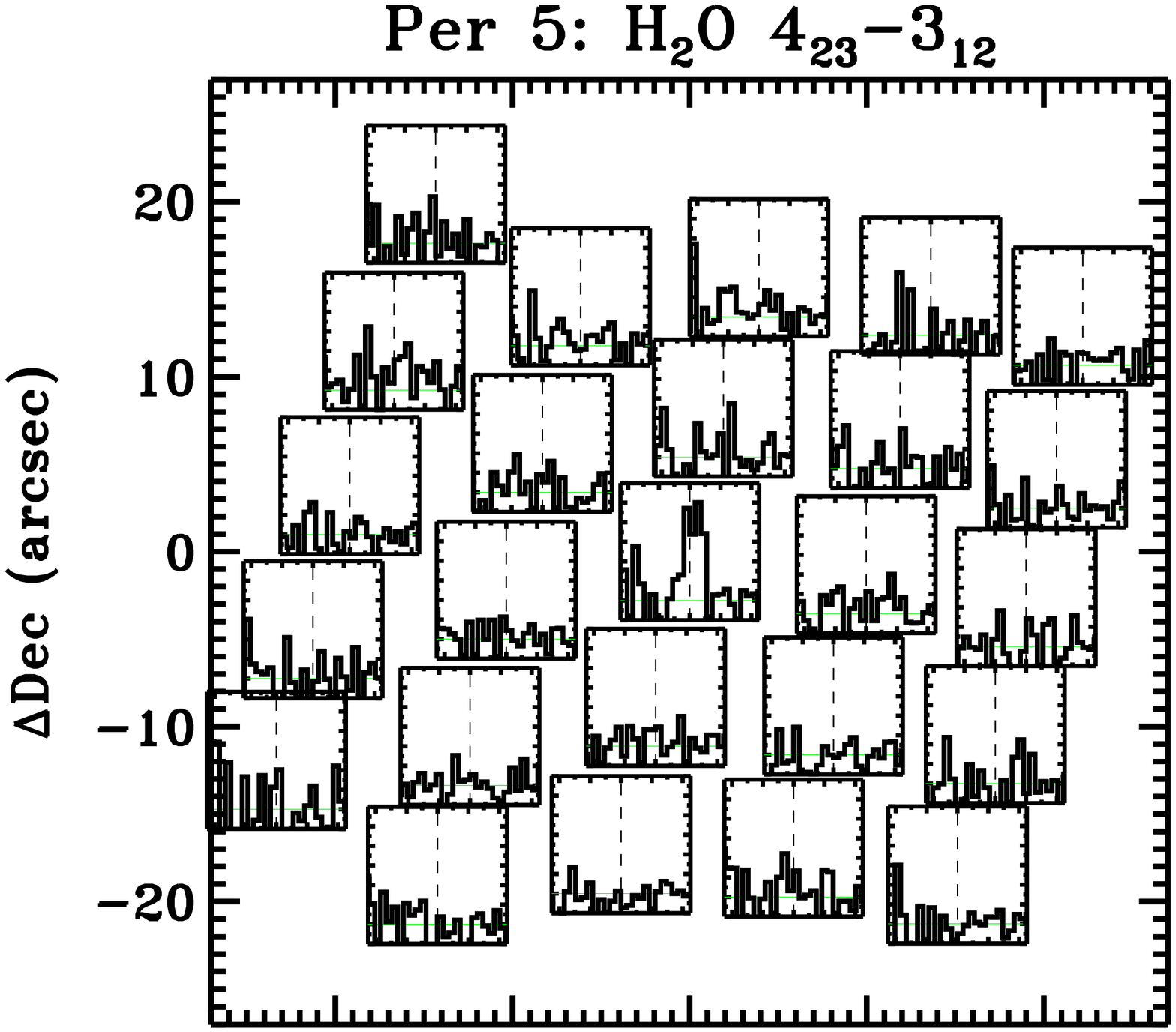}
       \vspace{-1cm}
       
       \hspace{+0.2cm}
       \includegraphics[angle=90,height=6.3cm]{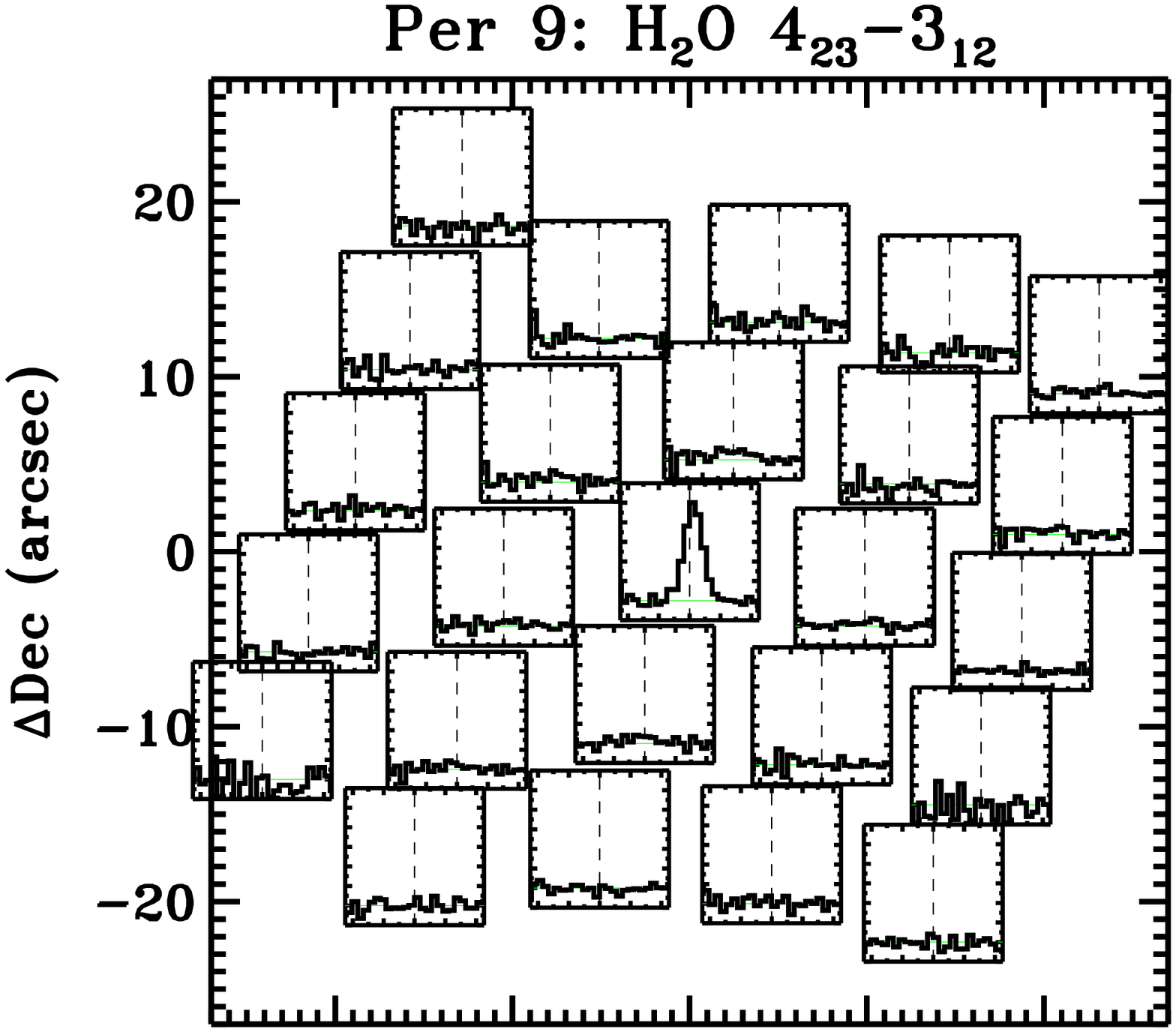}
              \vspace{-1cm}
              
          \hspace{+0.2cm}    
       \includegraphics[angle=90,height=6.3cm]{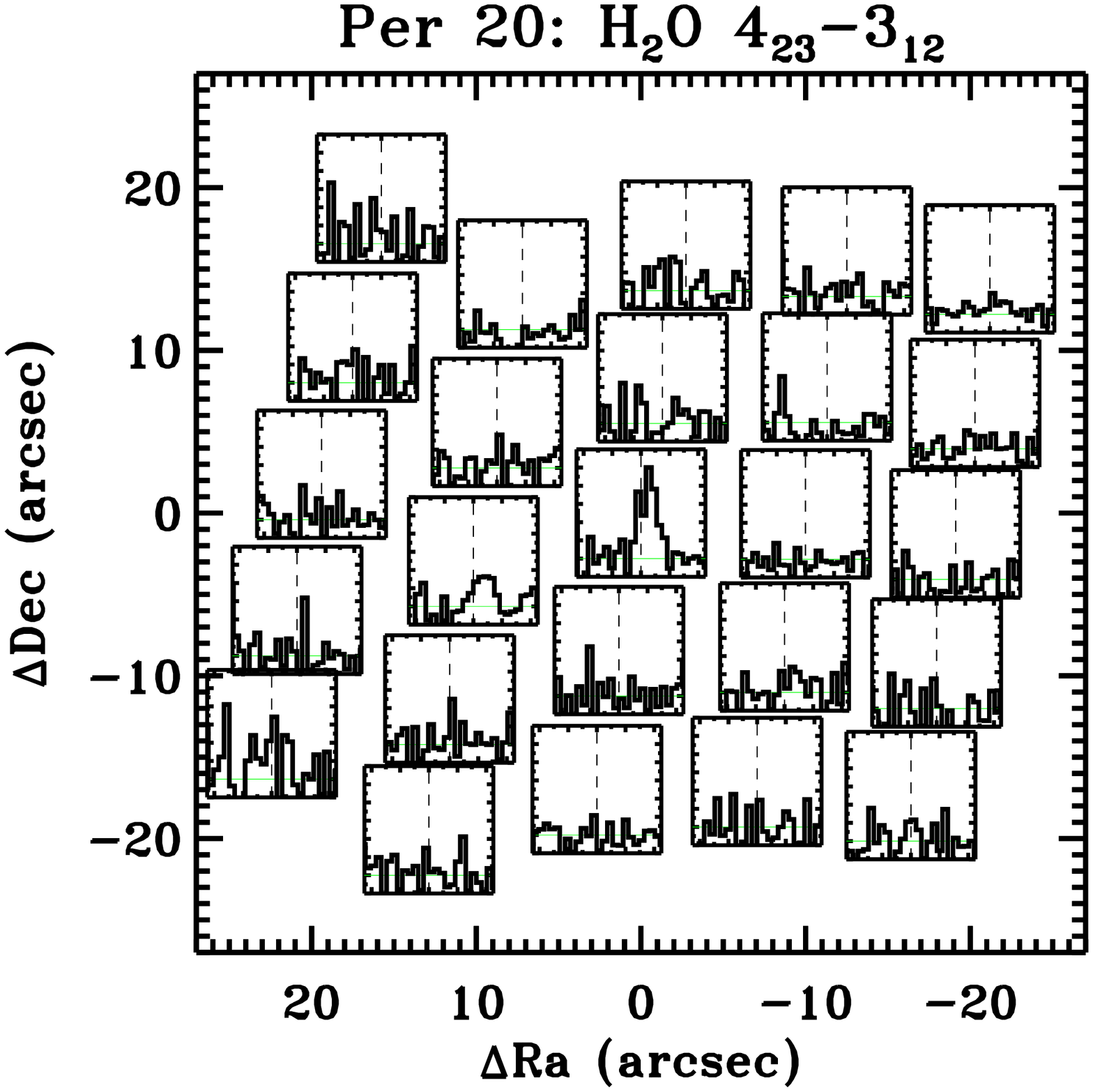}
                     \vspace{+0.2cm}
  \end{minipage}
  \hfill
  \begin{minipage}[t]{.3\textwidth}
 \hspace{-0.2cm} 
    \includegraphics[angle=90,height=6.3cm]{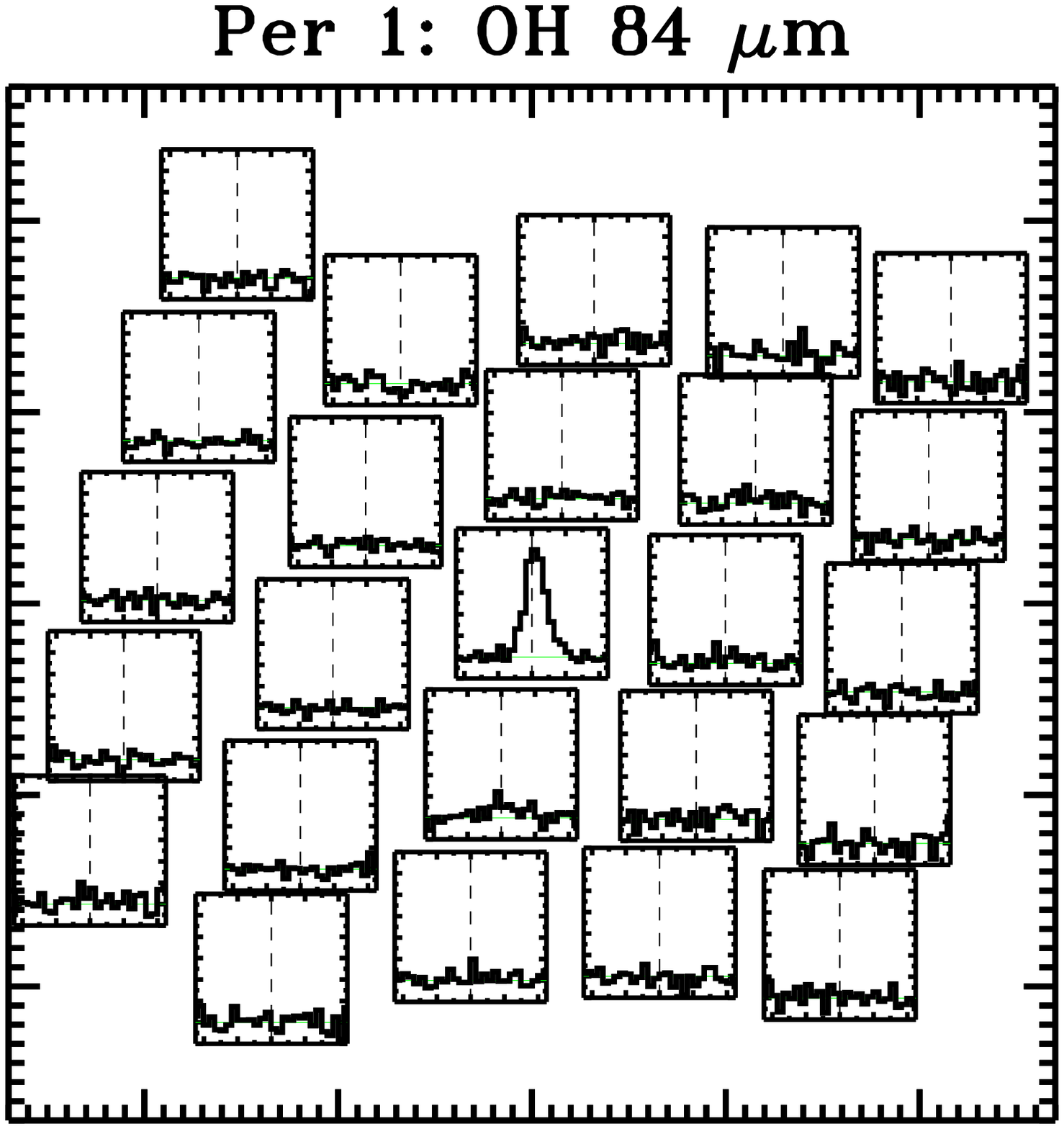} 
         \vspace{-1cm}
         
          \hspace{-0.2cm} 
     \includegraphics[angle=90,height=6.3cm]{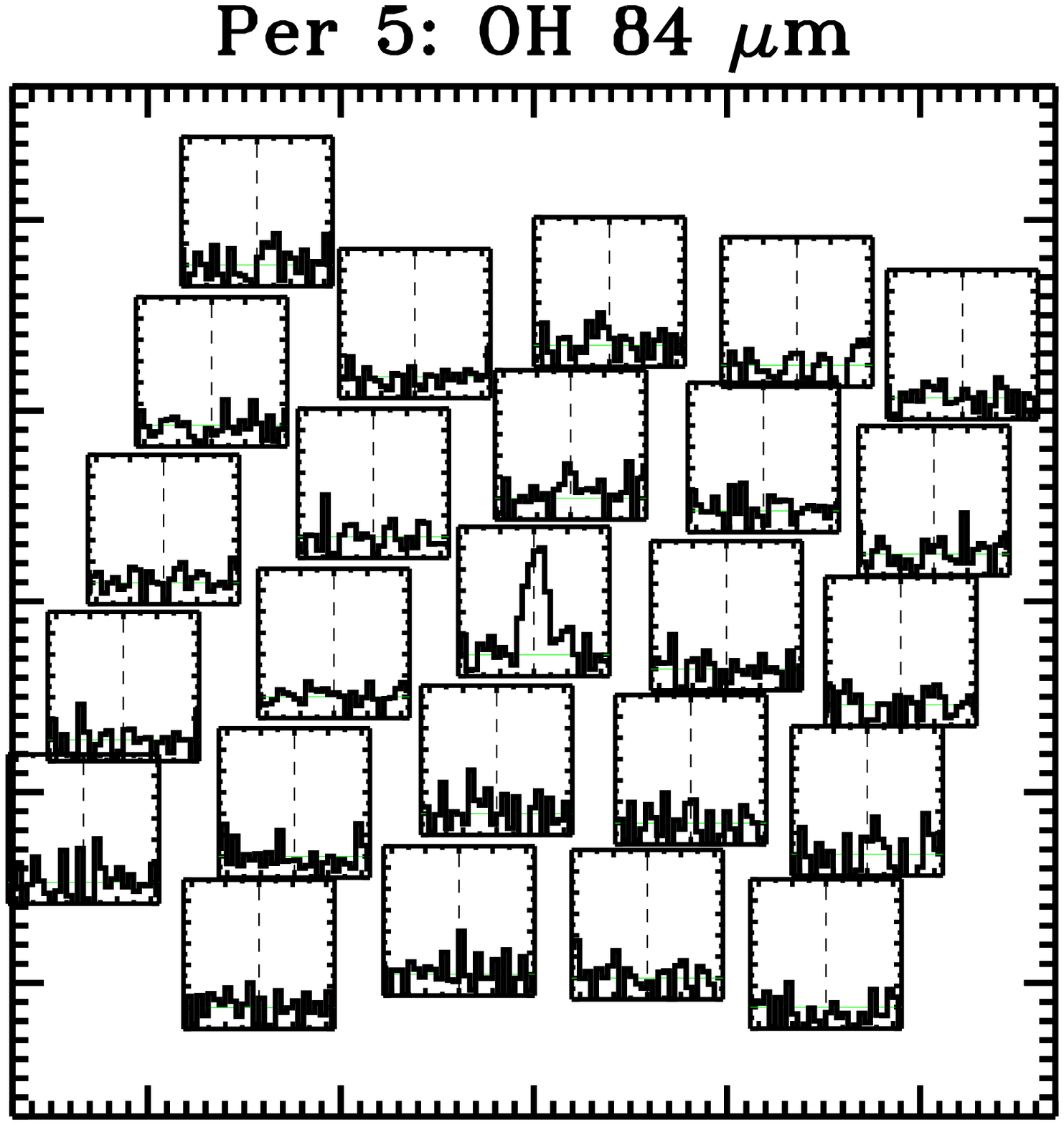}
              \vspace{-1cm}
              
         \hspace{-0.2cm}       
     \includegraphics[angle=90,height=6.3cm]{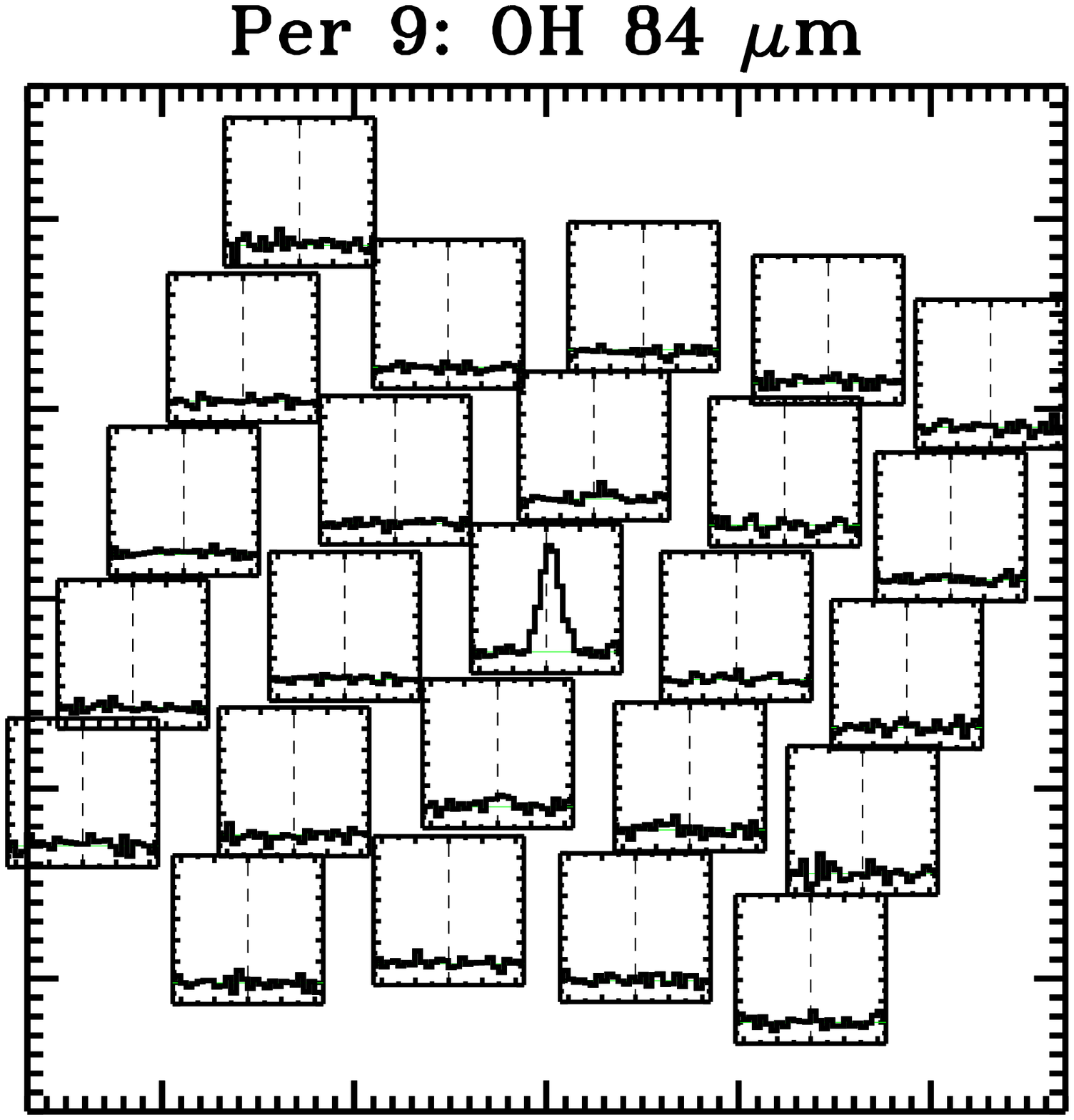}           
        \vspace{-1cm}
        
         \hspace{-0.2cm} 
       \includegraphics[angle=90,height=6.3cm]{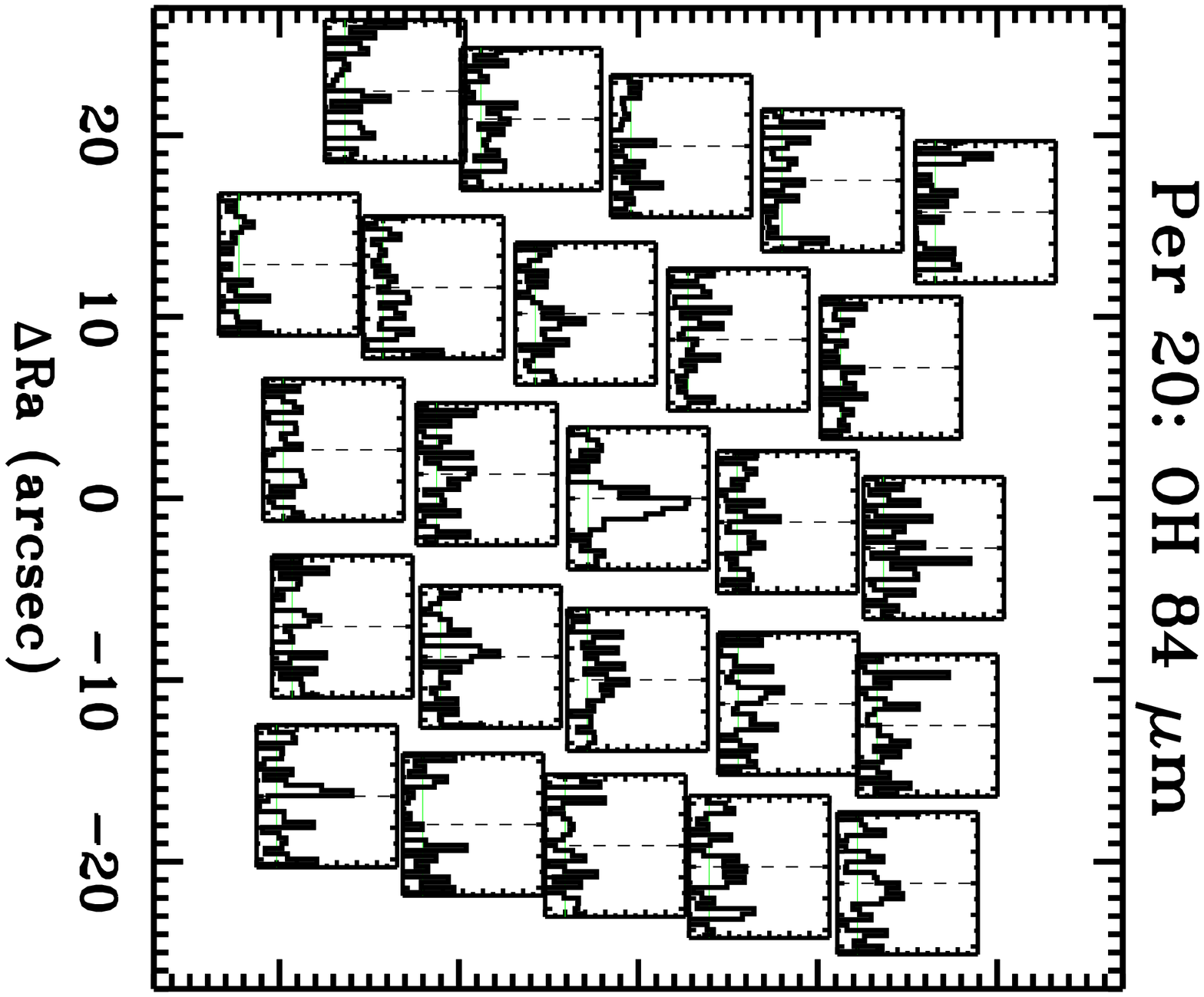}
                     \vspace{+0.2cm}
  \end{minipage}
    \hfill
  \begin{minipage}[t]{.3\textwidth}
  \hspace{-0.6cm}
     \includegraphics[angle=90,height=6.3cm]{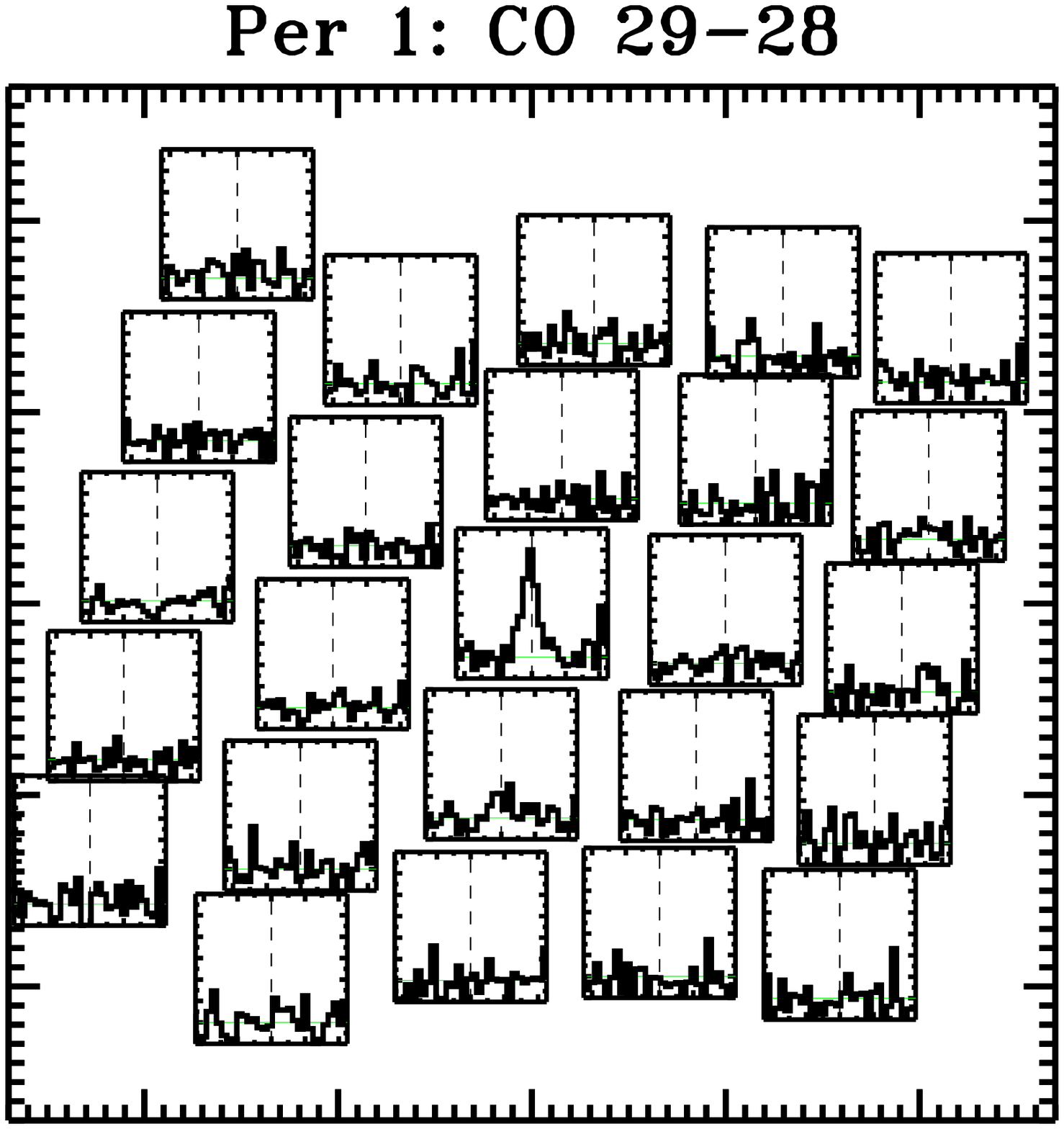}
          \vspace{-0.65cm}
          
          \hspace{-0.6cm}
     \includegraphics[angle=90,height=6.3cm]{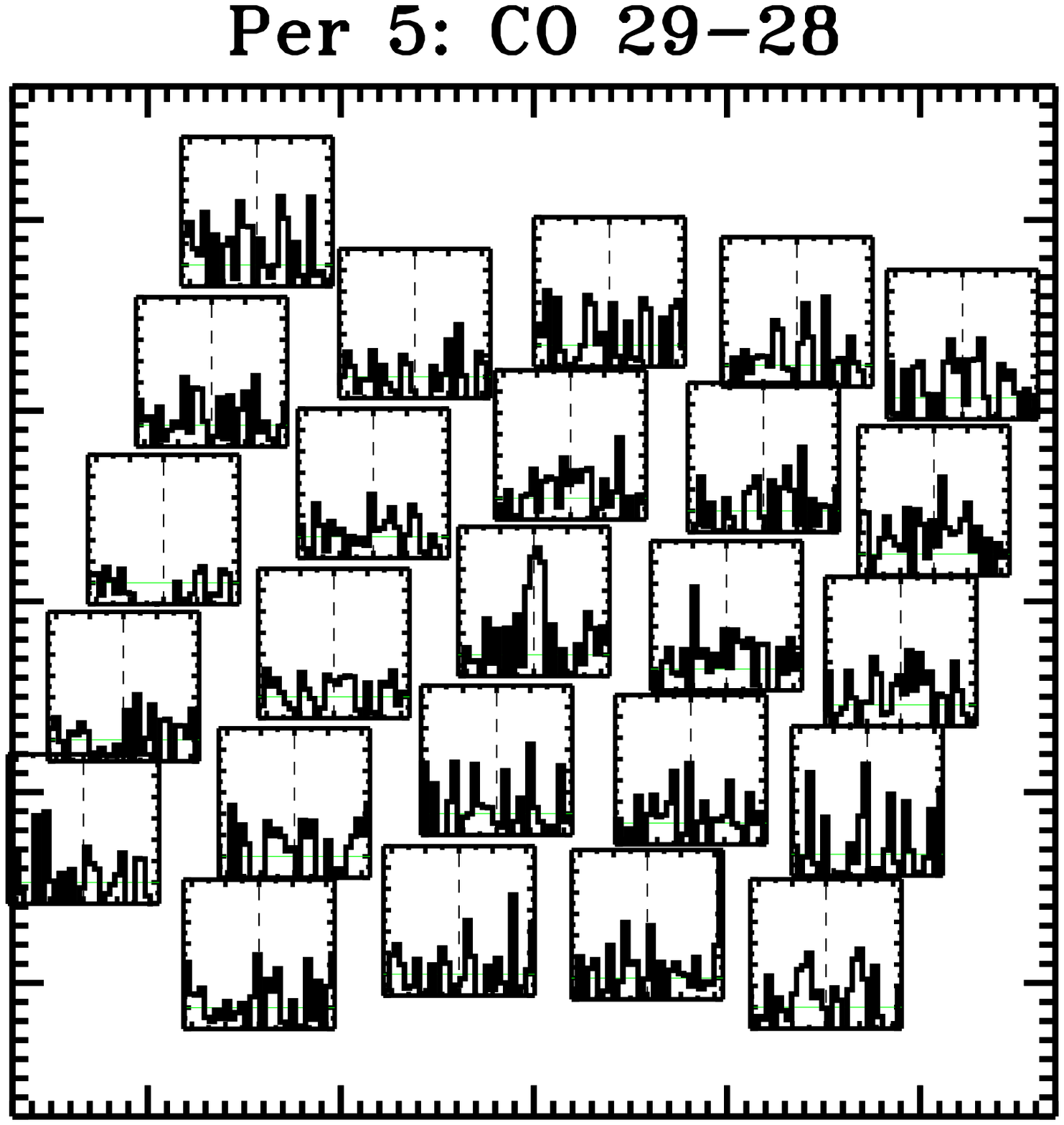}
               \vspace{-0.65cm}
               
       \hspace{-0.6cm}
     \includegraphics[angle=90,height=6.3cm]{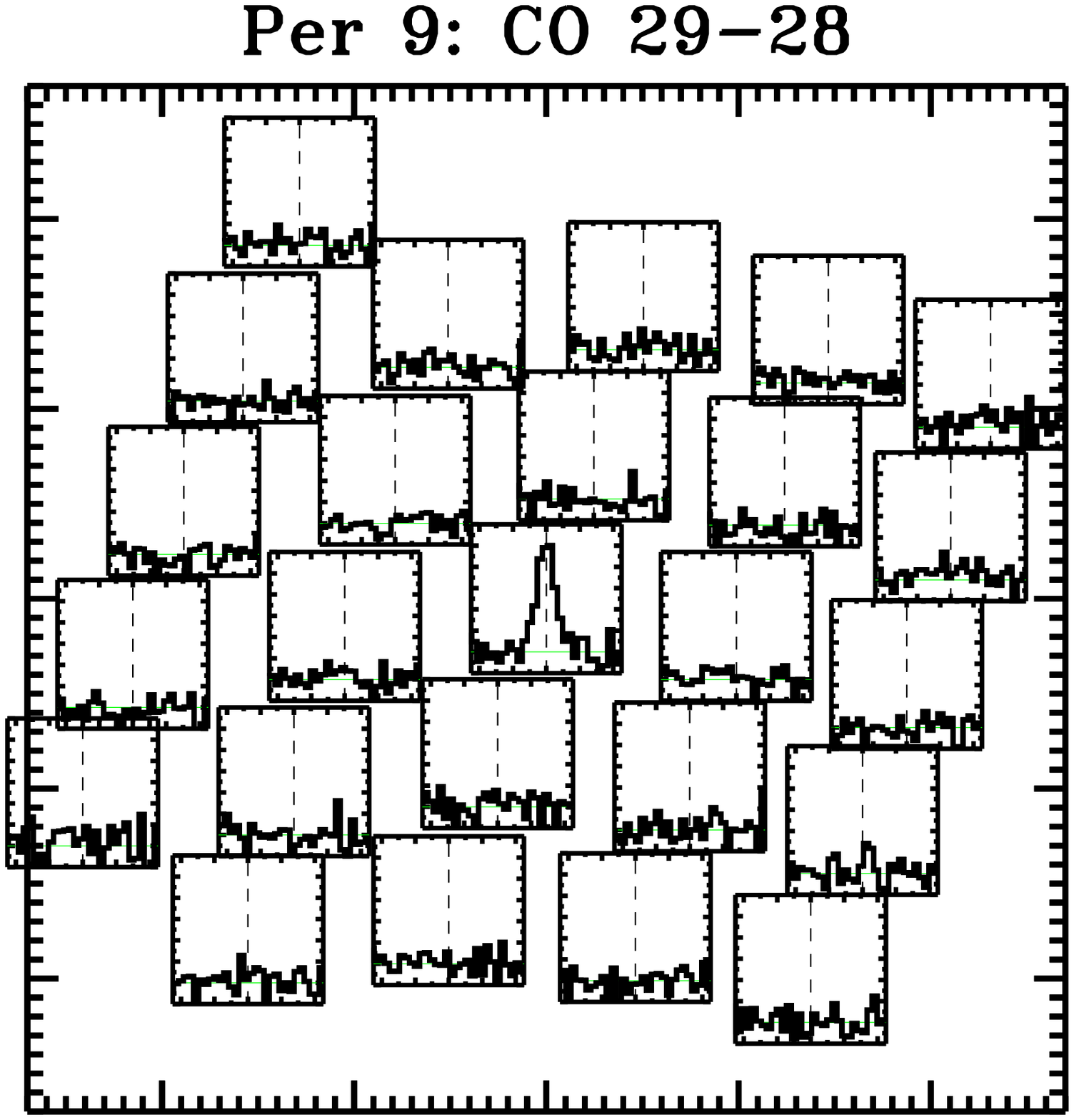}
                   \vspace{-0.65cm}
                   
     \hspace{-0.6cm}
       \includegraphics[angle=90,height=6.3cm]{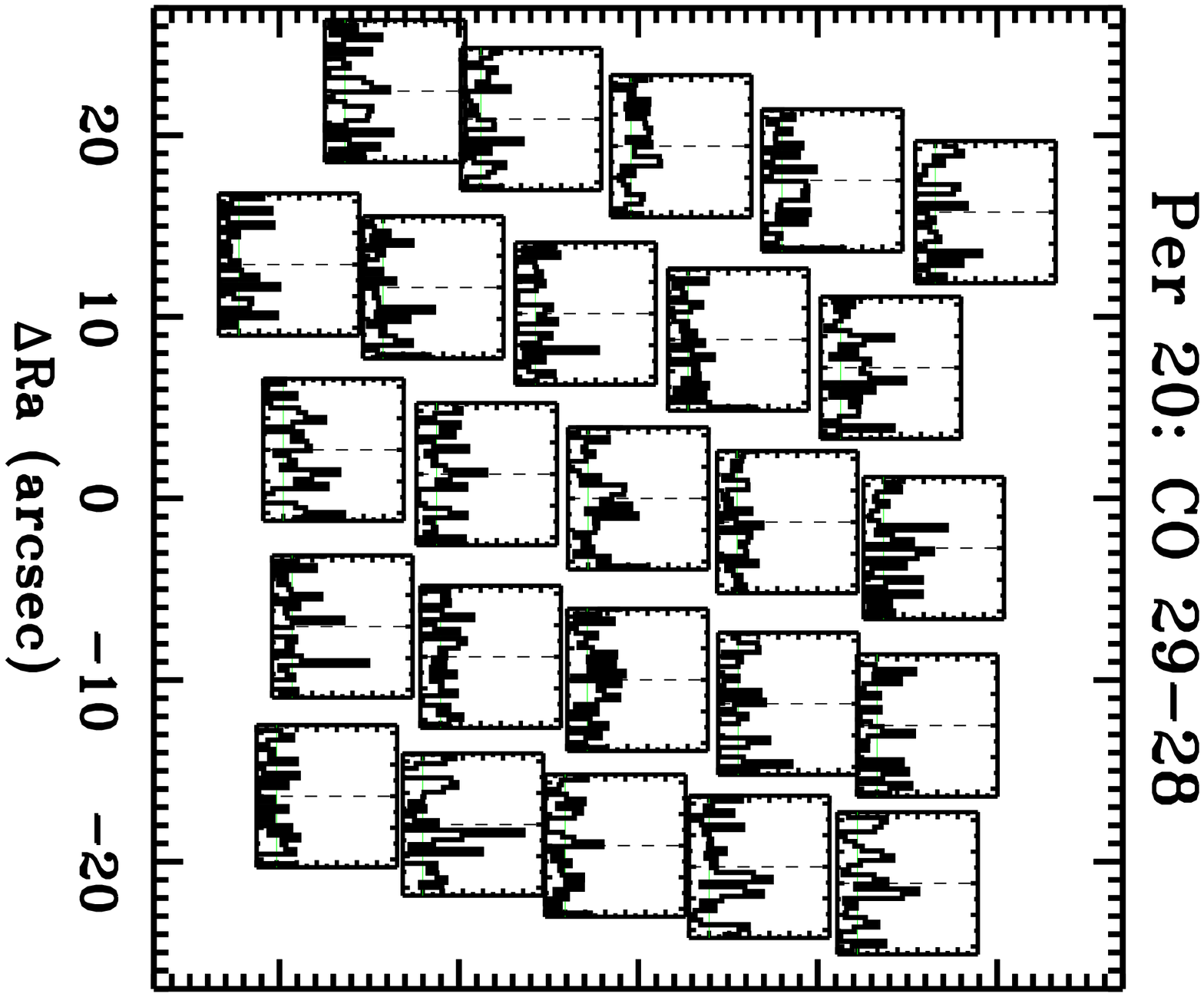}
                     \vspace{+0.2cm}
  \end{minipage} 
        \caption{\label{specmapmix} PACS spectral maps in the H$_2$O 4$_{23}$-3$_{12}$ 
        line at 78.74 $\mu$m, the OH line at 84.6 $\mu$m, and the CO 29-28 line at 90.16 $\mu$m for a few 
        sources with bright line emission and centrally peaked continuum. 
 The x-axis range covers -450 to 450 km\,s$^{-1}$ and the y-axis range covers -0.2 to 1.2, 
 with line emission normalized to the emission in the brightest spaxel.}
\end{figure*}

Figure \ref{specmapmix} show maps in the H$_2$O 4$_{23}$-3$_{12}$ 
        line at 78.74 $\mu$m, OH 84.6 $\mu$m, and CO 29-28 at 90.16 $\mu$m for Per1, Per5, Per9, 
        and Per20, all of which show bright line emission and centrally peaked continuum. 
        The lines are chosen to be located close in the wavelength so that the variations in the 
        PSF does not introduce significant changes in the emission extent.
        
Table \ref{tab:ratios} summarizes observed and modeled line ratios used in the Analysis section.

\onecolumn
\begin{landscape}
\begin{table*}
\centering 
\caption{\label{det} Line detections toward our Perseus sources}              
\renewcommand{\footnoterule}{}  
\begin{tabular}{lcccccccccccccccccccccccccccccccccccc}     
\hline\hline       
Wave ($\mu$m) & Species & Per 1 & 2 & 3 & 4 & 5 & 6 & 7 & 8 & 9 & 10 & 11 & 12 & 13 & 14 & 15 & 16 & 17 & 18 & 19 & 20 & 21 & 22 \\
\hline    
 63.184 & {[\ion{O}{i}]} &  $\checkmark$ & $\checkmark$ & $\checkmark$ & $\checkmark$  & $\checkmark$ & $\checkmark$ &  $\checkmark$   & $\checkmark$ & $\checkmark$ & $\checkmark$ & $\checkmark$ &  $\checkmark$    &      --      & $\checkmark$ & $\checkmark$ & $\checkmark$ &  $\checkmark$    & $\checkmark$ & $\checkmark$ & $\checkmark$ & $\checkmark$ & $\checkmark$ \\
 63.324 &    H$_2$O      &  $\checkmark$ & $\checkmark$ & $\checkmark$ &      --      &      --      & $\checkmark$ &      --      & $\checkmark$ & $\checkmark$ &      --      & $\checkmark$ &      --      &      --      & $\checkmark$ &      --      &      --      &      --      &      --      &      --      &      --      &      --      &      --      \\ 
 63.458 &    H$_2$O      &        --      &      --      & $\checkmark$ &      --      &      --      &      --      &      --      & $\checkmark$ & $\checkmark$ &      --      & $\checkmark$ &      --      &      --      &      --      &      --      &      --      &      --      &      --      &      --      &      --      &      --      &      --      \\
 78.742 &    H$_2$O      &   $\checkmark$ & $\checkmark$ & $\checkmark$ &      --      & $\checkmark$ & $\checkmark$ &      --      & $\checkmark$ & $\checkmark$ &      --      & $\checkmark$ &      --      &      --      & $\checkmark$ &      --      &      --      &      --      &      --      &      --      & $\checkmark$ & $\checkmark$ & $\checkmark$ \\
 78.928 &    H$_2$O      &   $\checkmark$ & $\checkmark$ & $\checkmark$ &      --      & $\checkmark$ &      --      &      --      &      --      & $\checkmark$ &      --      & $\checkmark$ &      --      &      --      &      --      &      --      &      --      &      --      &      --      &      --      &      --      &      --      &      --     \\
 79.120 &    OH          &   $\checkmark$ & $\checkmark$ & $\checkmark$ &      --      & $\checkmark$ & $\checkmark$ &      --      & $\checkmark$ & $\checkmark$ & $\checkmark$ & $\checkmark$ &      --      &      --      & $\checkmark$ &      --      &      --      &      --      &      --      & $\checkmark$ & $\checkmark$ & $\checkmark$ & $\checkmark$\\
 79.180 &    OH          &   $\checkmark$ & $\checkmark$ & $\checkmark$ &      --      & $\checkmark$ & $\checkmark$ &      --      & $\checkmark$ & $\checkmark$ & $\checkmark$ & $\checkmark$ &      --      &      --      & $\checkmark$ &      --      &      --      &      --      &      --      & $\checkmark$ & $\checkmark$ & $\checkmark$ & $\checkmark$\\
 79.360 &    CO          &   $\checkmark$ & $\checkmark$ & $\checkmark$ &      --      & $\checkmark$ &      --      &      --      &      --      & $\checkmark$ &      --      & $\checkmark$ &      --      &      --      & $\checkmark$ &      --      &      --      &      --      &      --      &      --      &      --      &      --      &      --     \\
 81.806 &    CO          &   $\checkmark$ & $\checkmark$ & $\checkmark$ &      --      &      --      & $\checkmark$ &      --      & $\checkmark$ & $\checkmark$ &      --      & $\checkmark$ &      --      &      --      & $\checkmark$ &      --      &      --      &      --      &      --      &      --      &      --      &      --      &      --     \\
 82.032 &    H$_2$O      &   $\checkmark$ & $\checkmark$ & $\checkmark$ &      --      & $\checkmark$ & $\checkmark$ &      --      & $\checkmark$ & $\checkmark$ &      --      & $\checkmark$ &      --      &      --      & $\checkmark$ & $\checkmark$ &      --      &      --      &      --      & $\checkmark$ &      --      & $\checkmark$ & $\checkmark$\\
 84.411 &    OH+CO       &   $\checkmark$ & $\checkmark$ & $\checkmark$ &      --      & $\checkmark$ & $\checkmark$ &      --      & $\checkmark$ & $\checkmark$ & $\checkmark$ & $\checkmark$ &      --      &      --      & $\checkmark$ & $\checkmark$ &      --      &      --      &      --      & $\checkmark$ & $\checkmark$ & $\checkmark$ & $\checkmark$\\
 84.600 &    OH          &   $\checkmark$ & $\checkmark$ & $\checkmark$ &      --      & $\checkmark$ & $\checkmark$ &      --      & $\checkmark$ & $\checkmark$ & $\checkmark$ & $\checkmark$ &      --      &      --      & $\checkmark$ & $\checkmark$ &      --      &      --      &      --      & $\checkmark$ & $\checkmark$ & $\checkmark$ & $\checkmark$\\
 84.767 &    H$_2$O      &        --      &      --      & $\checkmark$ &      --      &      --      &      --      &      --      &      --      &      --      &      --      &      --      &      --      &      --      &      --      &      --      &      --      &      --      &      --      &      --      &      --      &      --      &      --     \\
 89.988 &    H$_2$O      &   $\checkmark$ & $\checkmark$ & $\checkmark$ &      --      & $\checkmark$ &      --      &      --      & $\checkmark$ & $\checkmark$ &      --      & $\checkmark$ &      --      &      --      & $\checkmark$ &      --      &      --      &      --      &      --      &      --      &      --      & $\checkmark$ &      --     \\
 90.163 &    CO          &   $\checkmark$ & $\checkmark$ & $\checkmark$ &      --      & $\checkmark$ &      --      &      --      & $\checkmark$ & $\checkmark$ & $\checkmark$ & $\checkmark$ &      --      &      --      & $\checkmark$ &      --      &      --      &      --      &      --      &      --      &      --      & $\checkmark$ &      --     \\
108.073 &    H$_2$O      &   $\checkmark$ & $\checkmark$ & $\checkmark$ &      --      & $\checkmark$ & $\checkmark$ &      --      & $\checkmark$ & $\checkmark$ & $\checkmark$ & $\checkmark$ &      --      & $\checkmark$ & $\checkmark$ & $\checkmark$ &      --      &      --      & $\checkmark$ & $\checkmark$ & $\checkmark$ & $\checkmark$ & $\checkmark$\\
108.763 &    CO          &   $\checkmark$ & $\checkmark$ & $\checkmark$ &      --      & $\checkmark$ & $\checkmark$ &      --      & $\checkmark$ & $\checkmark$ & $\checkmark$ & $\checkmark$ &      --      &      --      & $\checkmark$ & $\checkmark$ &      --      &      --      & $\checkmark$ & $\checkmark$ & $\checkmark$ & $\checkmark$ & $\checkmark$\\
124.193 &    CO          &   $\checkmark$ & $\checkmark$ & $\checkmark$ &      --      & $\checkmark$ & $\checkmark$ &      --      & $\checkmark$ & $\checkmark$ & $\checkmark$ & $\checkmark$ &      --      &      --      & $\checkmark$ & $\checkmark$ & $\checkmark$ &      --      & $\checkmark$ & $\checkmark$ & $\checkmark$ & $\checkmark$ & $\checkmark$\\
125.354 &    H$_2$O      &   $\checkmark$ & $\checkmark$ & $\checkmark$ &      --      & $\checkmark$ & $\checkmark$ &      --      & $\checkmark$ & $\checkmark$ & $\checkmark$ & $\checkmark$ &      --      & $\checkmark$ & $\checkmark$ & $\checkmark$ &      --      &      --      &      --      & $\checkmark$ & $\checkmark$ & $\checkmark$ & $\checkmark$\\
162.812 &    CO          &   $\checkmark$ & $\checkmark$ & $\checkmark$ &      --      & $\checkmark$ & $\checkmark$ &      --      & $\checkmark$ & $\checkmark$ & $\checkmark$ & $\checkmark$ &      --      &      --      & $\checkmark$ & $\checkmark$ & $\checkmark$ &      --      & $\checkmark$ & $\checkmark$ & $\checkmark$ & $\checkmark$ & $\checkmark$\\
163.120 &    OH          &   $\checkmark$ & $\checkmark$ & $\checkmark$ &      --      &      --      & $\checkmark$ &      --      & $\checkmark$ & $\checkmark$ & $\checkmark$ & $\checkmark$ &      --      &      --      & $\checkmark$ &      --      &      --      &      --      & $\checkmark$ & $\checkmark$ & $\checkmark$ &      --      & $\checkmark$\\
163.400 &    OH          &   $\checkmark$ & $\checkmark$ & $\checkmark$ &      --      &      --      & $\checkmark$ &      --      & $\checkmark$ & $\checkmark$ & $\checkmark$ & $\checkmark$ &      --      &      --      & $\checkmark$ & $\checkmark$ &      --      &      --      & $\checkmark$ & $\checkmark$ & $\checkmark$ & $\checkmark$ &      --     \\
179.527 &    H$_2$O      &   $\checkmark$ & $\checkmark$ & $\checkmark$ &      --      & $\checkmark$ & $\checkmark$ &      --      & $\checkmark$ & $\checkmark$ & $\checkmark$ & $\checkmark$ &      --      &      --      & $\checkmark$ & $\checkmark$ &      --      &      --      & $\checkmark$ & $\checkmark$ & $\checkmark$ & $\checkmark$ & $\checkmark$\\
180.488 &    H$_2$O      &   $\checkmark$ & $\checkmark$ & $\checkmark$ &      --      & $\checkmark$ &      --      &      --      &      --      & $\checkmark$ &      --      & $\checkmark$ &      --      &      --      & $\checkmark$ &      --      &      --      &      --      & $\checkmark$ & $\checkmark$ & $\checkmark$ & $\checkmark$ & $\checkmark$\\
\hline                  
\end{tabular}
\end{table*}
\end{landscape}
 
\begin{table*}
\centering
\caption{Observed and model line ratios based on Kaufman \& Neufeld (1996)}
\label{tab:ratios}
\renewcommand{\footnoterule}{}  
\begin{tabular}{|ll|ccccc|ccc|ccc|ccc|}
\hline \hline
Line 1  & Line 2 & \multicolumn{5}{c}{Observed line ratios} &  \multicolumn{9}{c}{Models - $C$ shock (KN96)} \\
~ & ~ & \multicolumn{5}{c}{Line 1 / Line 2} & \multicolumn{3}{c}{log $n_\mathrm{H}$=4} & \multicolumn{3}{c}{log $n_\mathrm{H}$=5} & \multicolumn{3}{c}{log $n_\mathrm{H}$=6} \\
\hline
~ & ~ & N & Min & Max & Mean & Std & $\varv$=20 & 30 & 40 & 20 & 30 & 40 & 20 & 30 & 40 \\
\hline
\hline
CO 16-15 & 21-20 & 17 & 1.2 & 2.5 & 1.7 & 0.4 & 4.4 & 2.9 & 2.4 & 2.7 & 1.4 & 1.0 & 0.8 & 0.5 & 0.5 \\
CO 16-15 & 24-23 & 16 & 1.2 & 4.6 & 2.3 & 0.9 & 11.6 & 5.7 & 4.2 & 6.2 & 2.1 & 1.3 & 0.9 & 0.5 & 0.4 \\
CO 16-15 & 29-28 & 10 & 1.9 & 8.3 & 3.8 & 1.8 & 72.2 & 31.4 & 2.2 & 19.1 & 5.5 & 0.5 & 11.3 & 2.7 & 0.3 \\
CO 21-20 & 29-28 & 10 & 1.4 & 4.1 & 2.2 & 0.8 & 16.5 & 11.7 & 2.9 & 6.6 & 4.1 & 0.9 & 4.6 & 2.7 & 0.6 \\
\hline
H$_2$O 2$_{12}$-1$_{01}$ & 4$_{04}$-3$_{13}$ & 15 & 1.3 & 6.3 & 3.0 & 1.6 & 6.7 & 5.3 & 4.6 & 1.5 & 1.3 & 1.2 & 0.7 & 0.7 & 0.7 \\
H$_2$O 2$_{12}$-1$_{01}$ & 6$_{16}$-5$_{05}$ & 13 & 0.7 & 5.9 & 2.4 & 1.8 & 8.1 & 4.7 & 3.8 & 1.3 & 0.8 & 0.7 & 0.3 & 0.2 & 0.2 \\
H$_2$O 2$_{21}$-1$_{10}$ & 4$_{04}$-3$_{13}$ & 16 & 1.4 & 5.5 & 2.8 & 1.1 & 4.2 & 3.6 & 3.3 & 2.3 & 2.0 & 1.8 & 1.5 & 1.3 & 1.3 \\
H$_2$O 2$_{21}$-1$_{10}$ & 6$_{16}$-5$_{05}$ & 13 & 0.9 & 5.5 & 2.1 & 1.3 & 5.0 & 3.2 & 2.7 & 2.0 & 1.2 & 1.0 & 0.6 & 0.4 & 0.4 \\
\hline
OH 84 & OH 79 & 14 & 1.1 & 2.4 & 1.7 & 0.3 & 0.6 & 0.7 & 0.8 & 0.8 & 0.9 & 0.9 & 1.4 & 1.3 & 1.1 \\
\hline
\hline
H$_2$O 2$_{12}$-1$_{01}$ & CO 16-15 & 16 & 0.2 & 2.4 & 0.9 & 0.8 & 18.8 & 10.6 & 8.0 & 0.4 & 0.4 & 0.4 & 0.1 & 0.1 & 0.1 \\
H$_2$O 2$_{21}$-1$_{10}$ & CO 24-23 & 16 & 0.6 & 3.8 & 1.7 & 1.0  & 135.7 & 41.1 & 24.2 & 22.5 & 7.0 & 4.8 & 3.0 & 2.2 & 2.5 \\
H$_2$O 3$_{22}$-2$_{11}$ & CO 29-28 & 9 & 0.7 & 2.1 & 1.2 & 0.4 & 51.9 & 11.6 & 6.7 & 45.3 & 8.3 & 4.7 & 6.6 & 2.2 & 2.0 \\
H$_2$O 4$_{04}$-3$_{13}$ & CO 16-15 & 15 & 0.1 & 0.5 & 0.3 & 0.1 & 2.8 & 2.0 & 1.7 & 1.6 & 1.7 & 1.9 & 2.2 & 3.6 & 5.4 \\
H$_2$O 4$_{04}$-3$_{13}$ & CO 21-20 & 15 & 0.2 & 0.9 & 0.5 & 0.2 & 12.3 & 5.8 & 4.2 & 4.2 & 2.3 & 2.0 & 1.7 & 2.0 & 2.6 \\
H$_2$O 6$_{16}$-5$_{05}$ & CO 16-15 & 13 & 0.1 & 1.0 & 0.5 & 0.3 & 2.3 & 2.2 & 2.1 & 0.05 & 0.07 & 0.1 & 0.01 & 0.02 & 0.04 \\
H$_2$O 6$_{16}$-5$_{05}$ & CO 21-20 & 13 & 0.2 & 1.4 & 0.7 & 0.4 & 10.2 & 6.5 & 5.1 & 0.1 & 0.1 & 0.1 & 0.01 & 0.01 & 0.02 \\
H$_2$O 6$_{16}$-5$_{05}$ & CO 24-23 & 13 & 0.4 & 1.6 & 0.9 & 0.5 & 27.0 & 12.7 & 8.8 & 11.2 & 5.7 & 4.7 & 5.2 & 5.0 & 6.4 \\
H$_2$O 6$_{16}$-5$_{05}$ & CO 32-31 & 8 & 1.2 & 4.6 & 3.0 & 1.1 & 576.3 & 91.6 & 43.3 & 165.4 & 28.1 & 15.6 & 28.6 & 6.6 & 5.5 \\
\hline
H$_2$O 2$_{12}$-1$_{01}$ & OH 84 & 15 & 0.1 & 4.2 & 1.2 & 1.3 & 182.2 & 422.8 & 429.4 & 72.5 & 140.7 & 98.6 & 15.4 & 37.0 & 21.2 \\
H$_2$O 3$_{22}$-2$_{11}$ & OH 84 & 9 & 0.2 & 0.8 & 0.4 & 0.2 & 7.0 & 24.2 & 32.4 & 43.9 & 97.3 & 71.2 & 28.8 & 65.3 & 36.5 \\
\hline
CO 16-15 & OH 84 & 15 & 0.4 & 2.8 & 1.1 & 0.7 & 9.7 & 40.0 & 54.0 & 30.4 & 64.6 & 41.5 & 9.6 & 14.2 & 5.5 \\
CO 24-23 & OH 84 & 15 & 0.2 & 1.1 & 0.5 & 0.3 & 0.8 & 7.0 & 12.7 & 4.9 & 30.3 & 30.8 & 10.6 & 30.8 & 15.1 \\
\hline
\end{tabular}
\tablefoot{Number of YSOs with detections of the two lines in the ratio (N), minimum (Min) and maximum (Max) value of the ratio,
mean ratio value (Mean) and standard deviation (Std) is given for the observed line ratios. Model line ratios are calculated for 
three values of pre-shock densities, $n_\mathrm{H}$=10$^4$, 10$^5$, and 10$^6$ cm$^{-3}$, 
and three values of shock velocities, v=20, 30, and 40 km\,s$^{-1}$. See Table \ref{lines} for the full identifiers of the lines.}
\end{table*}
\twocolumn
\section{Correlations with source parameters}
\begin{figure*}[!tb]
\begin{center}
\includegraphics[angle=90,height=9cm]{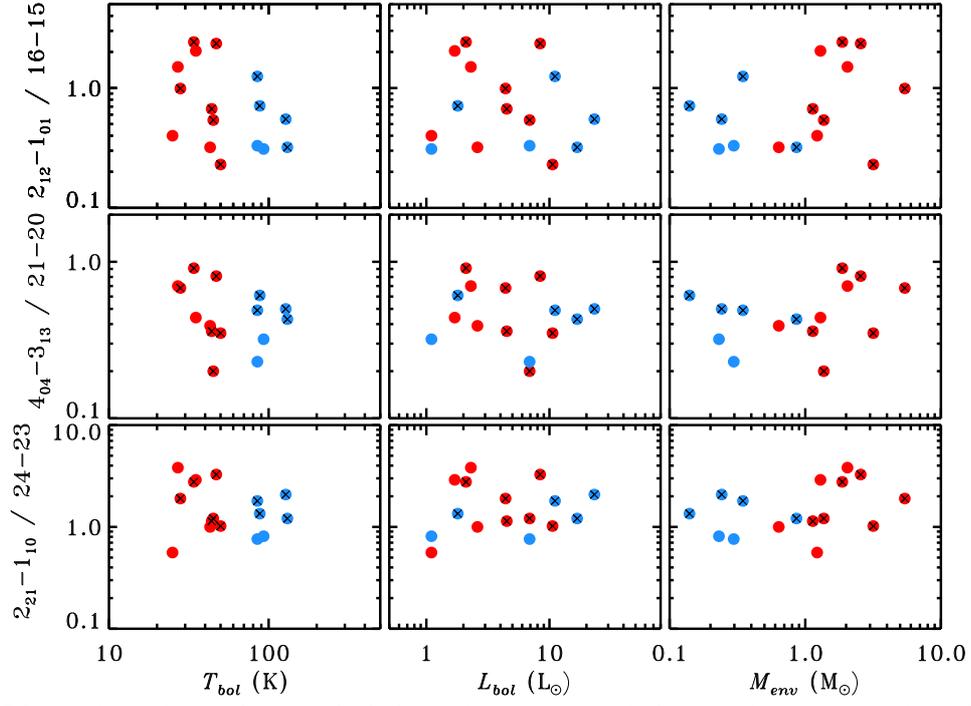}
\caption{\label{corr125} H$_\mathrm{2}$O-to-CO line ratios 
and, from left to right, bolometric temperature, bolometric luminosity, and envelope mass.
The ratios are calculated using lines at nearby wavelengths for which the same, limited to the central 
spaxel, extraction region is considered. Shown are line ratios of the H$_\mathrm{2}$O 2$_{12}$-1$_{01}$ line 
at 179 $\mu$m and CO 16-15 at 163 $\mu$m (top), the H$_2$O 4$_{04}$-3$_{13}$ 
and CO 21-20 lines at about 125 $\mu$m (middle), and the H$_2$O 2$_{21}$-1$_{10}$ and 
CO 24-23 lines at about 108 $\mu$m (bottom). Red and blue symbols correspond to 
sources with $T_\mathrm{bol}\leq70$ and $>70$ K, respectively. Black crosses are drawn for the 
sources with detections of CO 29-28 line.}
\end{center}
\end{figure*}

Figure~\ref{corr125} shows selected H$_2$O-to-CO line ratios as a function of source physical parameters 
(bolometric luminosity, temperature, and envelope mass). Lack of correlation is seen in all cases. 
\end{document}